\documentclass[3p,times]{elsarticle}

\usepackage{mathrsfs}

\usepackage{amssymb}

 \usepackage{graphicx}
 \usepackage{epstopdf}
 \usepackage{epsfig}
 \usepackage{booktabs}
 \usepackage{lscape} 
\usepackage{dashrule}

\usepackage[figuresright]{rotating}
\usepackage{amsmath,amssymb,amsthm,color}
\usepackage[pdftex, colorlinks]{hyperref}

\newcommand{\rmu}{{\rm u}}
\newcommand{\rmx}{{\rm x}}
\newcommand{\Ref}{{\rm ref}}
\newcommand{\rmeq}{{\rm eq}}
\newcommand{\beq}{{\rm(eq)}}
\newcommand{\rmd}{{\rm d}}
\newcommand{\rmi}{{\rm i}}

\newcommand{\Real}{{\rm Re}}
\newcommand{\Imag}{{\rm Im}}

\newcommand{\sbracket}[1]{{\left(#1\right)}}

\newcommand{\lbracket}[1]{{\left\{#1\right\}}}
\newcommand{\slbracket}[1]{{\left\{#1\right.}}

\begin{document}

\begin{frontmatter}


\title{Analysis of the absorbing layers for the weakly-compressible lattice Boltzmann schemes}
 \author{Hui Xu\corref{cor1}}%
 \ead{xuhuixj@gmail.com or xu@lmm.jussieu.fr}
\cortext[cor1]{Corresponding author. }
 \author{Pierre Sagaut}
 \ead{sagaut@lmm.jussieu.fr}
\address{Institut Jean le Rond d'Alembert, UMR CNRS 7190, Universit\'e Pierre et Marie Curie - Paris 6, 4 Place Jussieu case 162 Tour 55-65, 75252 Paris
Cedex 05, France}





\begin{abstract}
It has been demonstrated that Lattice Boltzmann schemes (LBSs) are very efficient for  Computational AeroAcoustics (CAA). In order to handle the issue of absorbing acoustic boundary conditions for LBS,   three kinds of damping terms are  proposed and added into the right hand sides of the governing equations of LBS. From the classical theory, these  terms play an important role to absorb and minimize the acoustic wave reflections from computational boundaries. Meanwhile, the corresponding macroscopic equations with the damping terms are recovered for analyzing the macroscopic behaviors of the these damping terms and determining the critical absorbing strength.  Further, in order to detect the dissipation and dispersion behaviors , the linearized LBS with the damping terms is derived and analyzed. The  dispersive and dissipative properties are explored in the wave-number spaces via the Von Neumann analysis. The related damping strength critical values and the optimal absorbing term are addressed. Finally, some benchmark problems are implemented to assess the theoretical results.
\end{abstract}


\end{frontmatter}


\section{Introduction}\label{Intro}
In the past decades,  there has been an increasing highlighted interest in developing and applying the LBS for aeroacoustic applications \cite{xusagaut,duboislallemand,buickgreatedcampbell,simondenispierre,
ricotmariesagautbailly,buickbuckleygreated}, because the LBS has been developed into an innovative mescoscopic numerical method for the computational modeling of a wide variety of complex fluid flows and acoustics \cite{xusagaut,duboislallemand,buickgreatedcampbell,simondenispierre,
ricotmariesagautbailly,buickbuckleygreated,chendoolen,lallemandluo,xu2,xu3}. In the field of the aeroacoustic applications, it has been demonstrated that the LBS possesses the low dispersion and low dissipation properties for capturing the weak acoustic pressure fluctuations \cite{buickgreatedcampbell,simondenispierre,ricotmariesagautbailly,buickbuckleygreated}. 
Recently, the MRT-LBM (multi-relaxation time lattice Boltzmann method \cite{lallemandluo,dhumieres2}) has been improved for the acoustic application with the nice dissipation and dispersion relations. The improvement of the MRT-LBM overcomes the drawback of the large bulk dissipation \cite{xusagaut}. The improvement has made the MRT-LBM appearing as a very well-suited method  for  acoustic simulations. 

In the simulations of the realistic flows, artificial computational boundaries are defined around the flow region of physical interest. The space outside the computational domain is neglected.  In some cases, the region of interest should extend to infinity, especially for the computational aeroacoustics \cite{colonius}. Because of the use of a finite computational domain and artificial boundary conditions (e.g. Dirichlet, Neumann or Robin types) the outcoming acoustic waves may be  reflected. The reflected waves often have a significant influence on the flow field and may overwhelm the physical acoustic waves \cite{colonius,bodony,orszag}.  The obvious ways to suppress the reflected waves are to introduce artificial dissipation (by upwinding) or  to increase the value of physical viscosity (or add hyperviscosity) \cite{colonius}. The simplest way is to define   sponge/absorbing layers is to add an additional penalty term  to the governing equations to suppress the computed solution and  to match a prescribed or precomputed solution \cite{bodony,orszag,freund}.   The most used and efficient  forcing term is given by \cite{orszag}
\begin{equation}\label{eq:sponge}
-\chi(\rmx)(q-q_\Ref),
\end{equation} 
where $q$ denotes the physical variables under consideration. Physically, the term (\ref{eq:sponge}) plays the role of a linear friction. Generally, the parameter $\chi$  
, called the sponge strength, is a function that  depends on  space only and $q_\Ref$  is a function of space and time (especially, $q_\Ref$ denotes any variable of the far fields or time-dependent mean value). In most cases $\chi$ will vary smoothly between zero in the physical domain and a positive value in the absorbing laryer \cite{bodony,orszag,freund}. Provided $\chi$ is large enough, in the classical theory of absorbing layers, the outgoing disburbances are exponentially attenuated when crossing the layer\cite{colonius,orszag,freund,sagautbook}. The numerical solutions inside absorbing layers need not be physical as long as the use of the region does not introduce significant reflection back into the physical domain. Meanwhile, the absorbing layer is numerically stable. Compared with perfectly matched layers, the absorbing layer techniques can be applied to a wider class of problems and are often coupled with the characteristic boundary condition \cite{sagautbook}.
 
The definition absorbing layers for LBS is an emerging resarch topic. It must be emphasized that there is a clear need for theoretical researches on that topic, to improce the potential of LBS for CAA applications.  From  the macroscopic equations standpoint, we are interested in the macroscopic variables which are obtained by the mesoscopic distribution statistics. We  propose in the present paper several possible absorbing terms which are enforced on LBS. By the classical Chapman-Enskog multi-scale expansion, the macroscopic systems are recovered in the absorbing layer. With the aid of the macroscopic systems,  the related analysis is implemented in spectral space. The emphasis is put on  the coupling between  $\sigma(\rmx)$ and LBS relaxation parameters, especially on finding the critical value for $\sigma(\rmx)$.  The critical expressions of $\sigma(\rmx)$ will offer us a guide for choosing the reasonable absorbing strength. This is different from the classical absorbing layer theory in Navier-Stokes equations (NSEs).  In the classical absorbing layer theory, $\sigma$ can be regarded as a penalized parameters, and theoretically, $\sigma$ can be chosen as any positive real number. However, in order to avoid a stiff problem, $\sigma$ will not be chosen very large. In LBS, $\sigma$ is a finite parameter dependent on the relaxation parameters, as shown hereafter.   

In next section, the fundamentals of  LBS are reviewed and the possible absorbing terms are proposed.  In the third section, in the spectral space,  the dispersion and dissipation relations are analyzed with the aid of Von Neumann method considering monochromatic wave solutions. In the fourth section,  the numerical simulations are performed to validate the proposed  absorbing  terms considering some benchmark problems.

\section{Fundamentals of LBS and absorbing terms}\label{Method}

In this section, the fundamental theory of LBS is briefly reviewed. Then, we give the linearized LBM  problem coupled with absorbing terms and establish the relation between the classical BGK-LBM and the Navier-Stokes equations  in absorbing layers.

\subsection{Lattice Boltzmann schemes}\label{subMRTLBM}

The governing equations of  the lattice Boltzmann schemes are described by the following universal form \cite{lallemandluo,dhumieres2}

\begin{equation}\label{blbs}
f_i(\rmx+v_i\delta t,t+\delta
t)=f_i(\rmx,t)+\Lambda_{ij}\left(f_j^{\rm(eq)}(\rmx,t)-f_j(\rmx,t)\right),\quad
0\leq i,j\leq N,
\end{equation}
where $v_i$ belongs to the discrete velocity set $\mathcal{V} $,
$f_i(\rmx,t)$ is the discrete single particle distribution function
corresponding to $v_i$ and $f_i^{\rm(eq)}$ denotes the discrete
single particle equilibrium distribution function. Generally,  $f_i^{\rm(eq)}$ can be expressed by a combination of a linear part $f_i^{\rm(L,\ eq)}(\rmx,t)$ and a quadratic part $f_i^{\rm(Q,\ eq)}(\rmx,t)$ \cite{xu2}

\begin{equation}
f_i^{\rm(eq)}(\rmx,t)=f_i^{\rm(L,\ eq)}(\rmx,t)+f_i^{\rm(Q,\ eq)}(\rmx,t).
\end{equation}
$\delta t$
denotes the time step and $N+1$ is the number of discrete
velocities. $\Lambda_{ij}$ is the generalized relation matrix.  
From here on, the repeated index  indicates that the Einstein summation is used except for some special explanations. Let $\mathcal{L}\in \mathbb{R}^d $ ($d$ denotes the spatial dimension)
denotes the lattice system, and the following condition is required \cite{duboislallemand}

\begin{equation}
\rmx+v_j\delta t\in \mathcal{L},
\end{equation}
that is to say, if $\rmx$ is a node of the lattice, $\rmx+v_j\delta t$ is
necessarily another node of the lattice. Generally, for BGK-LBM, the relaxation matrix is given by

\begin{equation}\label{eq_BGK_LBM_R}
\Lambda_{ij}=s\delta_{ij},
\end{equation}
where $s$ denotes the relaxation frequency of BGK-LBM. If the relaxation matrix $\Lambda$  is defined by

\begin{equation}
\Lambda=M^{-1}SM,
\end{equation}
where $S$ is a diagonal matrix which denotes the relaxation
parameters of MRT-LBM, which is given by

\begin{equation}
S = {\rm diag}(\{\underbrace{0,0,0}_{d+1},\underbrace{s_{d+1},\ldots,s_{N}}_{N-d-1}\}).
\end{equation}
 $M=\left(M_{ij}\right)_{0\leq i\leq N,0\leq
i\leq N}$ is the transformation matrix (see \ref{app:t} for details of D2Q9 and D3Q15), which satisfies the
following basic conditions \cite{lallemandluo} 

\begin{equation}
M_{0j}=1,M_{\alpha j}=v_j^{\alpha},(1\leq\alpha\leq d).
\end{equation}
The macroscopic quantities are defined by
\cite{duboislallemand,lallemandluo}

\begin{equation}\label{q-macro}
m_i=M_{ij}f_j,\quad m_i^{\rm (eq)}=M_{ij}f_j^{\rm(eq)}.
\end{equation}
By the simple algebra, the standard isothermal MRT-LBM is recovered in the following form
\cite{duboislallemand,lallemandluo}

\begin{equation}\label{eq_conservative}
m_i=W_i=m_i^{\rm(eq)},0\leq i\leq d,
\end{equation}
and
\begin{equation}\label{eq_noconservative}
m_i(x+\delta t v_j,t+\delta
t)=m_i(x,t)+s_i\left(m_i^{\rm(eq)}(x,t)-m_i(x,t)\right),d+1\leq
i\leq N.
\end{equation}

It is necessary to point out that for isothermal flows, the number of the conservative quantities is equal to $d+1$. According to the work of Lallemand and Luo \cite{lallemandluo}, the
relaxation parameters in Eq. (\ref{eq_noconservative}) should
satisfy the following stability constraints

\begin{equation}
s_i\in (0,2),\ d+1\leq i\leq N.
\end{equation}

\subsection{Absorbing terms}\label{sec:abc}

Following the theory of Israeli \& Orszag \cite{orszag}, Eq. (\ref{blbs}) coupled with the absorbing terms has the following form

\begin{equation}\label{blbs-abc-1}
f_i(\rmx+v_i\delta t,t+\delta
t)=f_i(\rmx,t)+\Lambda_{ij}\left(f_j^{\rm(eq)}(\rmx,t)-f_j(\rmx,t)\right)+\delta t\Gamma_{ij}\left({f}_j^{(\Ref)}(\rmx,t)-f_j^*(\rmx,t)\right),
\end{equation}
where $\Gamma_{ij}$ is the generalized absorbing strength defined by 

\begin{equation}\label{eq:abst}
\Gamma = M^{-1}\Sigma M, 
\end{equation}
and ${f}_i^{(\Ref)}$ denotes the reference  state of  $f_i$ and $f_j^*$ denotes the possible representations of mesoscopic distribution functions. From the classical theory, a natural choice of $f_j^{*}$ is $f_j$.  The matrix $\Sigma$ in the expression (\ref{eq:abst}) is defined by

\begin{equation}
\Sigma={\rm diag}\{\sigma_0,\ldots,\sigma_N\},
\end{equation}
where $\sigma_i$ is the absorbing coefficient for each  first-order moment of $f_i$.  In the expression (\ref{eq:abst}), the transformation matrix $M$ is given in   \ref{app:t}. The choice of the generalized absorbing strength expression (\ref{eq:abst}) is based on the ideal for the different modes, the different absorbing strength can be applied as indicated in \cite{bodony}.  For simplicity, we consider  $\sigma_0=\ldots=\sigma_N=\chi$,  and the matrix $\Gamma$ become the following diagonal form

\begin{equation} \label{eq:gamma}
\Gamma_{ij}=\delta_{ij}\chi.
\end{equation}
In this paper, the researches are focused on the absorbing strength with the form (\ref{eq:gamma}) .

Generally, the reference distribution function ${f}_j^{(\Ref)}(\rmx,t)$ is expressed by the time-averaged mean value \cite{sagautbook}.  Here, we define the expression of ${f}_j^{(\Ref)}(\rmx,t)$ by the equilibrium distribution function as follows

\begin{equation}\label{eq:reff}
{f}_j^{(\Ref)}(\rmx,t)={f}_j^{(\rmeq)}(\rho^{(\Ref)}(\rmx,t),\rmu^{(\Ref)}(\rmx,t),t),
\end{equation}
where $\rho^{(\Ref)}(\rmx,t)$ and $\rmu^{(\Ref)}(\rmx,t)$ denote the reference (or base) density and velocity, respectively. Especially, let $\rho^{(\Ref)}(\rmx,t)$ and $\rmu^{(\Ref)}(\rmx,t)$ denote the far fields. The influence of the absorbing terms on the macroscopic equations can be interpreted as the source terms of the macroscopic equations. Based on this consideration, the corresponding source terms can be obtained by the classical Chapman-Enskog procedure.

It is necessary to discuss the choices of $f_j^{\rm (ref)}$ and $f_j^*$. Intuitively, let $f_j^{\rm (ref)}$ and $f_j^*$  expressed by the equilibrium distribution functions. This consideration makes us easy to analysis the macroscopic behaviors of the absorbing layers. Meanwhile, from these choices, we only need to handle the macroscopic statistical quantities and reference quantities. Compared with handling the distribution functions, these choices are effective and simple.  Especially, provided you consider the time-dependent mean quantities as the reference state, the time-dependent mean of the macroscopic statistical quantities  is easier to be handled than the mesoscopic distribution functions from the view of saving memory and CPU cost.  Another more  careful consideration is that generally, for aeroacoustic problems, if the background flows or far fields are used, it is difficult to construct a suitable $f_j^{\rm (ref)}$.  If $f_j^{\rm (ref)}$ is defined by the equilibrium distribution functions,  this will lead to some potential instabilities when $f_j^*$ is defined by $f_j$. The analysis will be given in Sec. \ref{analysisabc}.  From above considerations, there exist three-kind simplest absorbing terms.

\subsubsection{Type I absorbing term}\label{sec:1tabc}
Combining Eq. (\ref{blbs-abc-1}) and Eq. (\ref{eq:reff}) and using the far field as the reference state, and considering $f_j^{(\rm ref)}(\rmx,t)=f_j^{(\rmeq)}(\rho^f,\rmu^f,t)$ and $f^*_j(\rmx,t)=f_j(\rmx,t)$ we have 

\begin{equation}\label{app:newbgka}
f_i(\rmx+v_i\delta t,t+\delta
t)=f_i(\rmx,t)+s^\prime\left(f_i^{\rm(eq)}(\rho^*,{\rmu^*},t)-f_i(\rmx,t)\right)+\delta t F_i(\rho^{f},{\rmu^f},\rho^*,{\rmu^*},t),
\end{equation}
where $s^\prime=s+\delta t\chi$, $F_i(\rho^{f},{\rmu^f},\rho^*,{\rmu^*},t)$ is defined by

\begin{equation}
F_i(\rho^{f},{\rmu^f},\rho^*,{\rmu^*},t)=\chi\left(f_i^{\rm(eq)}(\rho^{f},{\rmu^f},t)-f_i^{\rm(eq)}(\rho^*,{\rmu^*},t)\right).
\end{equation}
In Eq. (\ref{app:newbgka}), $\rho^*$ and $\rmu^*$ denote the density and velocity, which are defined by the following parameterized forms \cite{Guo}

\begin{equation}
\rho^*=\sum_{i}f_i+n\delta t\sum_{i}F_i,\quad \rho^*u^*_\alpha=\sum_{i}c_{i\alpha}f_i+m\delta t\sum_{i}c_{i\alpha}F_i.
\end{equation}
Then, we have the following macroscopic equations (see \ref{app:1} for details)

\begin{equation}\label{app:abs-eq1}
\slbracket{
\begin{array}{l}
\displaystyle\partial_{t}\rho^*+\partial_{\alpha}(\rho^* u^*_{\alpha}) =(1+n s^\prime)\chi(\rho^{f}-\rho^*)+\zeta_n\chi\partial_{t}(\rho^{f}-\rho^*)+\zeta_m\chi\partial_{\alpha}\left(\rho^{f}u^{f}_{\alpha}-\rho^{*}u^{*}_{\alpha}\right),\\[2mm] 
\displaystyle\partial_{t}(\rho^* u^*_{\alpha})-\partial_{\beta}(\nu^\prime\rho^*(\partial_{\alpha}u^*_{\beta}+\partial_{\beta}u^*_{\alpha}))+\partial_{\beta}\left(\rho^* u^*_\alpha u^*_\beta+p^*\delta_{\alpha\beta}\right) =(1+m s^\prime )\chi\left(\rho^{f}u^{f}_{\alpha}-\rho^{*}u^{*}_{\alpha}\right)+\\[2mm]
\displaystyle \zeta_m\chi\partial_{t}\left(\rho^{f}u^{f}_{\alpha}-\rho^{*}u^{*}_{\alpha}\right)-\left(\sigma+\frac{\delta t}{2}\right)\chi\partial_{\beta}\left(\rho^{f}u^{f}_{\alpha}u^{f}_{\beta}+p^{f}\delta_{\alpha\beta}-\rho^{*}u^{*}_{\alpha}u^{*}_{\beta}-p^{*}\delta_{\alpha\beta}\right),
\end{array}
}
\end{equation}
where $s^\prime$, $\nu^\prime$, $\sigma$, $\zeta_n$ and $\zeta_m$ are defined by

\begin{equation}
s^\prime=s+\delta t\chi, \nu^\prime=c_s^2\left(\frac{1}{s^\prime}-\frac{1}{2}\right)\delta t, \sigma=\left(\frac{1}{s^\prime}-\frac{1}{2}\right)\delta t,\zeta_n = n\delta t\left(1-\frac{s^\prime}{2}\right)-\frac{\delta t}{2},\ \zeta_m = m\delta t\left(1-\frac{s^\prime}{2}\right)-\frac{\delta t}{2}.
\end{equation}

According to the definition of the effective relaxation frequency $s^\prime$, it is known that for  Eq. (\ref{app:newbgka}),  an effective negative viscosity $\nu^\prime$ will appear,  if $\chi$ satisfies the following condition

\begin{equation}\label{effectrelax}
s^\prime=s+\delta t\chi> 2. 
\end{equation}

\subsubsection{Type II absorbing term}\label{sec:2tabc}
According to the deviation in Sec. \ref{sec:1tabc} and considering $f_j^{(\rm ref)}(\rmx,t)=f_j^{(\rmeq)}(\rho^f,\rmu^f,t)$ and $f^*_j(\rmx,t)=f_j^{(\rm eq)}(\rho^*,\rmu^*,t)$, we propse the second-type absorbing term for LBS

\begin{equation}\label{app:newbgka2}
f_i(\rmx+v_i\delta t,t+\delta
t)=f_i(\rmx,t)+s\left(f_i^{\rm(eq)}(\rho^*,{\rmu^*},t)-f_i(\rmx,t)\right)+\delta t F^{(\rmeq)}_i(\rho^{f},{\rmu^f},\rho^*,{\rmu^*},t),
\end{equation}
where $F_i(\rho^{f},{\rmu^f},\rho^*,{\rmu^*},t)$ is defined by

\begin{equation}\label{eq:F2}
F^{(\rmeq)}_i(\rho^{f},{\rmu^f},\rho^*,{\rmu^*},t)=\chi\left(f_i^{\rm(eq)}(\rho^{f},{\rmu^f},t)-f_i^{\rm(eq)}(\rho^*,{\rmu^*},t)\right),
\end{equation}
In Eq. (\ref{app:newbgka2}), $\rho^*$ and $\rmu^*$ denote the density and velocity, which are defined by the following parameterized forms \cite{Guo}

\begin{equation}
\rho^*=\sum_{i}f_i+n\delta t\sum_{i}F^{(\rmeq)}_i,\quad \rho^*u^*_\alpha=\sum_{i}c_{i\alpha}f_i+m\delta t\sum_{i}c_{i\alpha}F^{(\rmeq)}_i.
\end{equation}
\\
The corresponding macroscopic equation is similar to Eq. (\ref{app:abs-eq1}). It reads (see  \ref{app:1} for details )

\begin{equation}\label{app:abs-eq3}
\slbracket{
\begin{array}{l}
\displaystyle\partial_{t}\rho^*+\partial_{\alpha}(\rho^* u^*_{\alpha}) =(1+n s)\chi(\rho^{f}-\rho^*)+\zeta_n\chi\partial_{t}(\rho^{f}-\rho^*)+\zeta_m\chi\partial_{\alpha}\left(\rho^{f}u^{f}_{\alpha}-\rho^{*}u^{*}_{\alpha}\right),\\[2mm] 
\displaystyle\partial_{t}(\rho^* u^*_{\alpha})-\partial_{\beta}(\nu\rho^*(\partial_{\alpha}u^*_{\beta}+\partial_{\beta}u^*_{\alpha}))+\partial_{\beta}\left(\rho^* u^*_\alpha u^*_\beta+p^*\delta_{\alpha\beta}\right) =(1+m s )\chi\left(\rho^{f}u^{f}_{\alpha}-\rho^{*}u^{*}_{\alpha}\right)+\\[2mm]
\displaystyle \zeta_m\chi\partial_{t}\left(\rho^{f}u^{f}_{\alpha}-\rho^{*}u^{*}_{\alpha}\right)-\left(\sigma+\frac{\delta t}{2}\right)\chi\partial_{\beta}\left(\rho^{f}u^{f}_{\alpha}u^{f}_{\beta}+p^{f}\delta_{\alpha\beta}-\rho^{*}u^{*}_{\alpha}u^{*}_{\beta}-p^{*}\delta_{\alpha\beta}\right),
\end{array}
}
\end{equation}
\\
where $\sigma$, $\zeta_n$ and $\zeta_m$ are defined by

\begin{equation}
\nu=c_s^2\left(\frac{1}{s}-\frac{1}{2}\right)\delta t, \sigma=\left(\frac{1}{s}-\frac{1}{2}\right)\delta t,\zeta_n = n\delta t\left(1-\frac{s}{2}\right)-\frac{\delta t}{2},\ \zeta_m = m\delta t\left(1-\frac{s}{2}\right)-\frac{\delta t}{2}.
\end{equation}

The difference between Eq. (\ref{app:newbgka}) and Eq. (\ref{app:newbgka2}) lies in  the  relaxation frequencies $s^\prime$ and $s$. In Eq. (\ref{app:newbgka2}), the effective relaxation frequency $s$ is kept in the original form while, in Eq. (\ref{app:newbgka}), the original relaxation  was modified, possibily leading to a negative viscosity if $\delta t\chi$ is too large. 
This problem is now cured.

\subsubsection{Type III absorbing term}\label{sec:3tabc}
In the recovered macroscopic equations (\ref{app:abs-eq1}) and (\ref{app:abs-eq3}),  in the left hand side, there exist momentum-velocity coupled terms $\rho^*u_\alpha^* u_\beta^*$ in recovered equations. Considering $f_j^{(\rm ref)}(\rmx,t)=f_j^{({\rm L}, \rmeq)}(\rho^f,\rmu^f,t)$ and $f^*_j(\rmx,t)=f_j^{({\rm L},\rmeq)}(\rho^*,\rmu^*,t)$, these terms can be dropped by the following third type absorbing term 

\begin{equation}\label{app:bgk_no2ordor2}
f_i(\rmx+v_i\delta t,t+\delta
t)=f_i(\rmx,t)+s\left(f_i^{\rm(eq)}(\rho^*(\rmx,t),{\rmu^*(\rmx,t)},t)-f_i(\rmx,t)\right)+\delta t F^{\rm (L,\ eq)}_i\left({\rho^f(\rmx,t)},{\rmu^f(\rmx,t)},{\rho^*(\rmx,t)},{\rmu^*(\rmx,t)},t)\right),
\end{equation}
where ${\rm u^*(x,t)}$ is the equilibrium velocity.
 $F^{\rm(L,\ eq)}_i({\rmu_f(\rmx,t)},{\rmu^*(\rmx,t)},t)$ is defined by
 
\begin{equation}
F^{\rm(L,\ eq)}_i(\rho^{f},{\rmu^f(x,t)},\rho^*,{\rm u^*(x,t)},t)=\chi\left(f_i^{\rm(L,\ eq)}(\rho^{f},{\rmu^f(x,t)},t)-f_i^{\rm(L,\ eq)}(\rho^*,{\rm u^*(x,t)},t)\right).
\end{equation}
In Eq. (\ref{app:bgk_no2ordor2}), $\rho^*$ and $\rmu^*$ denote the density and velocity, which are defined by the following parameterized forms \cite{Guo}

\begin{equation}
\rho^*=\sum_{i}f_i+n\delta t\sum_{i}F^{\rm(L,\ eq)}_i,\quad \rho^*u^*_\alpha=\sum_{i}c_{i\alpha}f_i+m\delta t\sum_{i}c_{i\alpha}F^{\rm(L,\ eq)}_i.
\end{equation}
Then, the following macroscopic equations are obtained (see \ref{app:1} for details)

\begin{equation}\label{app:abs-eq4}
\left\{\begin{array}{l}
\partial_{t}\rho^*+\partial_{\alpha}(\rho^* u^*_{\alpha}) =(1+n s)\chi(\rho^{f}-\rho^*)+\zeta_n\chi\partial_{t}(\rho^{f}-\rho^*)+\zeta_m\chi\partial_{\alpha}\left(\rho^{f}u^{f}_{\alpha}-\rho^{*}u^{*}_{\alpha}\right),\\[2mm]
\partial_{t}(\rho^* u^*_{\alpha})-\partial_{\beta}(\nu\rho^*(\partial_{\alpha}u^*_{\beta}+\partial_{\beta}u^*_{\alpha}))+\partial_{\beta}\left(\rho^* u^*_\alpha u^*_\beta+p^*\delta_{\alpha\beta}\right) =(1+m s^\prime )\chi\left(\rho^{f}u^{f}_{\alpha}-\rho^{*}u^{*}_{\alpha}\right)+
\zeta_m\chi\partial_{t}\left(\rho^{f}u^{f}_{\alpha}-\rho^{*}u^{*}_{\alpha}\right),
\end{array}
\right.
\end{equation}
where the parameters $\nu$, $\zeta_n$ and $\zeta_m$ are defined as follows
\begin{equation}
\nu = c_s^2\left(\frac{1}{s}-\frac{1}{2}\right)\delta t, \ \zeta_n = n\delta t\left(1-\frac{s}{2}\right)-\frac{\delta t}{2},\ \zeta_m = m\delta t\left(1-\frac{s}{2}\right)-\frac{\delta t}{2}.
\end{equation}

In this section, the fundamental theories of LBS are reviewed.  Based on the classical absorbing layer strategies, the three types of absorbing terms are proposed for LBS, and the corresponding macroscopic equations are given. In the following section, the behaviors of the absorbing terms are studied.

\subsection{Profiles of absorbing strength $\sigma$}
\begin{figure}[!h]
\begin{center}
\scalebox{0.45}[0.45]{\includegraphics[angle=0]{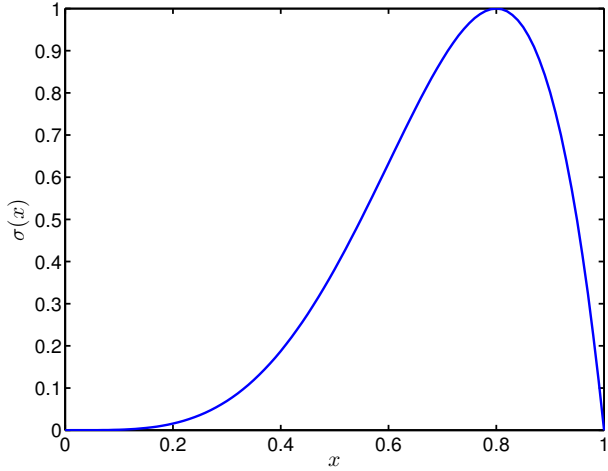}}
\caption{The normalized $\sigma(x)$ profile: $A=n=4$, $x_0=0$, $L=1$.}\label{fig:sigma}
\end{center}
\end{figure}

In the absorbing layers, the profile of $\sigma$ could not be specified as a uniformly distributed parameter. A uniform $\sigma$ distribution will induce a significant wave reflection from the interfaces  between the wave propagation domain and the sponge domain \cite{bodony,orszag}.  The most popular profiles of $\sigma$ are listed in Table \ref{tab:sigma}. In the current research, the first kind of $\sigma$ will be used which was first proposed in \cite{orszag}. For this kind of absorbing profile, $\sigma(\rmx)$ satisfies $\sigma(x_0)=\sigma(L)=0$. In the original paper of Israeli and Orszag \cite{orszag}, by taking $A = n =4$, they obtained the best results for the acoustic wave equation. In order to handle the current problem, we given following  normalized $\sigma_{\rmx}$ for $A=n=4$ as follows

\begin{equation}\label{normalizedsigma}
\displaystyle\tilde{\sigma}_{\rmx}=\frac{3125(L-x)(x-x_0)^4}{256(L-x_0)^5}.
\end{equation}
The normalized $\tilde{\sigma}(x)$ profile for $x_0=0$ and $L=1$ is given in Fig. \ref{fig:sigma}

\begin{table}[!htpb]
\caption{Newtonian cooling function: possible profiles of $\sigma$ from the literature \cite{orszag,durran,renaut}.}\label{tab:sigma}
\begin{tabular*}{\textwidth}{@{\extracolsep{\fill}}ccp{6cm}}\toprule[1pt]
Damping Types & $\sigma(\rmx)$ & Parameters   \\ \midrule[1pt]
Polynomial    & $\displaystyle A\frac{(x-x_0)^n(L-x)(n+1)(n+2)}{(L-x_0)^{n+2}}$  & $x_0$: Position of sponge layer; $A$: Absorbing strength; $L$: Width of sponge layer; for most choices, $n=4$.\\ [3mm]
&$\displaystyle\eta\cdot\sbracket{\frac{x-x_0}{L-x_0}}^n $& $\eta$: Absorbing strength; $L$: Width of sponge layer; $n=3$.\\[3mm]
Reyleigh  & $\alpha [1-{\rm cos}(8\pi(x-x_0-1))]$ &  $x_0$:  Position of sponge layer; $\alpha$: Absorbing strength.\\
\bottomrule[1pt]
\end{tabular*}
\end{table}

\section{Analysis of LBS coupled with absorbing terms}\label{analysisabc}

In this section, the influences of the absorbing terms on the dissipation/dispersion properties  are explored looking at the linearized forms of the recovered macroscopic equations and LBS.

\subsection{The linearized macroscopic equation with the absorbing terms}\label{macroabc}

If we consider $n=m$, Eqs. (\ref{app:abs-eq1}) and (\ref{app:abs-eq3}) can be expressed by

\begin{equation}\label{app:abs-eq7}
\slbracket{
\begin{array}{l}
\displaystyle\partial_{t}\rho^*+\partial_{\alpha}(\rho^* u^*_{\alpha}) =(1+n \tilde{s})\chi(\rho^{f}-\rho^*)+\zeta_n\chi\partial_{t}(\rho^{f}-\rho^*)+\zeta_n\chi\partial_{\alpha}\left(\rho^{f}u^{f}_{\alpha}-\rho^{*}u^{*}_{\alpha}\right),\\[2mm] 
\displaystyle\partial_{t}(\rho^* u^*_{\alpha})-\partial_{\beta}(\tilde{\nu}\rho^*(\partial_{\alpha}u^*_{\beta}+\partial_{\beta}u^*_{\alpha}))+\partial_{\beta}\left(\rho^* u^*_\alpha u^*_\beta+p^*\delta_{\alpha\beta}\right) =(1+n \tilde{s} )\chi\left(\rho^{f}u^{f}_{\alpha}-\rho^{*}u^{*}_{\alpha}\right)+\\[2mm]
\displaystyle \zeta_n\chi\partial_{t}\left(\rho^{f}u^{f}_{\alpha}-\rho^{*}u^{*}_{\alpha}\right)-\left(\sigma+\frac{\delta t}{2}\right)\chi\partial_{\beta}\left(\rho^{f}u^{f}_{\alpha}u^{f}_{\beta}+p^{f}\delta_{\alpha\beta}-\rho^{*}u^{*}_{\alpha}u^{*}_{\beta}-p^{*}\delta_{\alpha\beta}\right),
\end{array}
}
\end{equation}
where $\tilde{\nu}$ denotes ${\nu}$ or ${\nu}^{\prime}$, $\tilde{s}$ denotes $s$ or ${s}^{\prime}$, and  Eq. (\ref{app:abs-eq4}) can be rewritten as follows

\begin{equation}\label{app:abs-eq8}
\left\{\begin{array}{l}
\displaystyle \partial_{t}\rho^*+\partial_{\alpha}(\rho^* u^*_{\alpha}) =(1+n \tilde{s})\chi(\rho^{f}-\rho^*)+\zeta_n\chi\partial_{t}(\rho^{f}-\rho^*)+\zeta_n\chi\partial_{\alpha}\left(\rho^{f}u^{f}_{\alpha}-\rho^{*}u^{*}_{\alpha}\right),\\[2mm] \displaystyle
\partial_{t}(\rho^* u^*_{\alpha})-\partial_{\beta}(\nu\rho^*(\partial_{\alpha}u^*_{\beta}+\partial_{\beta}u^*_{\alpha}))+\partial_{\beta}\left(\rho^* u^*_\alpha u^*_\beta+p^*\delta_{\alpha\beta}\right) =(1+n \tilde{s})\chi\left(\rho^{f}u^{f}_{\alpha}-\rho^{*}u^{*}_{\alpha}\right)+\zeta_n\chi\partial_{t}\left(\rho^{f}u^{f}_{\alpha}-\rho^{*}u^{*}_{\alpha}\right),
\end{array}
\right.
\end{equation}

In order to detect the dissipative behavior,  we consider the following simplified 1D case of Eq. (\ref{app:abs-eq7}) (the flow propagation is only along x-axis) 
 
\begin{equation}\label{app:abs-eq5}
\slbracket{
\begin{array}{l}
\displaystyle\partial_{t}\rho^*+\partial_{x}(\rho^* u^*_x) =(1+n \tilde{s})\chi(\rho^{f}-\rho^*)+\zeta_n\chi\partial_{t}(\rho^{f}-\rho^*)+\zeta_n\chi\partial_x\left(\rho^{f}u^{f}_x-\rho^{*}u^{*}_x\right),\\[2mm] 
\displaystyle\partial_{t}(\rho^* u^*_x)-\partial_x(\tilde{\nu}\rho^*(\partial_xu^*_x+\partial_xu^*_x))+\partial_{x}\left(\rho^* u^*_x u^*_x+p^*\right) =(1+n \tilde{s})\chi\left(\rho^{f}u^{f}_x-\rho^{*}u^{*}_x\right)+\zeta_n\chi\partial_{t}\left(\rho^{f}u^{f}_x-\rho^{*}u^{*}_x\right)\\[2mm]
\displaystyle-\left(\sigma+\frac{\delta t}{2}\right)\chi\partial_x\left(\rho^{f}u^{f}_xu^{f}_x+p^{f}-\rho^{*}u^{*}_xu^{*}_x-p^{*}\right).
\end{array}
}
\end{equation}

We consider the following decomposition of $\rho^*$ and $u^*_x$
\begin{equation}\label{eq:decom}
\rho^*=\rho^0+\rho^\prime,u^*_x=u^0_x+u^\prime_x.
\end{equation} 
For the sake of simplicity, we take $u_x^f=u_x^0=0$ and $\rho^0=\rho^f=const=1$, leading to 
the following linearized system

\begin{equation}\label{app:abs-eq6}
\slbracket{
\begin{array}{l}
\displaystyle\partial_{t}\rho^\prime+\partial_{x}u^\prime_x =-(1+n \tilde{s})\chi\rho^{\prime}-\zeta_n\chi\partial_{t}\rho^{\prime}-\zeta_n\chi\partial_xu^{\prime}_x,\\[2mm] 
\displaystyle\partial_{t}u^\prime_x-2\tilde{\nu}\partial_x^2u^\prime_x+c_s^2\partial_{x}\rho^\prime =-(1+n \tilde{s})\chi u^{\prime}_x-\zeta_n\chi\partial_{t}u^{\prime}_x
+c_s^2\left(\sigma+\frac{\delta t}{2}\right)\chi\partial_x\rho^\prime.
\end{array}
}
\end{equation}
From the above equations, we obtain

\begin{equation}
\displaystyle  A\cdot(1+\zeta_n \chi)\partial_t^2\rho+(B\chi+A\cdot B\chi+B\zeta_n\chi^2)\partial_t\rho-2A\tilde{\nu}\partial_x^2(\partial_t\rho)-(2\tilde{\nu}B\chi+Ac_s^2-Ac_s^2 C\chi) \partial_x^2\rho+B^2\chi^2\rho=0,
\end{equation}
where $A=1+\zeta_n\chi $, $B=1+n\tilde{s}$ and $C=\sigma+\delta t/2$.

Now, considering a  monochromatic wave solution $\rho^\prime=\rho_0^\prime {\rm exp}[{\rm i}(k_x\cdot x-\omega t)]$, we get the following $k-\omega$ relation

\begin{equation}
\displaystyle  A\cdot(1+\zeta_n \chi)\omega^2+\rmi\cdot(B\chi+2A\tilde{\nu}k_x^2+A\cdot B\chi+B\zeta_n\chi^2)\omega-2\tilde{\nu}B\chi k_x^2-Ac_s^2 k_x^2-B^2\chi^2+Ac_s^2 C\chi k_x^2=0,
\end{equation}

Considering a monochromatic solution, the similar $k-\omega$ relation for the linearized form of Eq. (\ref{app:abs-eq6})
can be obtained as follows

\begin{equation}\label{komega2k}
\displaystyle  A\cdot(1+\zeta_n \chi)\omega^2+\rmi\cdot(B\chi+2A\tilde{\nu}k_x^2+A\cdot B\chi+B\zeta_n\chi^2)\omega-2\tilde{\nu}B\chi k_x^2-Ac_s^2 k_x^2-B^2\chi=0.
\end{equation}

If $\tilde{\nu}$ vanishes in Eq. (\ref{app:abs-eq6}),  then,  we are left with the following 1D form 
\begin{equation}\label{app:abs-eq6a}
\slbracket{
\begin{array}{ll}
\displaystyle\partial_{t}\rho^\prime+\partial_{x}u^\prime_x &=-(1+n \tilde{s})\chi\rho^{\prime}-\zeta_n\chi\partial_{t}\rho^{\prime}-\zeta_n\chi\partial_xu^{\prime}_x,\\[2mm] 
\displaystyle\partial_{t}u^\prime_x+c_s^2\partial_{x}\rho^\prime& =\displaystyle -(1+n \tilde{s})\chi u^{\prime}_x-\zeta_n\chi\partial_{t}u^{\prime}_x
+c_s^2\left(\sigma+\frac{\delta t}{2}\right)\chi\partial_x\rho^\prime.
\end{array}
}
\end{equation}
The above equation is different from the classical 1D model

\begin{equation}\label{app:abs-eq6c}
\slbracket{
\begin{array}{ll}
\displaystyle\partial_{t}\rho^\prime+\partial_{x}u^\prime_x &=-\sigma_\rho(\rmx)\rho^{\prime},\\[2mm] 
\displaystyle\partial_{t}u^\prime_x+c_s^2\partial_{x}\rho^\prime& =-\sigma_u(\rmx)u^{\prime}_x.
\end{array}
}
\end{equation}

Mathematically, the above classical absorbing strategy is similar to the penalty method \cite{bodony}. As indicated, if $(\rho^\prime, u_x^\prime)$ is suitably smooth and the norm $\|(\sigma_\rho,\sigma_u)\|$ suitably large, the spatial derivatives of   $(\rho^\prime, u_x^\prime)$ will be negligible compared with the right hand side. Eq.(\ref{app:abs-eq6c}) and the solution decays exponentially with the time scale $\mathcal{T}$ \cite{bodony}. If $t>\mathcal{T}$, the right hand side terms will not dominate Eq. (\ref{app:abs-eq6c})  and the convergence property will depend on the initial condition and the coercivity of the convection operator.  From the viewpoint of penalty methods, $\sigma_\rho$ and $\sigma_u$ could be chosen as any positive large numbers.  Now, let us rewrite Eq. (\ref{app:abs-eq6a}) as follows

\begin{equation}\label{app:abs-eq6b}
\slbracket{
\begin{array}{ll}
\partial_{t}\rho^\prime+\partial_{x}u^\prime_x &=\displaystyle -\frac{(1+n \tilde{s})\chi}{1+\zeta_n\chi}\rho^{\prime}\\[4mm] 
\displaystyle\partial_{t}u^\prime_x+\frac{\sbracket{c_s^2-c_s^2\left(\sigma+{\delta t}/{2}\right)\chi}}{1+\zeta_n\chi}\partial_{x}\rho^\prime& =\displaystyle -\frac{(1+n \tilde{s})\chi}{1+\zeta_n\chi}u^{\prime}_x.
\end{array}
}
\end{equation}

We now consider $\tilde{n}=1/2$ and $\delta t=1$, $\zeta_n = -{\tilde{s}}/{4}$.  It is clear that $\zeta_n$ is negative. That means if the following inequality is satisfied

\begin{equation}\label{criticalchi}
1+\zeta_n\chi<0\ \mbox{or}\  1-\chi\tilde{s}/4<0,
\end{equation}
the effective coefficient  $\sigma^{\rm eff}$ defined by 

\begin{equation}\label{effectivesigma}
\displaystyle \sigma^{\rm eff}=\frac{(1+n \tilde{s})\chi}{(1+\zeta_n\chi)}
\end{equation}
 will become negative and LBS will be unstable. This  phenomenon is different from that in Eq. (\ref{app:abs-eq6c}). Further, by $\sigma=(1/\tilde{s}-1/2)\delta t$ and $\delta t=1$, we have
 
\begin{equation}
\tilde{c}_s^2=\frac{\sbracket{c_s^2-c_s^2\left(\sigma+1/{2}\right)\chi}}{1+\zeta_n\chi}=\frac{c_s^2(1-\chi/\tilde{s})}{1-\chi\tilde{s}/4}.
\end{equation}
If $\tilde{s}=2$, the correct effective sound speed will be recovered. As indicated in Sec.\ref{Intro},  LBS solution in absorbing layers is not required to be physical, so  it is not necessary for $\tilde{s}$ to be equal to 2. Clearly, because of the collision term in LBS, the classical absorbing layer theory can not be used completely and directly. 

As indicated in Sec. \ref{sec:1tabc}, for the type I absorbing term, because of the definition of the effective relaxation $s^\prime$,  if the condition (\ref{effectrelax}) is satisfied, the instability will grow very fast.  In practical applications dealing with  acoustic problems, because of the nearly vanishing viscosity, $s$ is close to 2 and  it is impossible to choose suitable and  positive  $\chi$.  Therefore, the type I absorbing term is not well suited for absorbing acoustic waves.  

In the same way, we can get the following equation for the type III absorbing term

\begin{equation}\label{app:abs-eq1third}
\slbracket{
\begin{array}{ll}
\partial_{t}\rho^\prime+\partial_{x}u^\prime_x &=\displaystyle -\frac{(1+n \tilde{s})\chi}{1+\zeta_n\chi}\rho^{\prime}\\[4mm] 
\displaystyle\partial_{t}u^\prime_x+\frac{c_s^2}{1+\zeta_n\chi}\partial_{x}\rho^\prime& =\displaystyle -\frac{(1+n \tilde{s})\chi}{1+\zeta_n\chi}u^{\prime}_x.
\end{array}
}
\end{equation}

If $1+\zeta_n\chi=1$, in order to keep the sound speed $c_s$ uniform and $\chi$ different from 0,  the choice of $s$ must satisfy the following equality
\begin{equation}\label{s3rd}
\displaystyle s=\frac{2n-1}{n}.
\end{equation}
However, this requirement is not necessary for the sound speed. The  critical value of $\chi$ for stability is given by

\begin{equation}\label{s3rdchi}
\displaystyle \chi=\frac{2}{(1-2n+ns)\delta t}.
\end{equation}

The positivity constraint on $\chi$ requires the following inequality should be satisfied

\begin{equation}
\displaystyle n<\frac{1}{2-s}.
\end{equation} 
From the above inequality, it is easy to see that  if $s$ is close to 2, the choice of $n$ will become very optional. If we set $n=1/2$, the constraint on $s$ is natural ($s>0$).  

The numerical investigations will be presented in Sec \ref{vonnumann}, which are mainly focused on the type II and type III absorbing terms.

\subsection{The linearized LBS with the absorbing terms}

In order to give the linearized LBM with the absorbing terms, we consider the following general form

\begin{equation}\label{app:Common}
f_i(\rmx+v_i\delta t,t+\delta
t)=f_i(\rmx,t)+\tilde{s}\left(f_i^{\rm(eq)}(\rho^*(\rmx,t),{\rmu^*(\rmx,t)},t)-f_i(\rmx,t)\right)+\delta t \tilde{F}^{\rm (eq)}_i\left({\rho^f(\rmx,t)},{\rmu^f(\rmx,t)},{\rho^*(\rmx,t)},{\rmu^*(\rmx,t)},t)\right),
\end{equation}
where the absorbing term is given by

\begin{equation}
\tilde{F}^{\rm(eq)}_i(\rho^{f}(\rmx,t),{\rm u^f(\rmx,t)},\rho^*,{\rm u^*(x,t)},t)=\chi\left(\tilde{f}_i^{\rm(eq)}(\rho^{f}(\rmx,t),{\rmu^f(\rmx,t)},t)-\tilde{f}_i^{\rm(eq)}(\rho^*(\rmx,t),{\rm u^*(\rmx,t)},t)\right).
\end{equation}

Regarding ${\rho^f(\rmx,t)}$ and ${\rmu^f(\rmx,t)}$ as the uniform reference states, the reference $f_i^{(\Ref)}$ is defined by

\begin{equation}
f_i^{(\Ref)} = f_i^{\beq}(\rho^f,{\rmu^f}).
\end{equation}
The fluctuating quantity $\delta f_i$ is defined by

\begin{equation}
\delta f_i = f_i-f_i^{(\Ref)}.
\end{equation}
Using the definitions of  $f_i^{(\Ref)}$ and $\delta f_i$, and considering $f_j^{\rm(eq)}(\rho^*,{\rmu^*},t)=f_i^{\rm(eq)}\sbracket{\lbracket{f_k}_{0\leq k\leq
N}}$, the linearized version of Eq. (\ref{app:Common}) is depicted by

\begin{equation}\label{eq:linearized}
\displaystyle \delta f_i(\rmx+v_i\delta t,t+\delta
t)=\delta f_i(\rmx,t)+\tilde{s}\left(\left.\frac{\partial f_i^{\rm(eq)}\sbracket{\lbracket{f_k}_{0\leq k\leq
N}}}{\partial f_j}\right|_{f_j=f_j^f}-\delta_{ij}\right)\delta f_j+\delta t \delta\tilde{F}^{\rm (eq)}_i\left({\rho^f(\rmx,t)},{\rmu^f(\rmx,t)},{\rho^*(\rmx,t)},{\rmu^*(\rmx,t)},t)\right),
\end{equation}
where $ \delta\tilde{F}^{\rm (eq)}_i\left({\rho^f(\rmx,t)},{\rmu^{\it f}(\rmx,t)},{\rho^*(\rmx,t)},{\rmu^*(\rmx,t)},t)\right)= \delta\tilde{F}^{\rm (eq)}_i\left(\{f^f_k\}_{0\leq k\leq
N},\lbracket{f_k}_{0\leq k\leq
N}\right)$ is calculated by

\begin{equation}
\displaystyle \delta\tilde{F}^{\rm (eq)}_i\left(\{f^f_k\}_{0\leq k\leq
N},\lbracket{f_k}_{0\leq k\leq
N}\right) = -\chi\left.\frac{\partial \tilde{f}_i^{\rm(eq)}\sbracket{\lbracket{f^f_k}_{0\leq k\leq N}}}{\partial f_j}\right|_{f_j=f_j^f}\cdot\delta f_j.
\end{equation}
Then, considering a plane wave solution of the linearized equation

\begin{equation}\label{perturb-w-f}
\delta f_j=A_j{\rm exp}[\mathrm{i}({\bf k}\cdot {\bf x}-\omega t)],
\end{equation}
we get the following eigenvalue problem for the
L-MRT-LBM in the frequency-wave number space
\cite{simondenispierre,ricotmariesagautbailly,lallemandluo}

\begin{equation}\label{sec:matrix}
e^{-\mathrm{i}\omega}{\bf \delta f}=M\cdot{\bf \delta f}.
\end{equation}
The matrix $M$ in Eq. (\ref{sec:matrix}) is defined by

\begin{equation}
M_{ij}=\delta_{ij}+\tilde{s}\cdot\sbracket{\left.\frac{\partial f_i^{\rm(eq)}\sbracket{\lbracket{f_k}_{0\leq k\leq
N}}}{\partial f_j}\right|_{f_j=f_j^f}-\delta_{ij}}-\delta t\chi\left.\frac{\partial \tilde{f}_i^{\rm(eq)}\sbracket{\lbracket{f^f_k}_{0\leq k\leq N}}}{\partial f_j}\right|_{f_j=f_j^f}.
\end{equation}

For the sake of  comparison with the linearized isentropic Navier-Stokes equations, it is recalled that the analytical acoustic modes $\omega^{\pm}(\bf k)$ and shear
modes $\omega^s(\bf k)$ are given by \cite{landaulifshitz}

\begin{equation}\label{th:d}
\left\{\begin{array}{l} \Real[\omega^{\pm}({\bf k})]=|{\bf k}|(\pm
c_s+|{\bf u}|{\bf
cos}(\widehat{\bf k\cdot u})),\\ \displaystyle
\Imag[\omega^{\pm}({\bf k})]=-|{\bf
k}|^2\frac{1}{2}\left(\frac{2d-2}{d}\nu+\eta\right),\\
\Real[\omega^s(\bf k)]=|{\bf k}||{\bf u}|{\rm cos}(\widehat{\bf k\cdot u}),\\
\Imag[\omega^s(\bf k)]=-|{\bf k}|^2\nu,
\end{array}\right.
\end{equation}
where $\nu$ is the shear viscosity and $\eta $ is the bulk
viscosity.

\subsection{Von Neumann Analysis for D2Q9 model in spectral spaces}\label{vonnumann}

In the above section, the linearized LBS with the absorbing terms was given. In this section, combining the analysis presented in Sec. \ref{macroabc} and the linearized form (\ref{sec:matrix}) in spectral spaces, the numerical dissipative and dispersive properties will be analyzed thanks to  Von Neumann theory. The parameter $\delta t$ for the standard LBS is equal to 1. For the sake of convenience, and without loss of gnerality, let us choose $s=1.9$ and $s=2$  in Eqs. (\ref{app:abs-eq7}) and (\ref{app:abs-eq8}). For $s=2$, the molecular viscosity vanishes. One also consider $n=m=1/2$. In order to keep the numerical investigation identical to the theory given in Sec. \ref{macroabc},  let $\rho^f=1$ and $\rmu^f=0$. The wavenumber in Eq. (\ref{perturb-w-f}) is defined by

\begin{equation}\label{wavenumber}
k_x={\bf k}\cdot{\rm cos}(\theta),\  k_y={\bf k}\cdot{\rm sin}(\theta).
\end{equation}

\begin{figure}[!htbp]
\begin{center}
\scalebox{0.45}[0.45]{\includegraphics[angle=0]{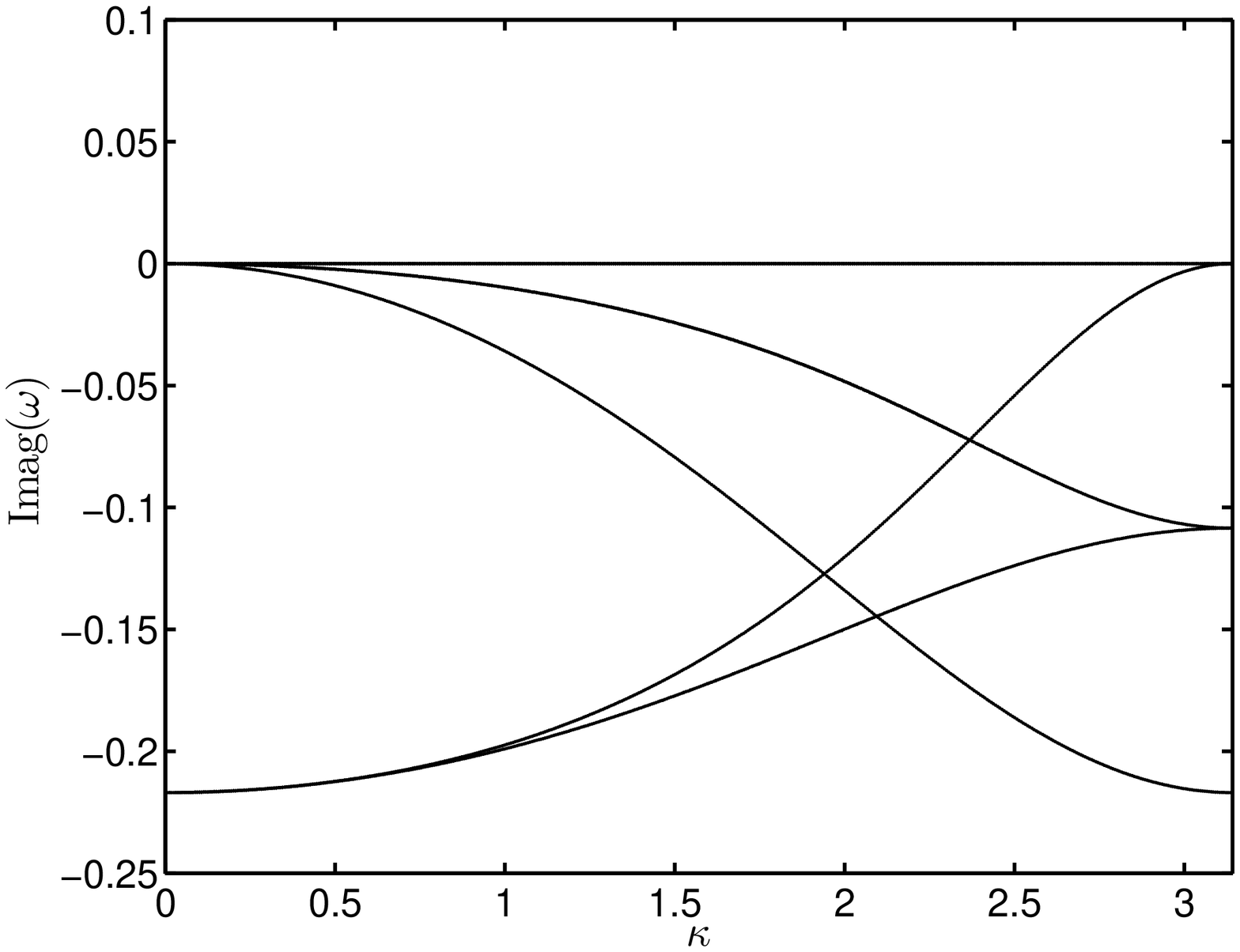}}
\scalebox{0.45}[0.45]{\includegraphics[angle=0]{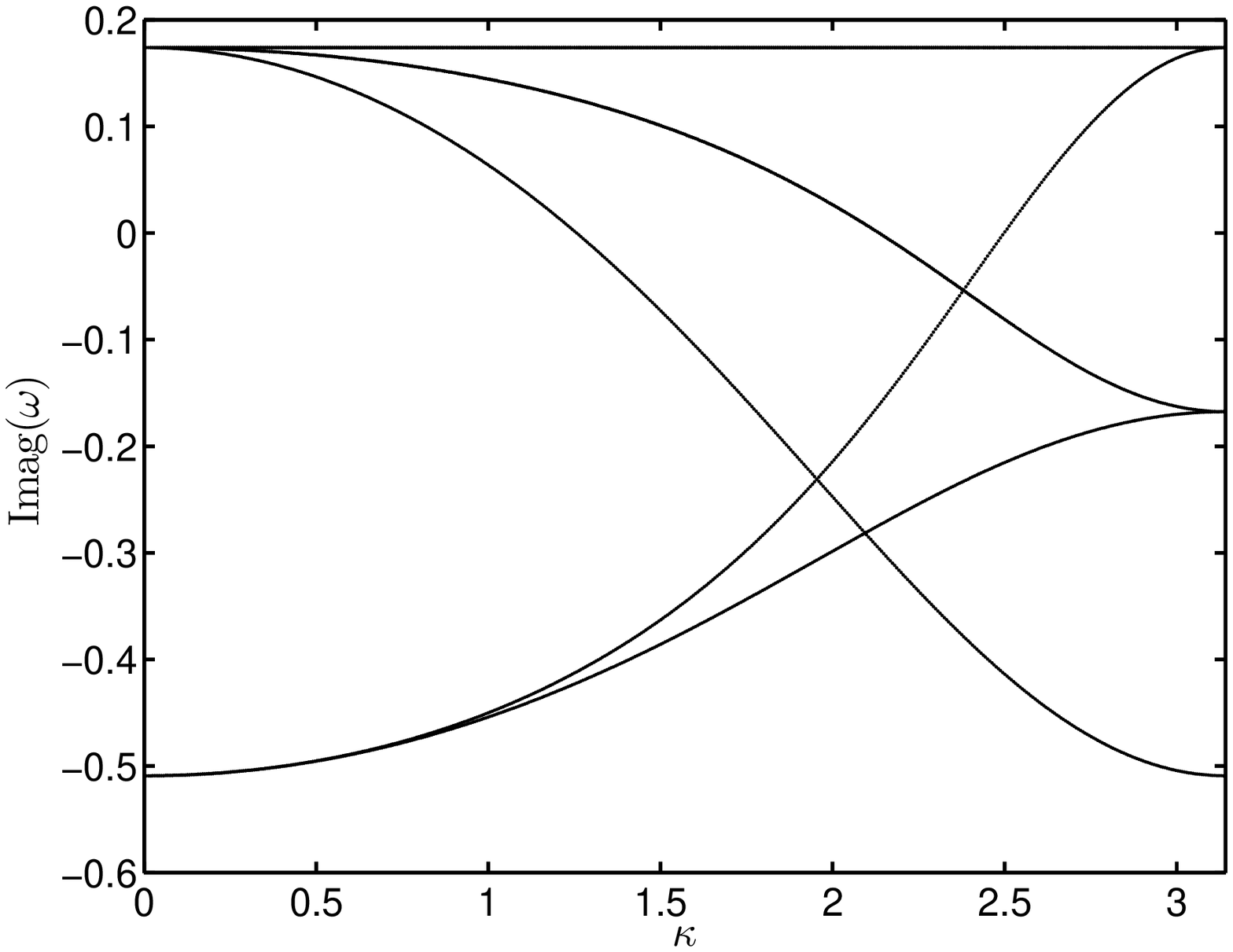}}\\
{\centering\hspace{1.cm}(a) $s=1.9$ \& $\chi=0.1$\hspace{5.cm}(b) $s=1.99$ \& $\chi=0.2$}
\caption{Stability properties of the type I absorbing term: $\theta=0$ , $n=m=1/2$.}\label{fig:1}
\end{center}
\end{figure}

Now, let us discuss  the stability of the type I absorbing term.  In Fig. \ref{fig:1}, the dissipation relations for  the type I absorbing term are shown for two different values of the parameter $\chi$. As indicated by the inequality (\ref{effectrelax}), in Fig. \ref{fig:1}(a), the effective relaxation parameter $s^\prime$  is equal to 2 and the LBS is  stable. When the effective relaxation parameter $s^\prime$ is larger than 2, it can be observed in  Fig. \ref{fig:1}(b) that there exist some modes which become unstable. Therefore,   type I absorbing term is useless for LBS. 

\begin{figure}[!h]
\begin{center}
\scalebox{0.445}[0.445]{\includegraphics[angle=0]{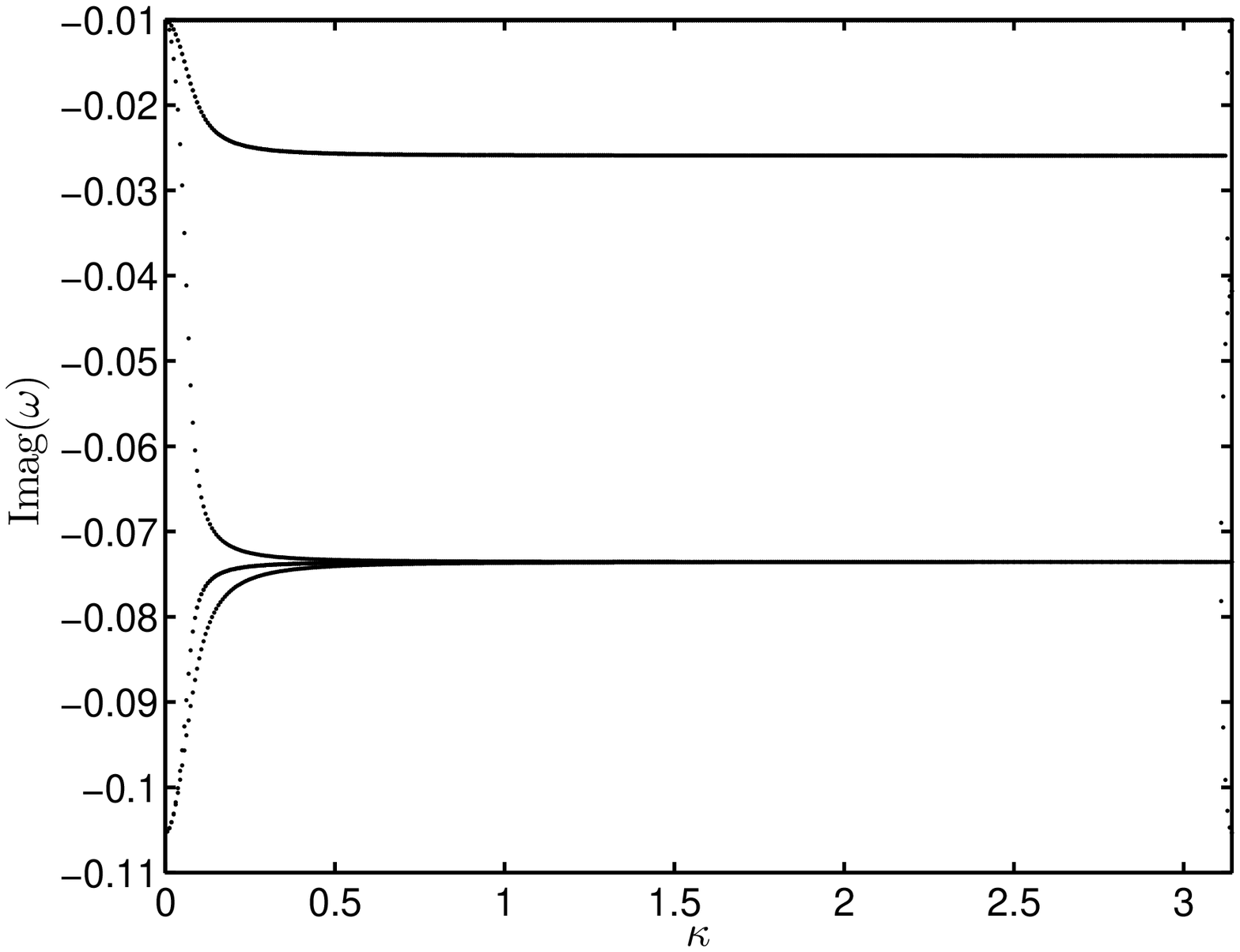}}
\scalebox{0.445}[0.445]{\includegraphics[angle=0]{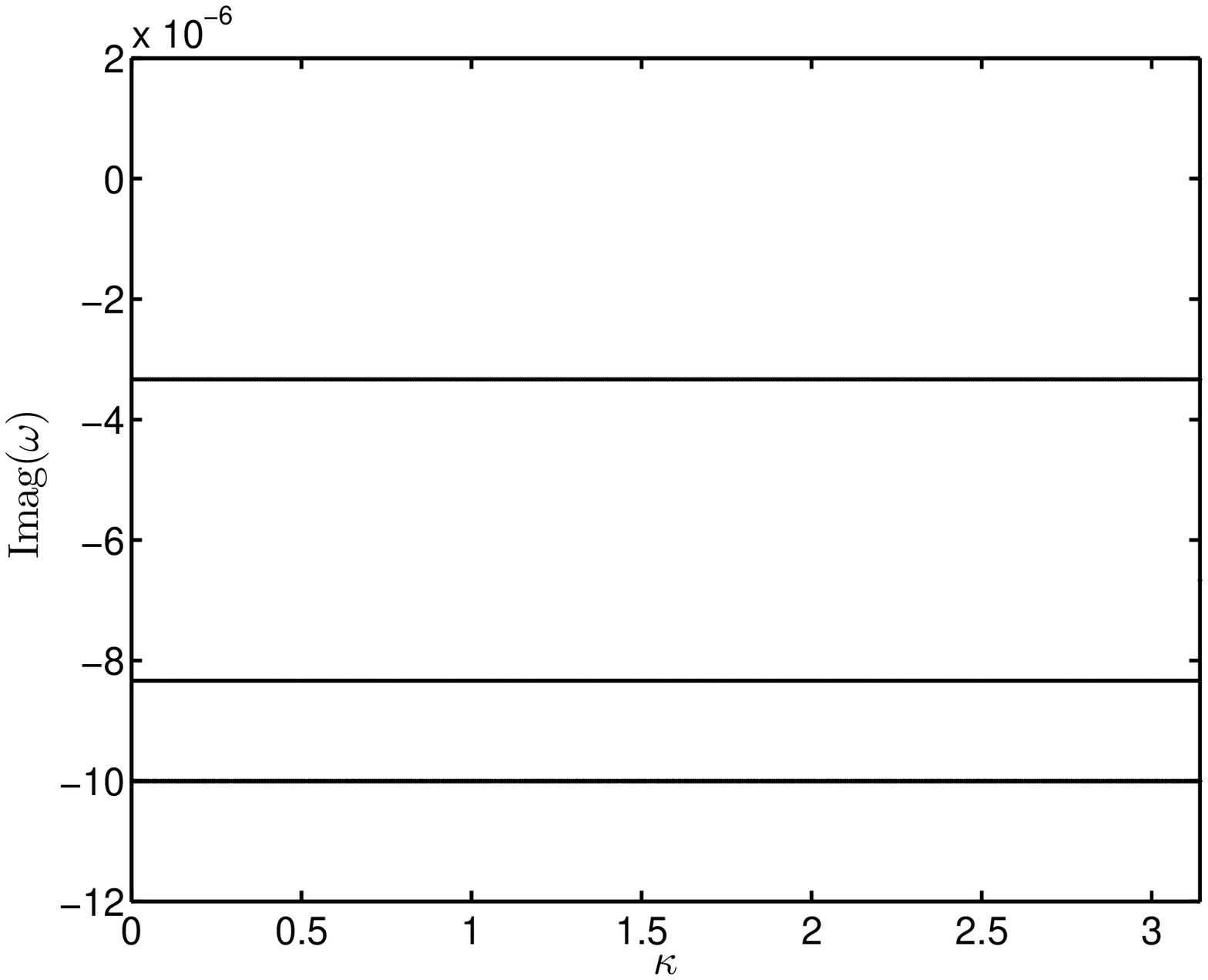}}\\
{\centering\hspace{1.cm}(a) $s=1.99$ \& $\chi=4/s$\hspace{5.cm}(b) $s=1.99999$ \& $\chi=4/s$}
\caption{Stability properties of the type II absorbing term: $\theta=0$ , $n=m=1/2$.}\label{fig:2}
\end{center}
\end{figure}

\begin{figure}[!h]
\begin{center}
\scalebox{0.445}[0.445]{\includegraphics[angle=0]{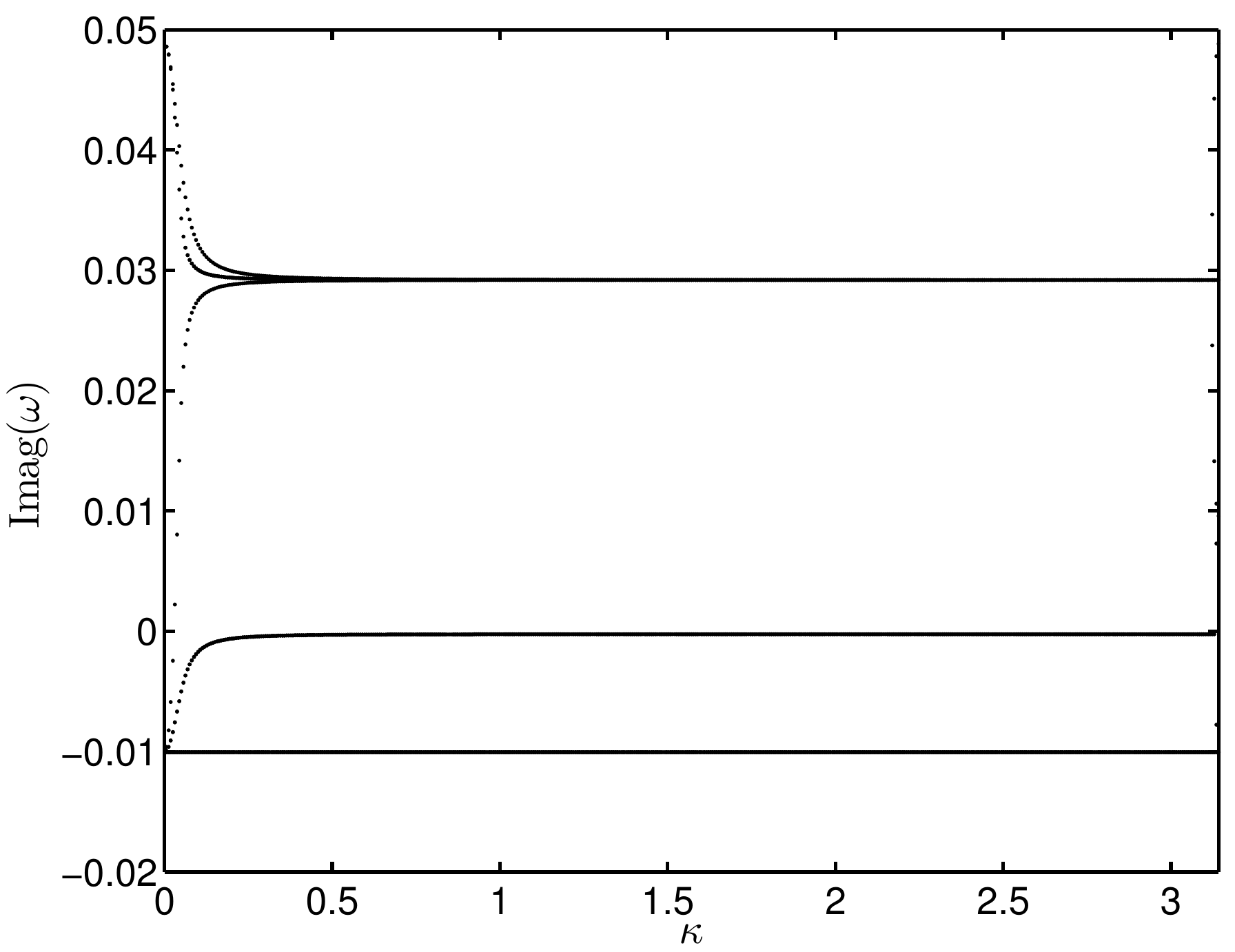}}
\scalebox{0.445}[0.445]{\includegraphics[angle=0]{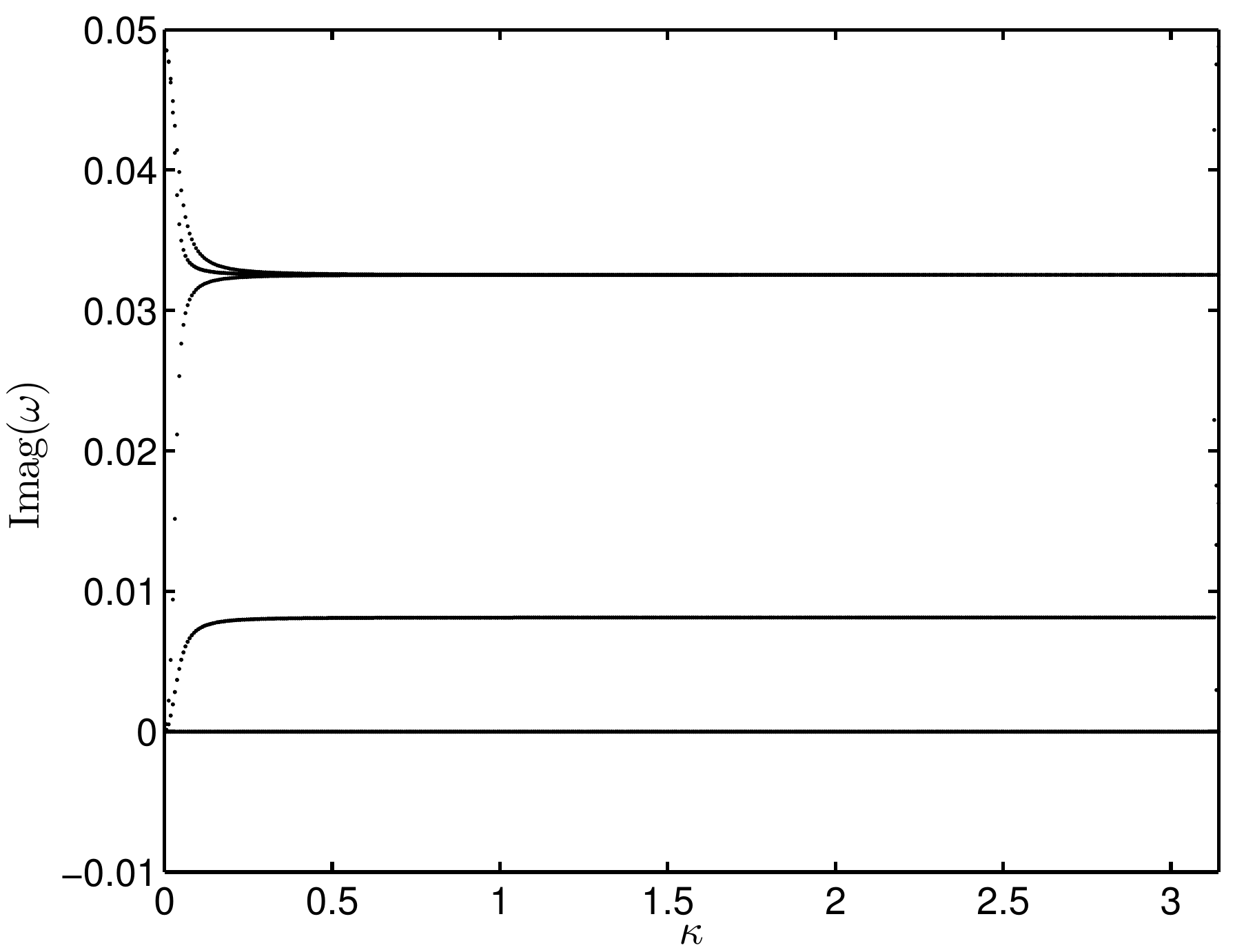}}\\
{\centering\hspace{1.cm}(a) $s=1.99$ \& $\chi=4/s+0.1$\hspace{5.cm}(b) $s=1.99999$ \& $\chi=4/s+0.1$}
\caption{Stability properties of the type II absorbing term: $\theta=0$ , $n=m=1/2$.}\label{fig:3}
\end{center}
\end{figure}

\begin{figure}[!htbp]
\begin{center}
\scalebox{0.445}[0.445]{\includegraphics[angle=0]{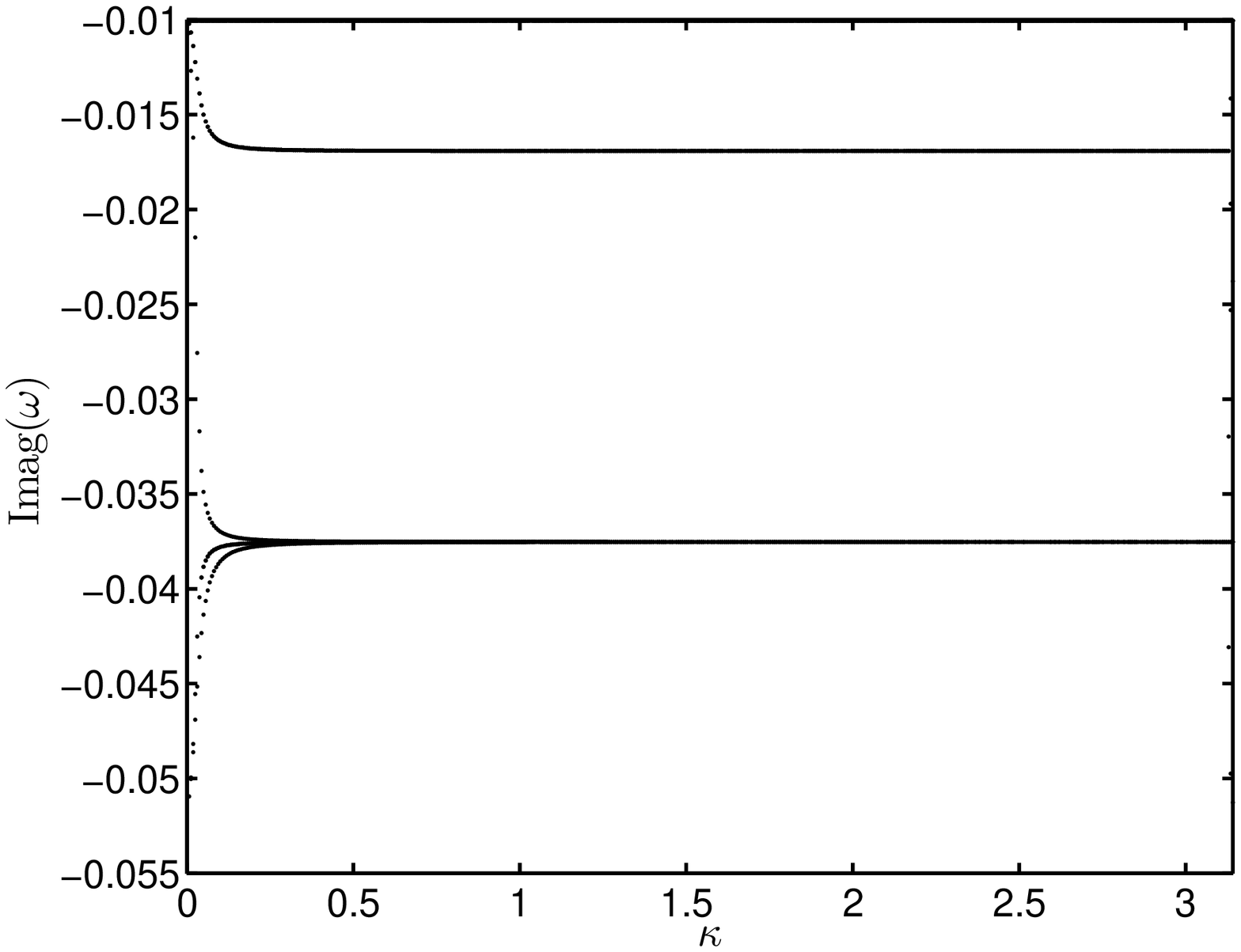}}
\scalebox{0.445}[0.445]{\includegraphics[angle=0]{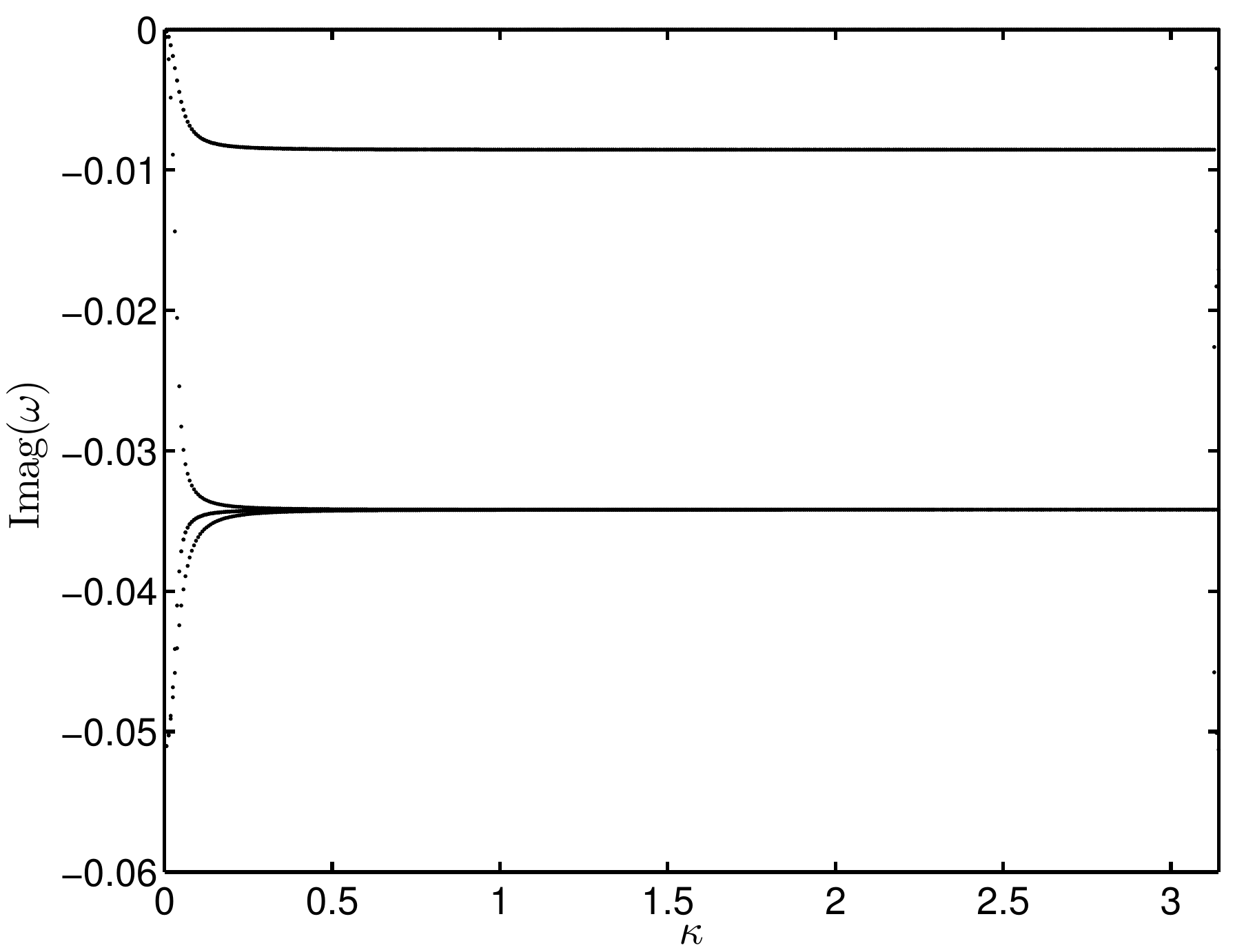}}\\
{\centering\hspace{1.cm}(a) $s=1.99$ \& $\chi=4/s-0.1$\hspace{5.cm}(b) $s=1.99999$ \& $\chi=4/s-0.1$}
\caption{Stability properties of the type II absorbing term: $\theta=0$ , $n=m=1/2$.}\label{fig:4}
\end{center}
\end{figure}

\begin{figure}[!htbp]
\begin{center}
\scalebox{0.445}[0.445]{\includegraphics[angle=0]{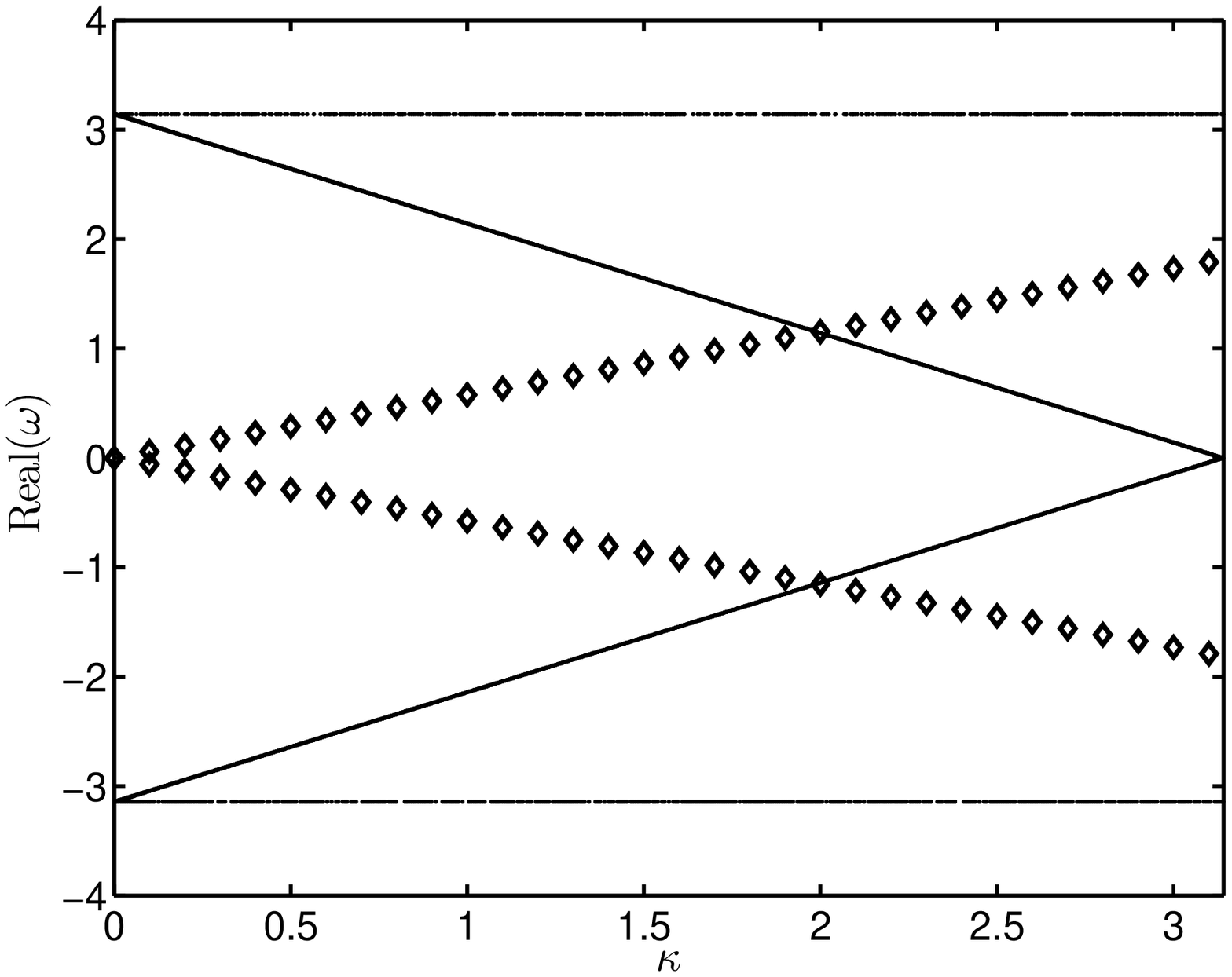}}
\scalebox{0.445}[0.445]{\includegraphics[angle=0]{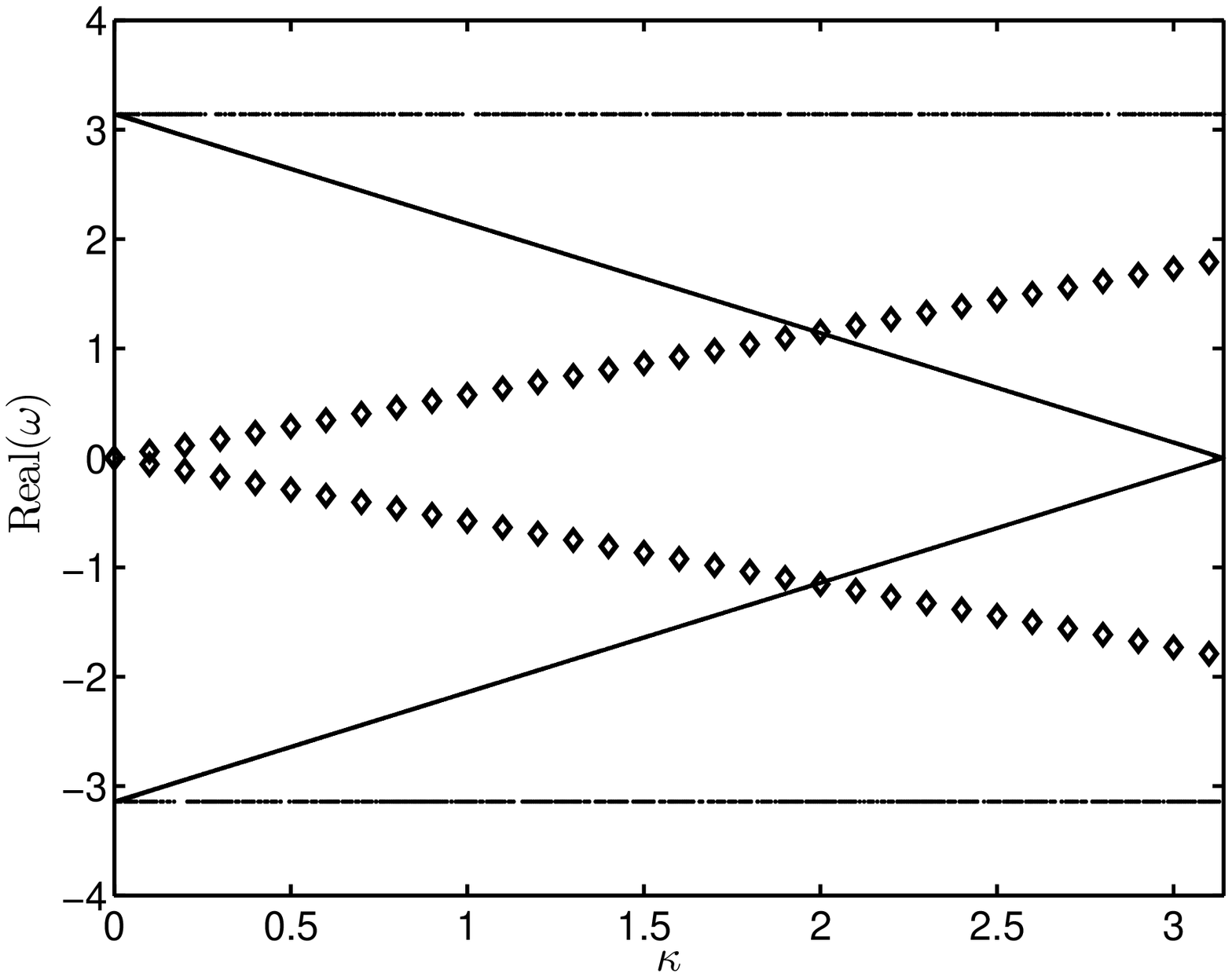}}\\
{\centering\hspace{1.cm}(a) $s=1.99$ \& $\chi=4/s-0.1$\hspace{4.cm}(b) $s=1.99999$ \& $\chi=4/s-0.1$}
\caption{Stability properties of the type II absorbing term: $\theta=0$ , $n=m=1/2$. $\diamondsuit$: the exact dispersion relations.}\label{fig:5}
\end{center}
\end{figure}

\begin{figure}[!htbp]
\begin{center}
\scalebox{0.445}[0.445]{\includegraphics[angle=0]{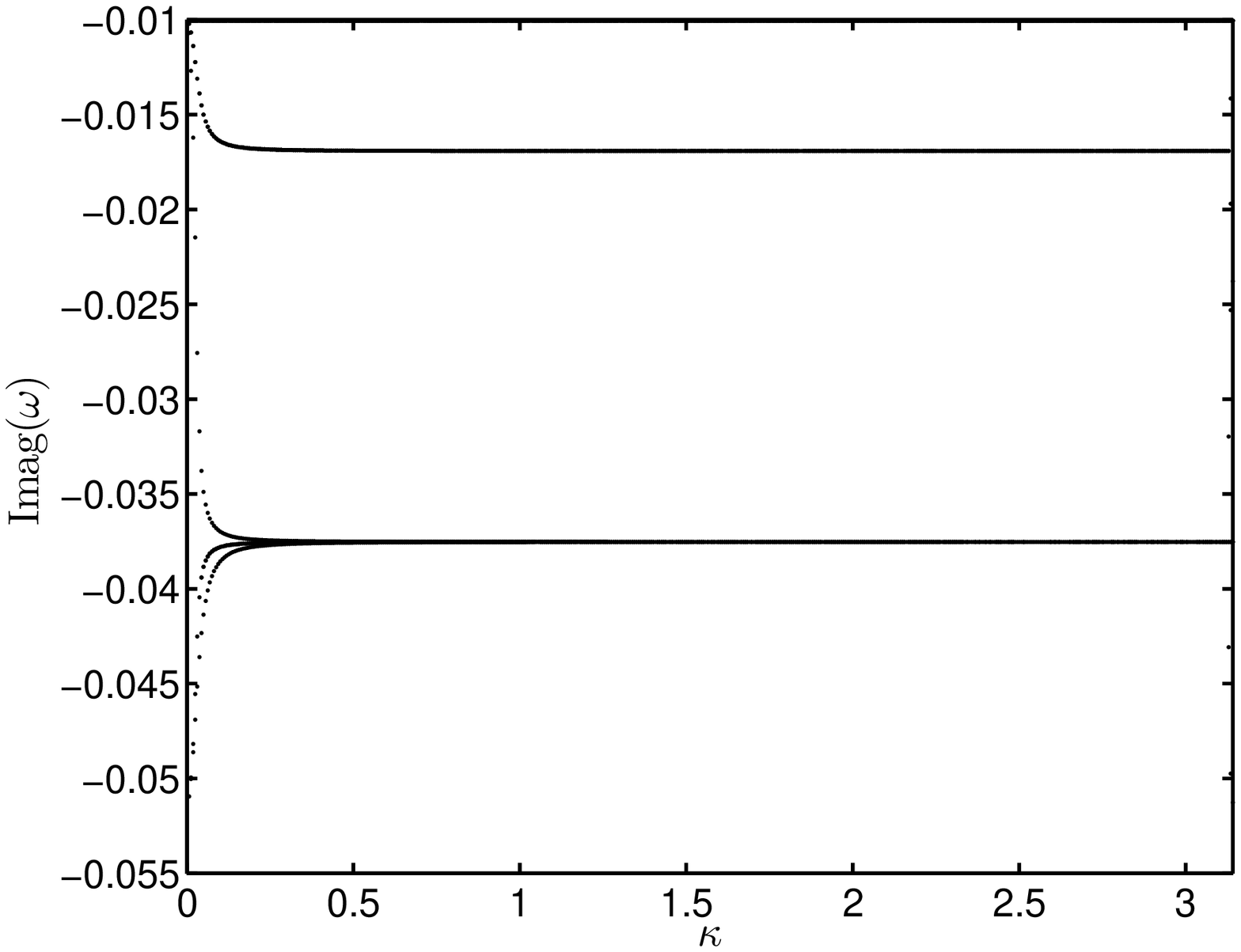}}
\scalebox{0.445}[0.445]{\includegraphics[angle=0]{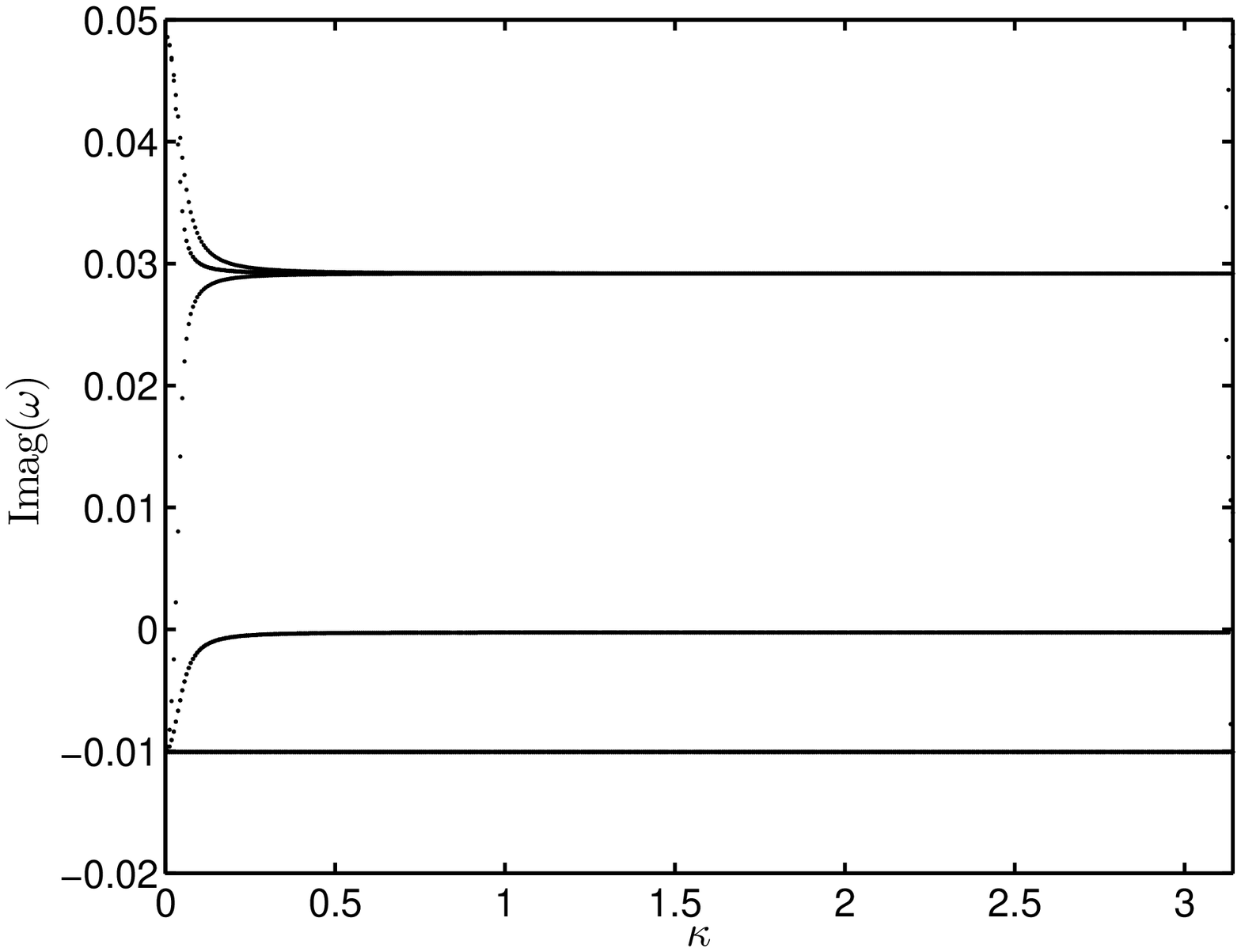}}\\
{\centering\hspace{1.cm}(a) $s=1.99$ \& $\chi=4/s-0.1$\hspace{4.cm}(b) $s=1.99$ \& $\chi=4/s+0.1$}
\caption{Stability properties of the type III absorbing term: $\theta=0$ , $n=m=1/2$.}\label{fig:6}
\end{center}
\end{figure}

We now investigate the second and third type absorbing terms. First, let us assess the critical value of the absorbing strength $\chi$ in Eq. (\ref{criticalchi}). If $\chi=4/s$, the effective absorbing strength $\tilde{\sigma}$ in Eq. (\ref{effectivesigma}) will become singular and the corresponding dissipation relations are illustrated in Fig. \ref{fig:2}. It is easy to observe that under this condition, for two values  of $s$  ($s=1.99$ or $s=1.99999$),  the corresponding LBS is stable. 
Further, if $\chi$ is larger than $4/s$,  we can observe from Fig. \ref{fig:3} that most  modes are unstable. These observations validate the theoretical  analysis presented in Sec. \ref{macroabc}. 
The dispersion relations for  smaller $\chi$ are displayed in Fig. \ref{fig:4}. From these results, it is seen that all  modes are stable. 
From Fig. \ref{fig:5}, the dispersion errors disappear.  Certainly, if $\chi$ becomes small, the dispersion error will appears again.  The suggestion for the choice of $\chi$ is that for nearly vanishing viscosity acoustic problems, the magnitude of $\chi$  is in the range [0,2].  Theoretically, there exist  similar results for the type II and type II absorbing terms. Further, let us investigate the type II absorbing term by the von Neumann method for several cases . In Fig. \ref{fig:6} ,  dissipation is shown for two choices of $\chi$. It is easy to conclude that when $\chi$ is larger than the critical $\chi=4/s$, LBS is unstable; otherwise, LBS is stable.

\begin{figure}[!htbp]
\begin{center}
\scalebox{0.445}[0.445]{\includegraphics[angle=0]{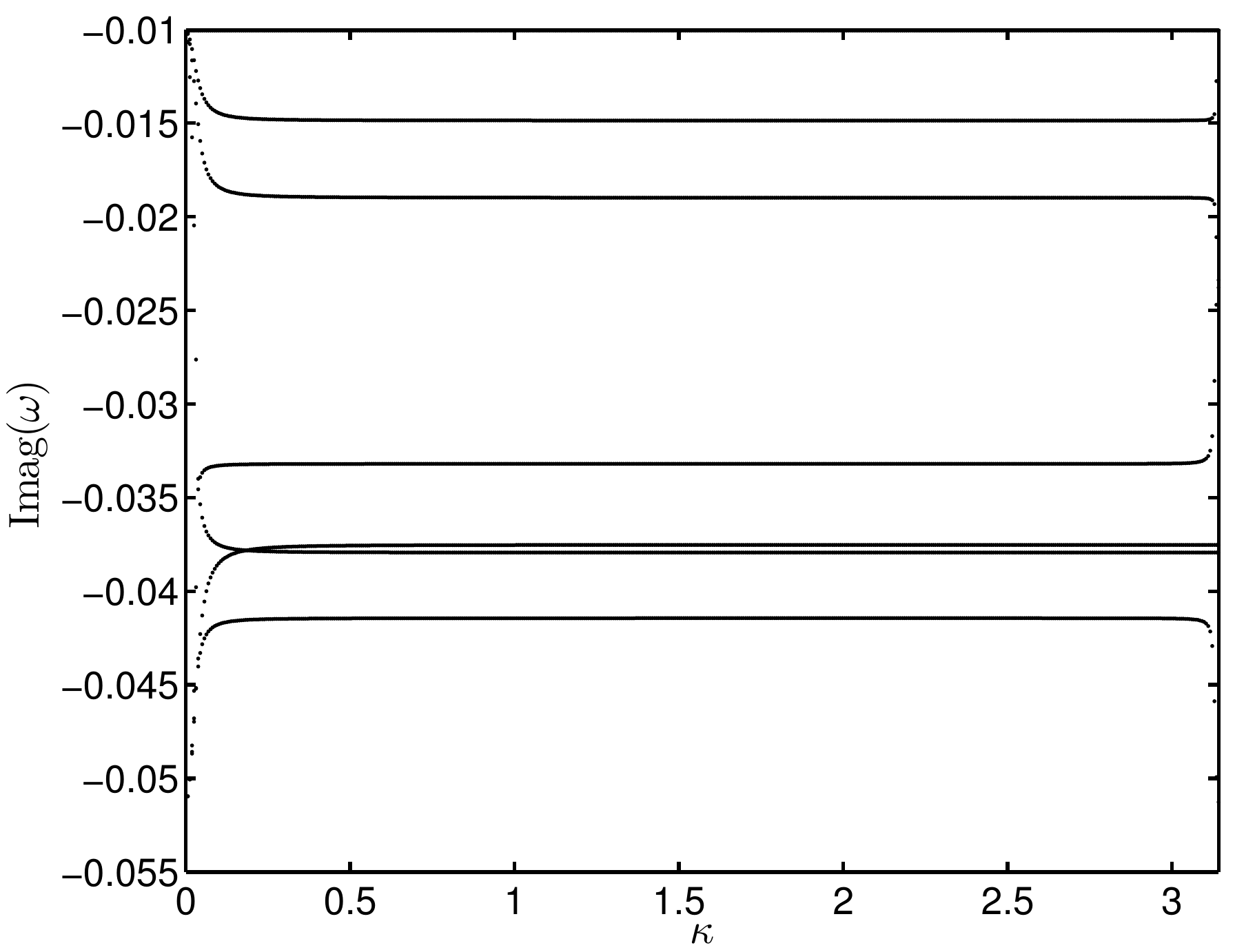}}
\scalebox{0.445}[0.445]{\includegraphics[angle=0]{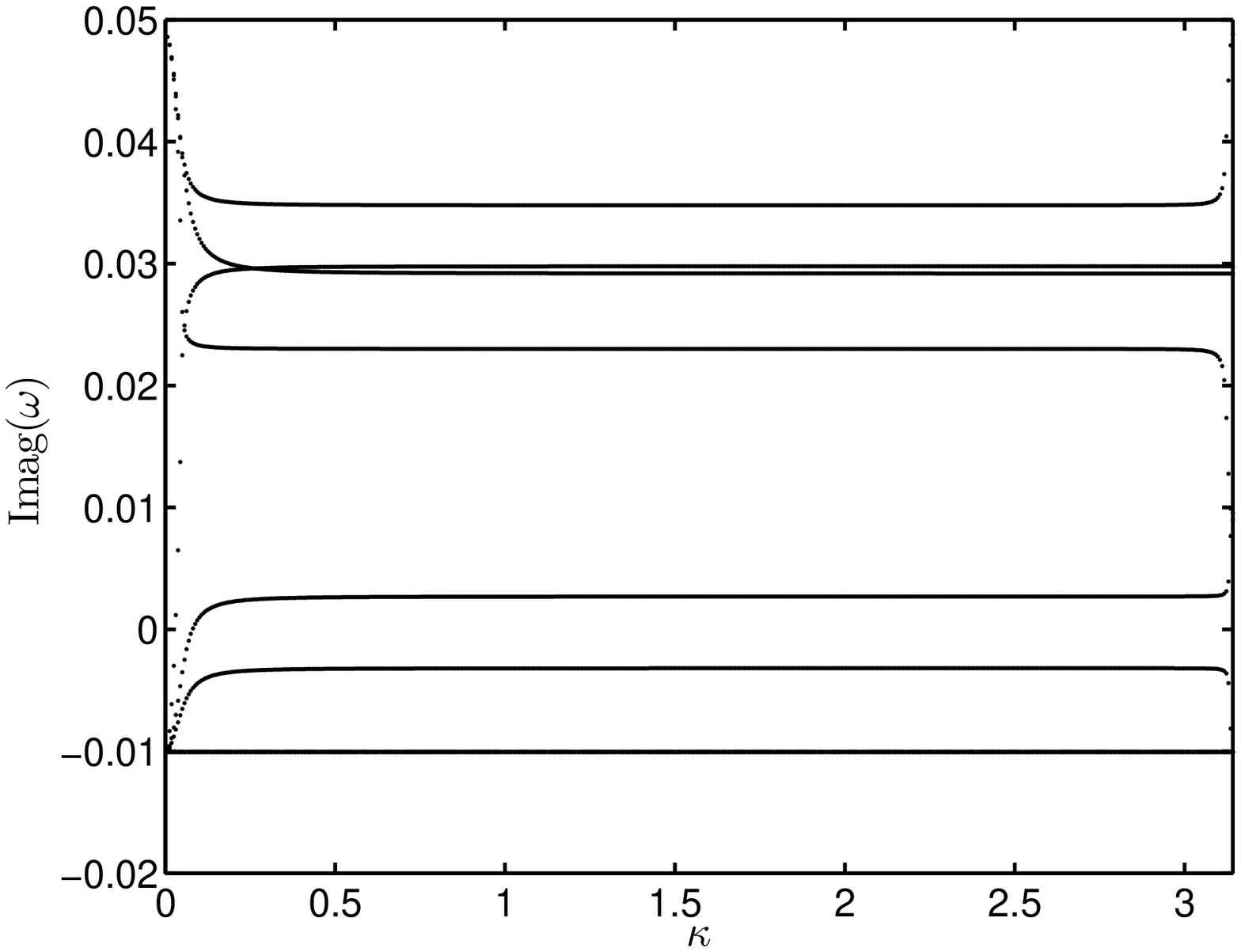}}\\
{\centering\hspace{1.cm}(a) $s=1.99$ \& $\chi=4/s-0.1$\hspace{4.cm}(b) $s=1.99$ \& $\chi=4/s+0.1$}
\caption{Stability properties of the type III absorbing term: $\theta=0$ , $n=m=1/2$. $(u_x^f,u_y^f)=(0.1,0)$.}\label{fig:8}
\end{center}
\end{figure}

\begin{figure}[!htbp]
\begin{center}
\scalebox{0.445}[0.445]{\includegraphics[angle=0]{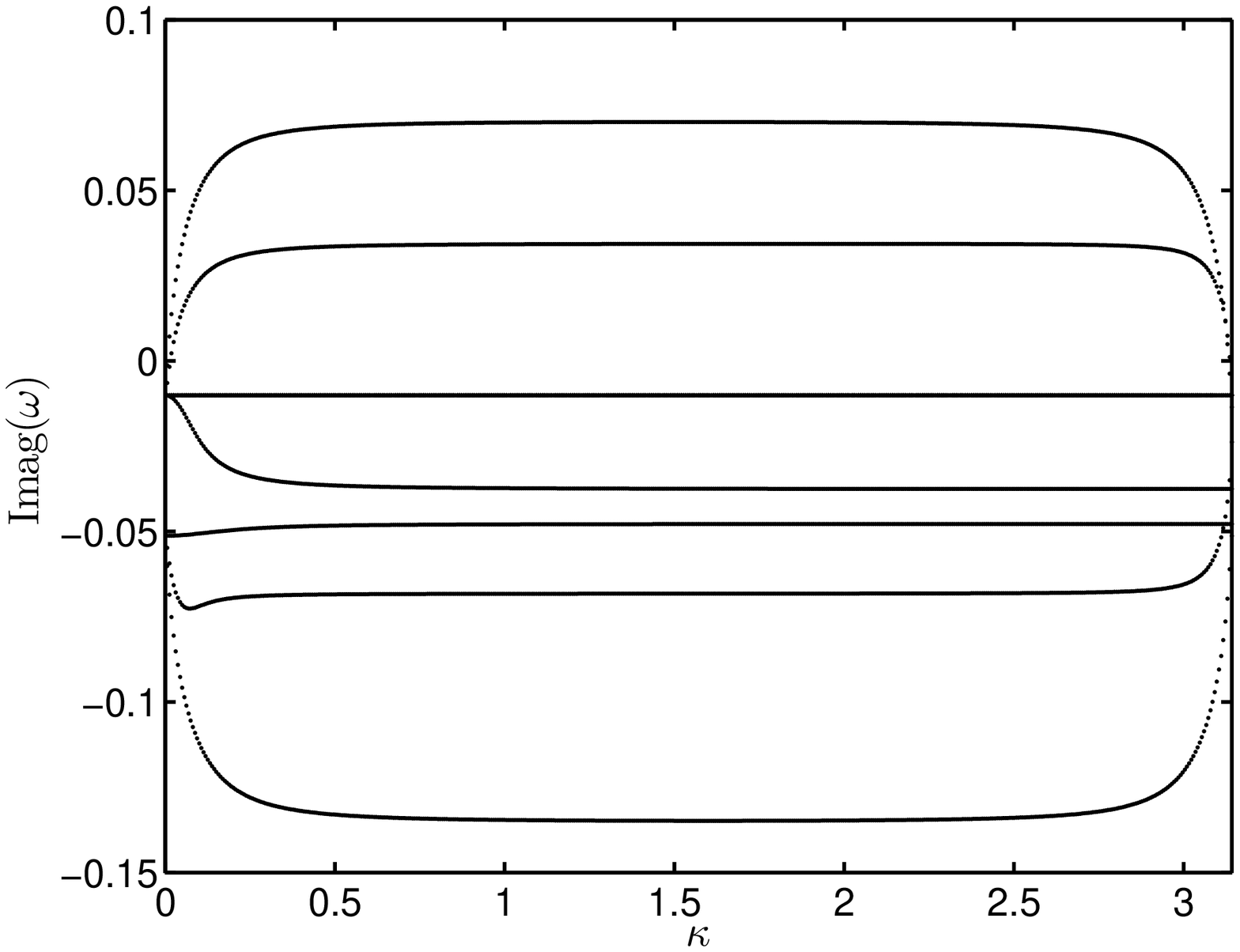}}
\scalebox{0.445}[0.445]{\includegraphics[angle=0]{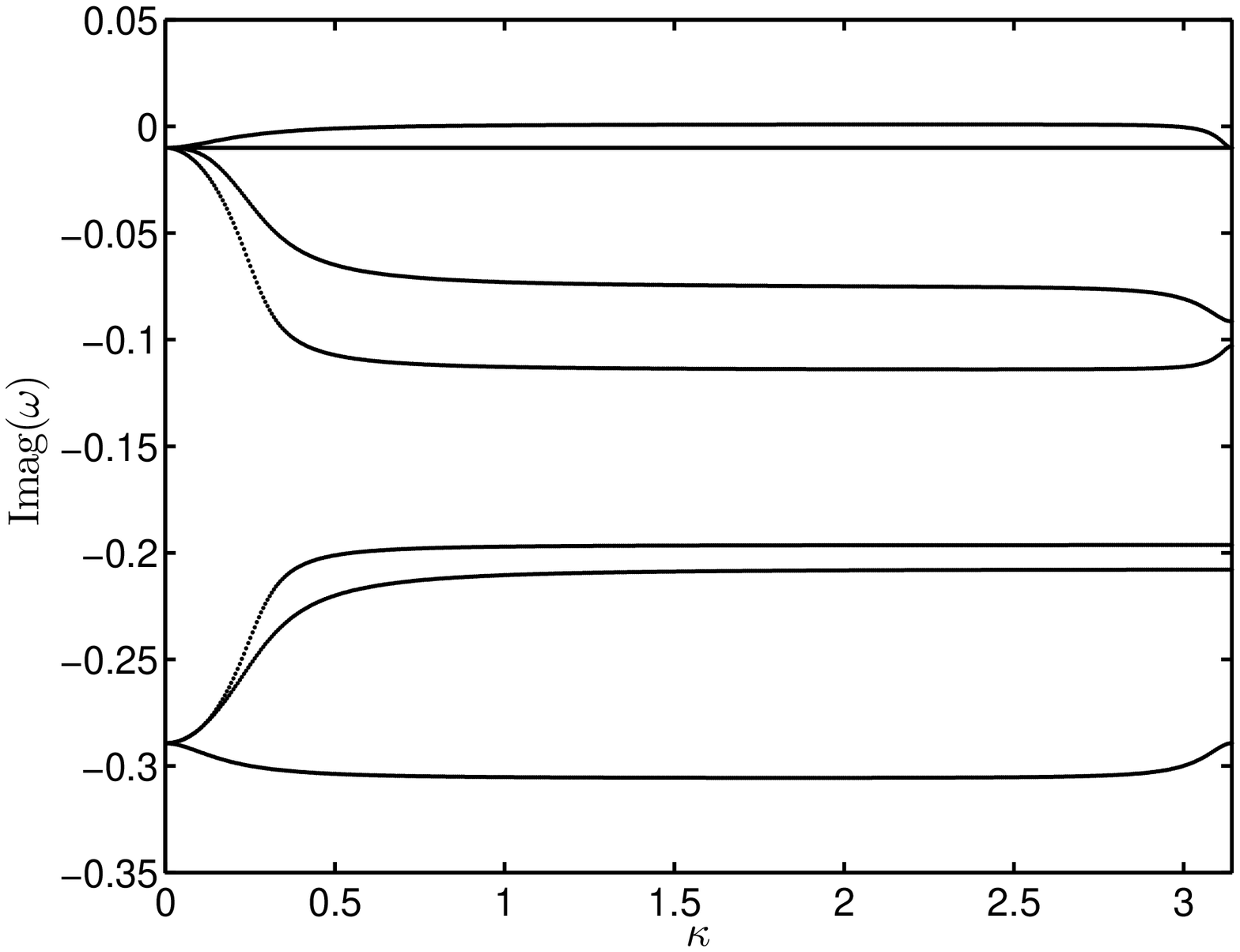}}\\
{\centering\hspace{1.cm}(a) $s=1.99$ \& $\chi=4/s-0.1$\hspace{4.cm}(b) $s=1.99$ \& $\chi=(4-1)/s$}
\caption{Stability properties of the type III absorbing term: $\theta=0$ , $n=m=1/2$. $(u_x^f,u_y^f)=(0.1,0)$.}\label{fig:9}
\end{center}
\end{figure}

We consider $u^f\neq 0$ and $(u_x^f,u_y^f)=(0.1,0)$. In Fig. \ref{fig:8}, $\rmu$ is parallel to $\bf k$ and the dissipation is given for the type II absorbing term. Looking at these  results, it is clear that the critical $\chi$ of $u^f= 0$ is suitable for that of $u^f\neq 0$.  The same phenomenon can be observed. However, this kind of phenomenon can not be observed for the type III absorbing term. Corresponding results are displayed in Fig. \ref{fig:9}. $\chi=4/s$ is not a critical value for the type III absorbing term. Numerical validation shows that the critical value of $\chi$ is smaller than $3/s$. So, it can be concluded that the type II absorbing term is the best suited absorbing term for LBS. Meanwhile, we can obtain that for LBS, when the far field $\rmu^f$ is not equal to 0, the absorbing terms should damp all possible physical quantities $\sbracket{\partial_t \rho, \partial_\alpha (\rho u^*_\alpha) \mbox{ and } \partial_\beta(\rho u^*_\alpha u^*_\beta+\delta_{\alpha\beta}p^*)}$ in the macroscopic systems. This is different from the classical absorbing layer for compressible Navier-Stokes equations.

 \begin{figure}[!htbp]
\begin{center}
\scalebox{0.445}[0.445]{\includegraphics[angle=0]{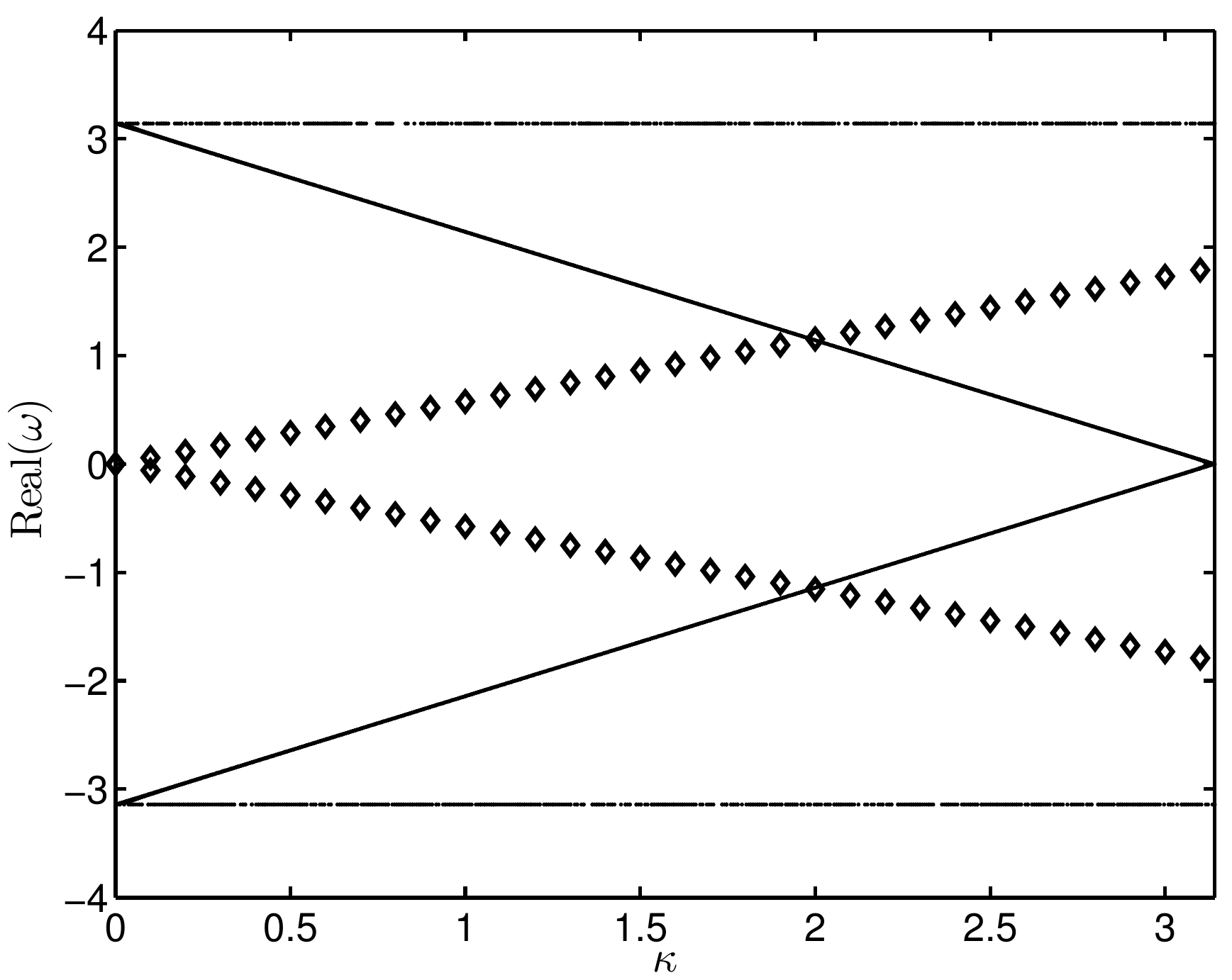}}
\scalebox{0.445}[0.445]{\includegraphics[angle=0]{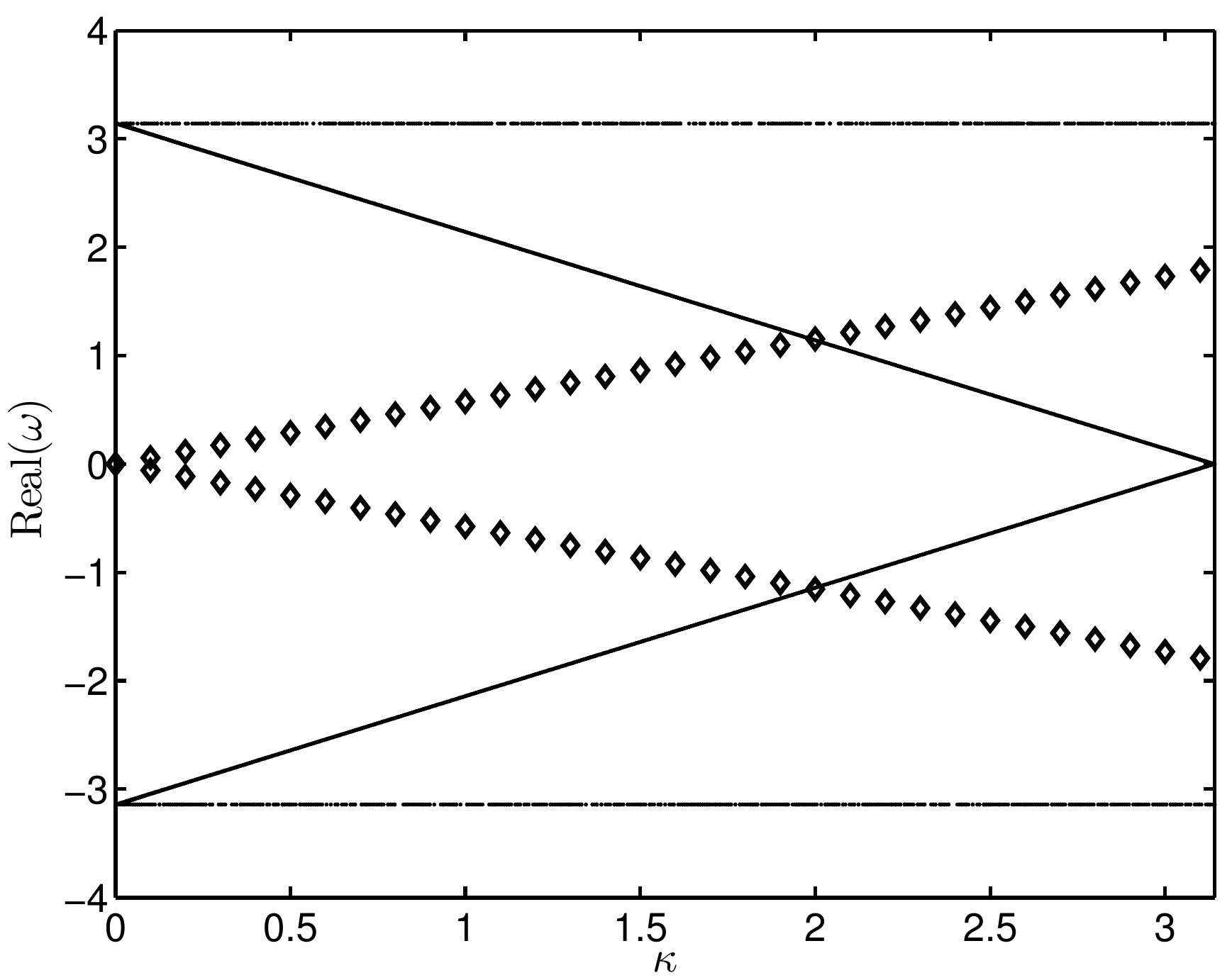}}\\
{\centering\hspace{1.cm}(a) $s=1.99$ \& $\chi=4/s+0.1$\hspace{4.cm}(b) $s=1.99$ \& $\chi=4/s-0.1$}
\caption{Dispersive properties of the type II absorbing term: $\theta=0$ , $n=m=1/2$. $\diamondsuit$: the exact dispersion relations.}\label{fig:7}
\end{center}
\end{figure}

\section{Numerical assessment}
In this section, the type II absorbing term is assessed by solving numerically some classical acoustic problems.  All numerical investigations were implemented in PalaBos \cite{palabos}.

\subsection{2D acoustic pulse}
We first consider a 2D acoustic pulse source. Assuming that  the viscosity effect is negligible on acoustic waves, the acoustic pulse problem possesses an analytical solution \cite{xusagaut,tam}.  The initial profile is given as follows

\begin{equation}\label{pulseprofile}
\left\{\begin{array}{ll}
\rho_0&=1+\rho\prime\\
u_0 &=U_0\\
v_0 & =0\\
\end{array}\right.
\end{equation}
where $\rho\prime$, $\epsilon$, $\alpha$, $r$ and $U_0$ are defined by 

\begin{equation}\label{3:para0}
\rho\prime=\epsilon{\rm exp}(-\alpha\cdot r^2),\ \epsilon=10^{-3},\ \alpha={\rm ln}(2)/b^2,\ r=\sqrt{(x-x_0)^2+(y-y_0)^2}。
\end{equation}
The computational domain is a $[0,1]\times [0,1]$ square. And the characteristic length is equal to 1. The parameter $b$ in Eq. (\ref{3:para0}) is equal to 1/20. The exact solution of $\rho\prime$ (if $(x_0,y_0)=(0,0)$) is given by \cite{tam}

\begin{equation}
\rho\prime(x,y,z)=\frac{\epsilon}{2\alpha}\int_{0}^{\infty}{\rm exp}\left(-\frac{\xi^2}{4\alpha}\right){\rm cos}(c_st\xi){\rm J}_0(\xi\eta)\xi{\rm d}\xi
\end{equation}
where $\eta=\sqrt{(x-U_0t)^2+y^2+z^2}$ and ${\rm J}_0(\cdot)$ is the Bessel zeroth-order function of the first kind. The adopted lattice resolution is $200^2$.  For the sake of convenience, we define the following rescaled time unit

\begin{equation}
\tilde{t} =1/(2t\cdot c_s).
\end{equation}

Theoretically, when a sponge layer is enforced around the computational domain, 
after several time units, the wave propagate outside the domain and the fluid flow field should converge to the steady reference flow.  The time for this phenomena is equal to $\tilde{t}_{\rm out} =1/(\sqrt{2}t\cdot c_s)$. The ideal state is that the density fluctuation in the computational domain vanishes. In order to address this phenomena, one defines the $L^2$ relative error as
 
\begin{equation}
E_{L^2}(t,\chi)=\sqrt{\frac{\sum_{i=1}^{N_{\rm nodes}}(\rho({\rm x}_i,t)-\rho_{\Ref}({\rm x}_i,t))^2}{\sum_{i=1}^{N_{\rm nodes}}(\rho_{\Ref}({\rm x}_i,t))^2}}.
\end{equation}
For  the current investigations, $\rho_{\Ref}=1$  in the initial condition (\ref{pulseprofile}).

 \begin{figure}[!h]
\begin{center}
\scalebox{0.45}[0.45]{\includegraphics[angle=0]{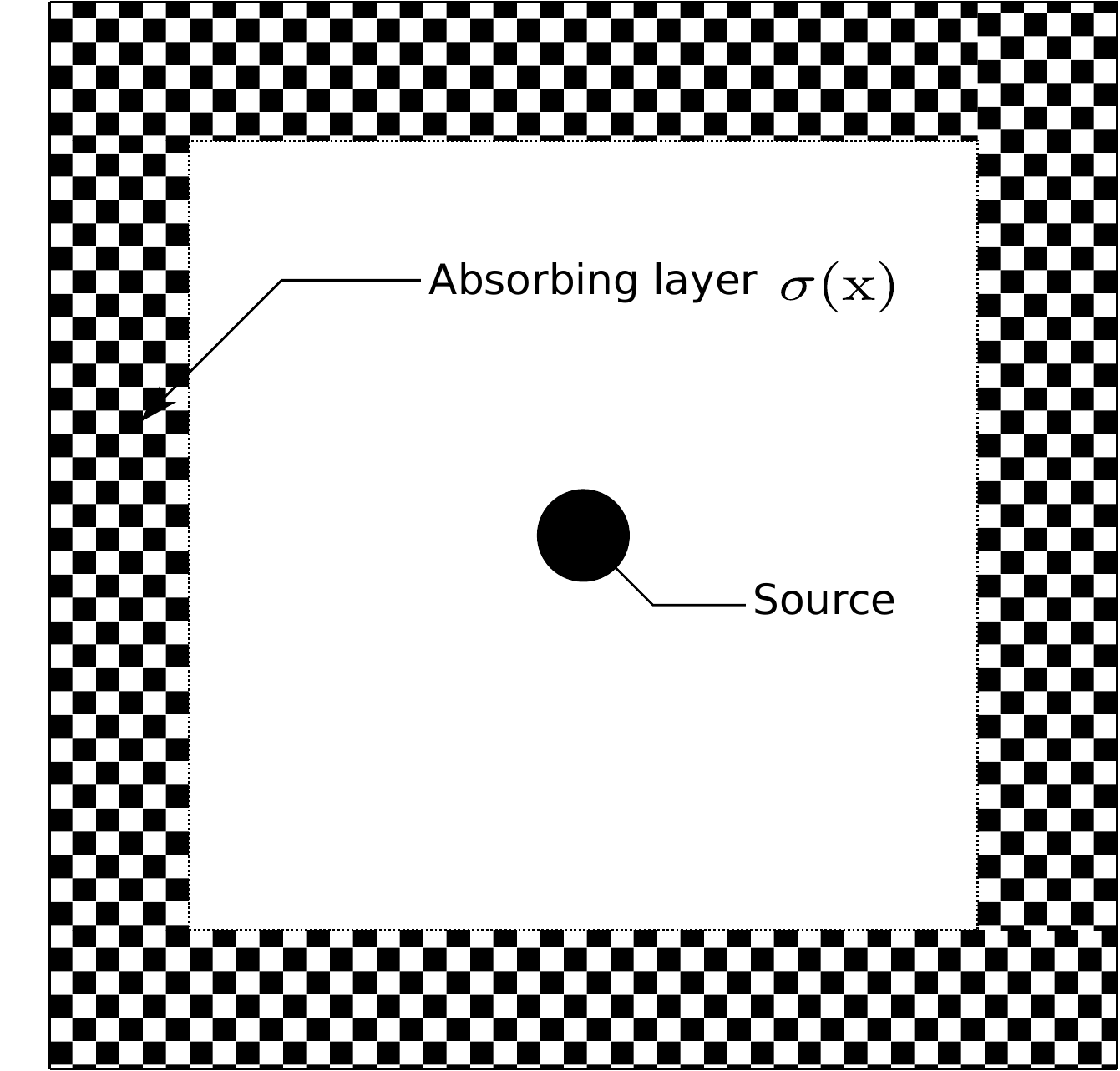}}
\caption{The schematic figure of the computational domain with the absorbing layer.}\label{fig:abc}
\end{center}
\end{figure}

The computational domain with the absorbing layer is shown schematically in Fig.\ref{fig:abc}.  The normalized $\tilde{\sigma}(\rmx)$ denoted by the formula (\ref{normalizedsigma}) will be used. The absorbing strength $\chi$ is chosen equal to $4/s-\varepsilon$, where $\varepsilon=0.001$ denote a small positive constant. The new $\sigma$ for this problem is defined by

\begin{equation}
\sigma(\rmx)=\chi\tilde{\sigma}(\rmx).
\end{equation}

In order to validate influence of the absorbing layer thickness, the thickness of the absorbing layer is chosen equal to $m\cdot b$ (where $m$ is a positive integer). The current investigations will be compared with the viscosity absorbing strategy,  which consists in linearly decreasing the relaxation frequency  from the physical value to 1 in the sponge region. The Reynolds number of this problem is defined by $Re=1/\nu=10^7$ and the viscosity nearly vanishes. For all simulations, the Dirichlet boundary condition is used.

\begin{figure}[!hbtp]
\begin{center}
\scalebox{0.3}[0.3]{\includegraphics[angle=0]{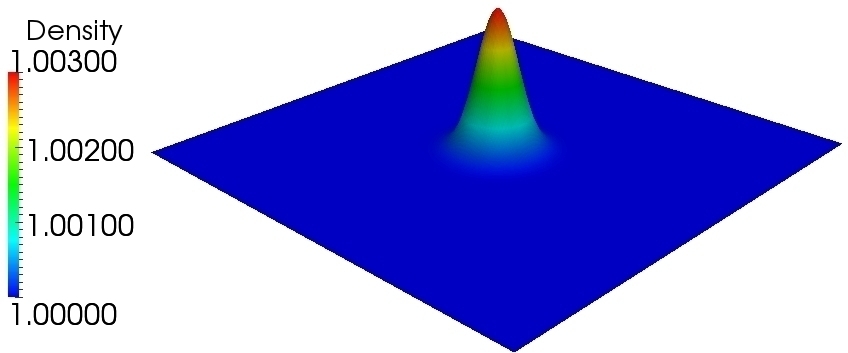}}
\caption{The initial profile of the 2D acoustic pulse.}\label{fig:n2}
\end{center}
\end{figure}

\begin{figure}[!hbtp]
\begin{center}
\scalebox{0.3}[0.3]{\includegraphics[angle=0]{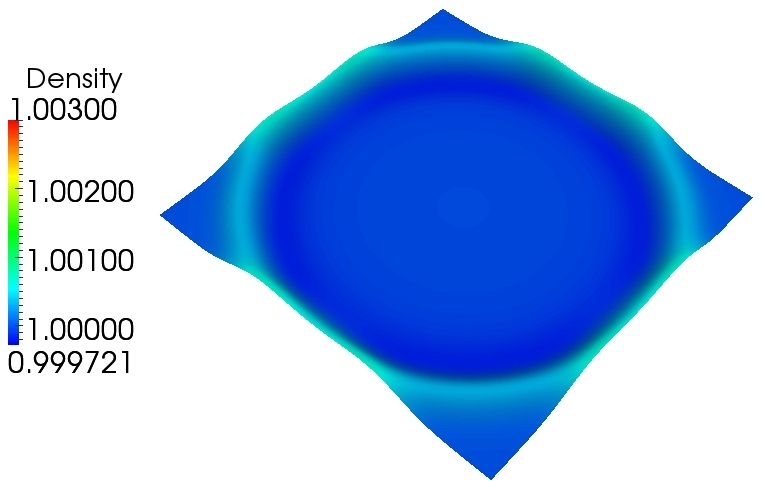}}
\scalebox{0.3}[0.3]{\includegraphics[angle=0]{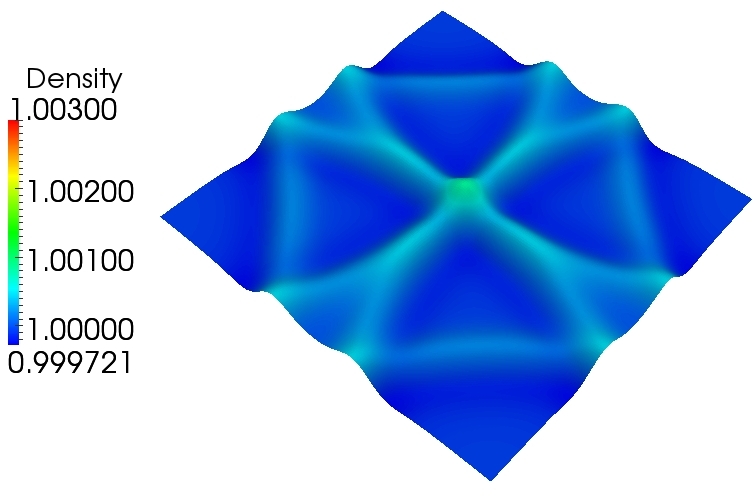}}\\
{\centering (a-1) At the time unit $T=\tilde{t}$ \hspace{4cm} (a-2) At the time unit $T=2\tilde{t}$}\\
\scalebox{0.3}[0.3]{\includegraphics[angle=0]{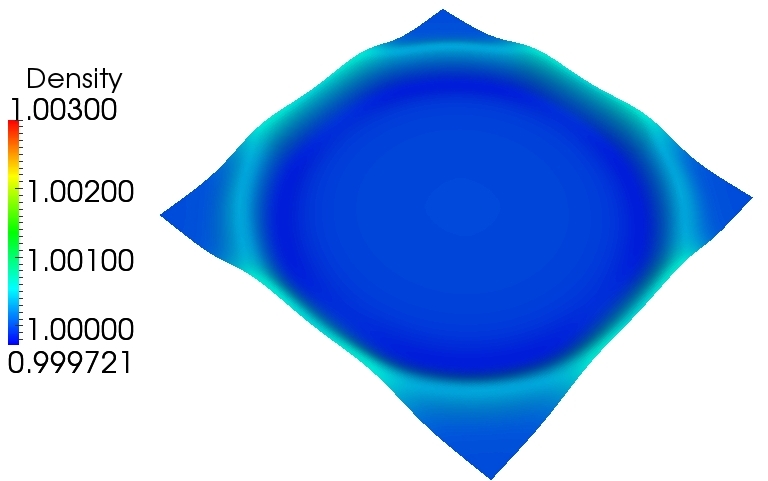}}
\scalebox{0.3}[0.3]{\includegraphics[angle=0]{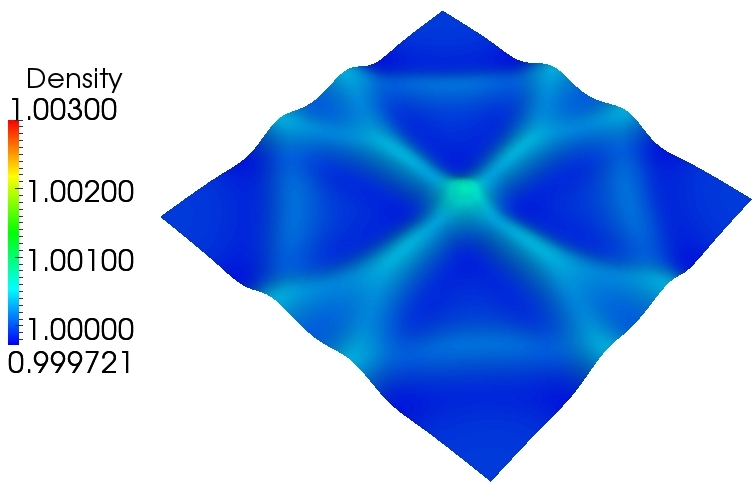}}\\
{\centering (b-1) At  the time unit $T=\tilde{t}$ \hspace{4cm} (b-2) At  the time unit $T=2\tilde{t}$}\\
\scalebox{0.3}[0.3]{\includegraphics[angle=0]{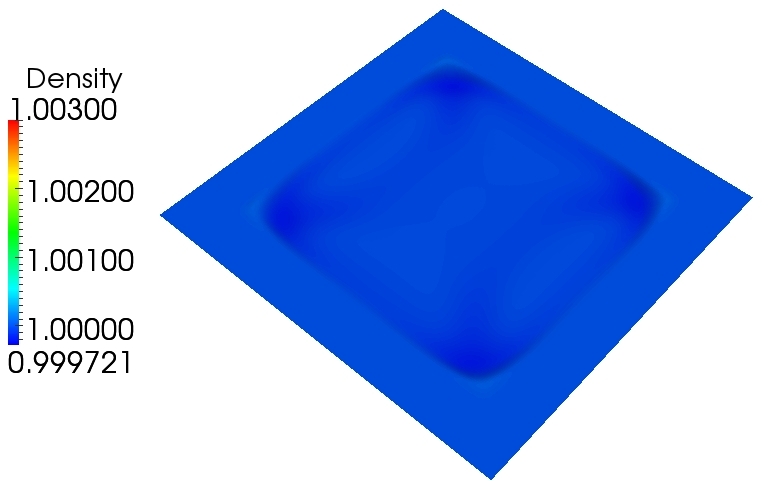}}
\scalebox{0.3}[0.3]{\includegraphics[angle=0]{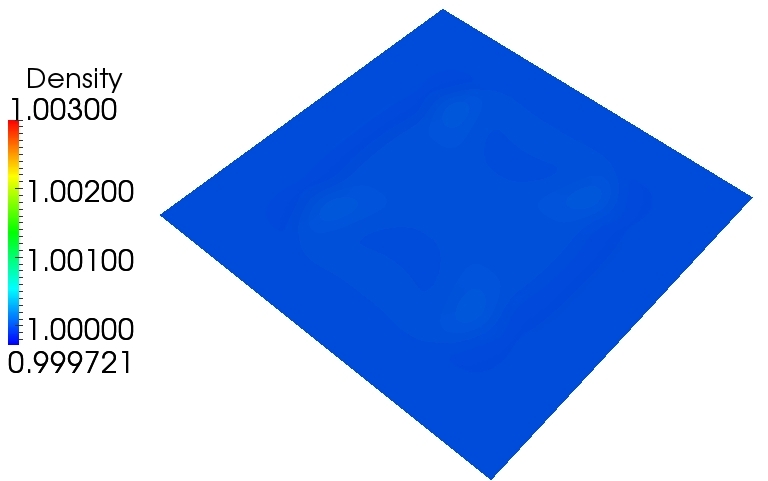}}\\
{\centering (c-1) At  the time unit $T=\tilde{t}$ \hspace{4cm} (c-2) At  the time unit $T=2\tilde{t}$}\\
\caption{The wave propagation at the different  times: (a) LBS without absorbing strategy; (b) LBS coupled with the viscosity damping strategy; (c) LBS coupled with the optimal absorbing term (\ref{app:newbgka2}). At the  time, the wave crests propagate toward the boundary. At the second time, the waves are reflected from the boundaries. The thickness of the absorbing layer is equal to $4b$. $\rmu^f=(U_0,0)=(0,0)$.}\label{fig:n3}
\end{center}
\end{figure}

\begin{figure}[!hbtp]
\begin{center}
\scalebox{0.6}[0.6]{\includegraphics[angle=0]{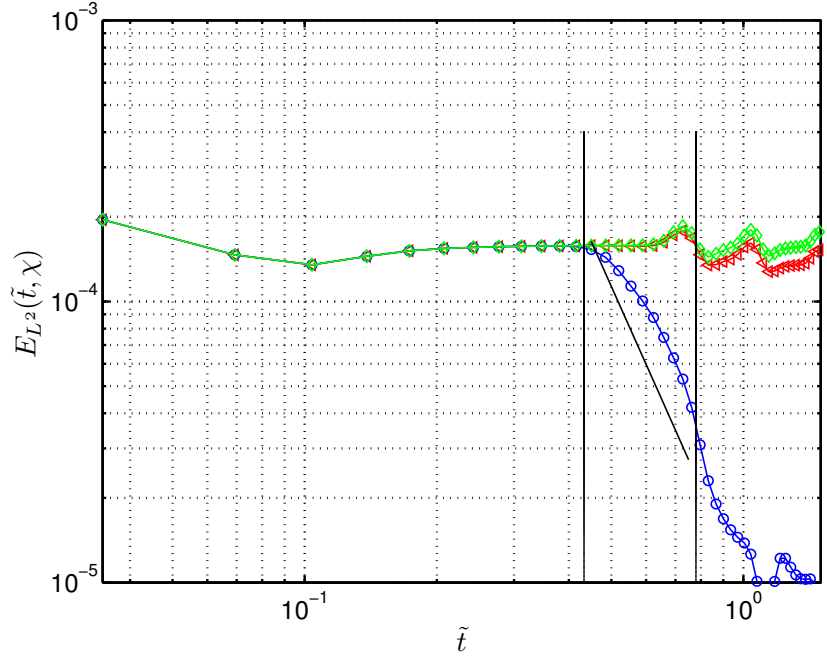}}
\caption{The $L^2$ relative error $E_{L^2}(\tilde{t},\chi)$ : $-{\diamond}-$: LBS without any absorbing strategy; $-\triangleleft-$: LBS coupled with viscosity damping strategy; $-{\rm o}-$ : LBS coupled with the optimal absorbing term (\ref{app:newbgka2}). The first vertical line indicates waves propagated to the absorbing layers. The second vertical line indicates  waves propagated to the boundaries. $\rmu^f=(U_0,0)=(0,0)$.}\label{fig:n4}
\end{center}
\end{figure}

The initial density profile is displayed in Fig.\ref{fig:n2}. In Fig.\ref{fig:n3}, the density profiles obtained using the three different strategies  are shown at different times for a zero mean flow. One can see that  without any absorbing strategies a  clear wave reflection from the boundaries is present. With the linearly decreasing relaxation parameter, there still exits  a significant wave reflection. With type II absorbing term,  only a  very weak wave reflection is observed. Especially, when the wave propagates across the absorbing layer, it is hardly possible to observe the wave trace. In Fig.\ref{fig:n4},  a quantitative comparison based on $E_{L^^2}(\tilde{t},\chi)$ is shown. Some clear differences can be observed and type II term is the best one. Meanwhile, it can be concluded that using the viscosity damping strategy, the absorbing  performance is weak.   From the given reference line, it can be concluded that the wave in the proposed absorbing layer exhibits a fast decay with a decay exponent about $-3.5$.

\begin{figure}[!hbtp]
\begin{center}
\scalebox{0.27}[0.27]{\includegraphics[angle=0]{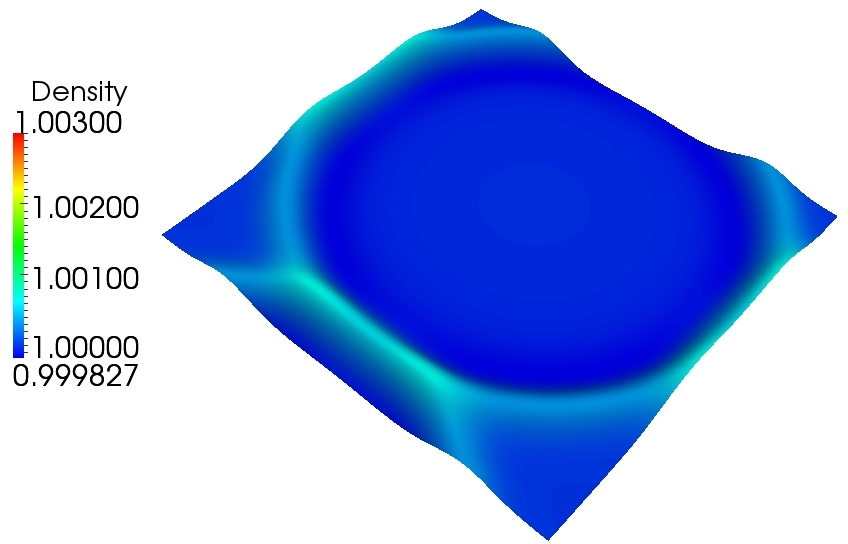}}
\scalebox{0.27}[0.27]{\includegraphics[angle=0]{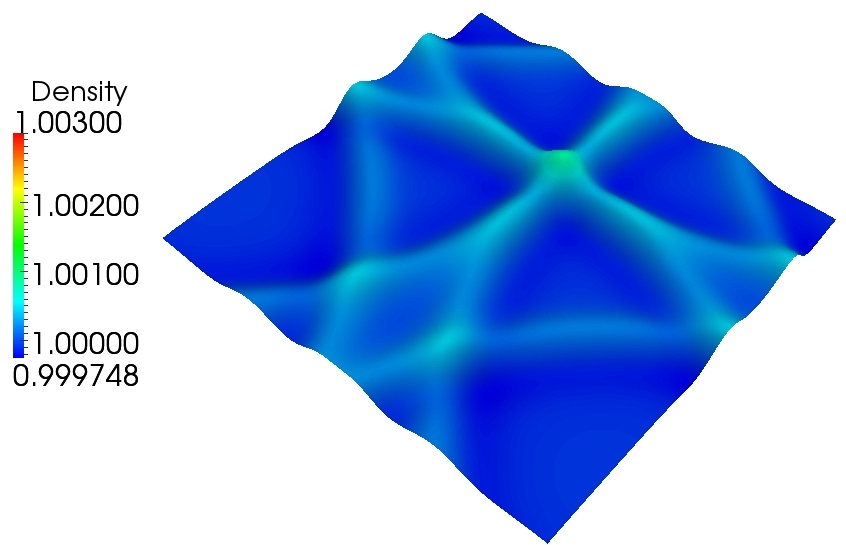}}\\
{\centering (a-1) At the time unit $T=\tilde{t}$ \hspace{4cm} (a-2) At the time unit $T=2\tilde{t}$}\\
\scalebox{0.27}[0.27]{\includegraphics[angle=0]{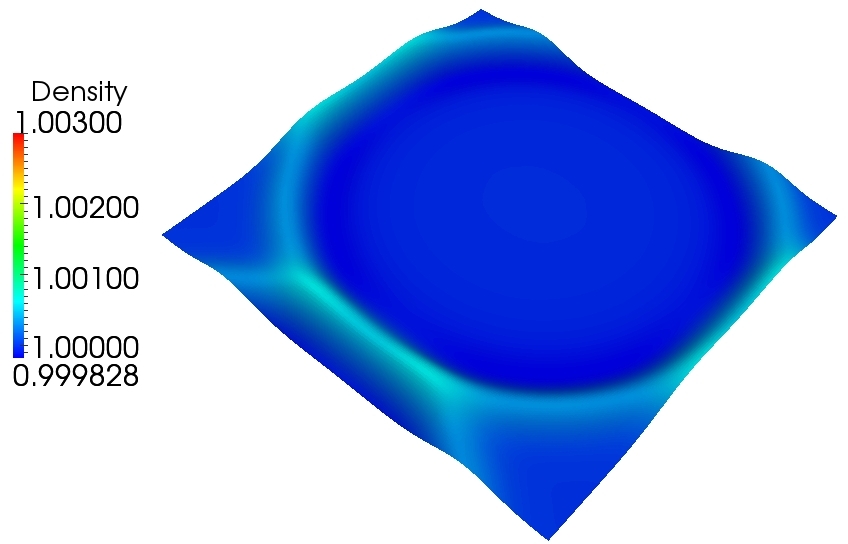}}
\scalebox{0.27}[0.27]{\includegraphics[angle=0]{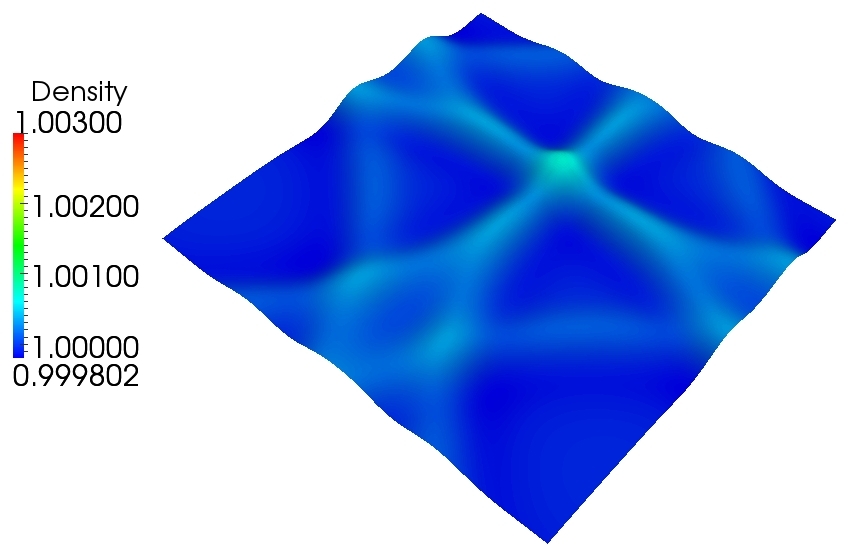}}\\
{\centering (b-1) At  the time unit $T=\tilde{t}$ \hspace{4cm} (b-2) At  the time unit $T=2\tilde{t}$}\\
\scalebox{0.27}[0.27]{\includegraphics[angle=0]{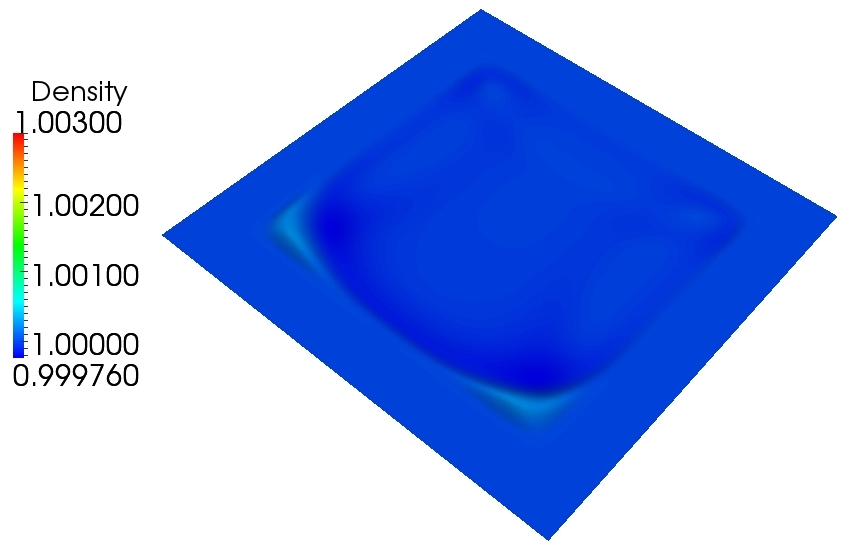}}
\scalebox{0.27}[0.27]{\includegraphics[angle=0]{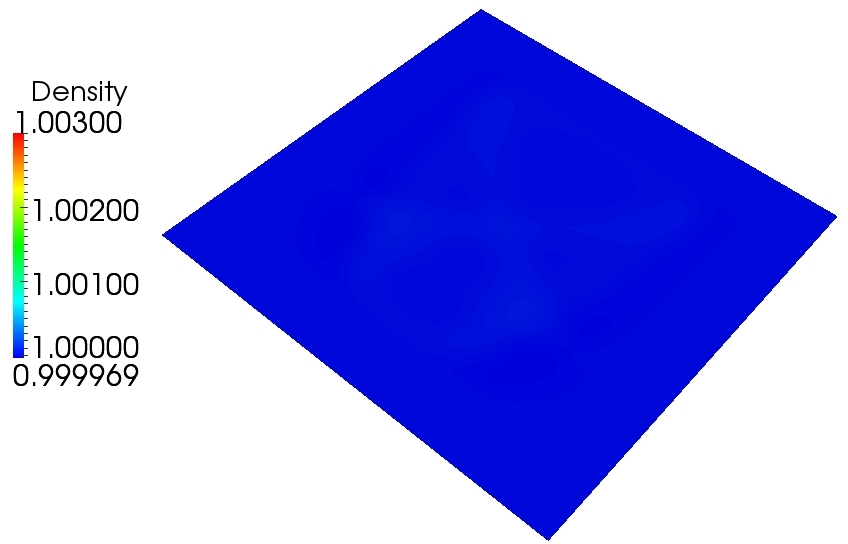}}\\
{\centering (c-1) At  the time unit $T=\tilde{t}$ \hspace{4cm} (c-2) At  the time unit $T=2\tilde{t}$}\\
\caption{The wave propagation at the different times: (a) LBS without absorbing strategy; (b) LBS coupled with the viscosity damping strategy; (c) LBS coupled with the optimal absorbing term (\ref{app:newbgka2}). At the first time, the wave crests propagate toward the boundary. At the second time, the waves are reflected from the boundaries. The thickness of the absorbing layer is equal to $4b$. $\rmu^f=(U_0,0)=(0.1,0)$}\label{fig:n5}
\end{center}
\end{figure}

\begin{figure}[!hbtp]
\begin{center}
\scalebox{0.6}[0.6]{\includegraphics[angle=0]{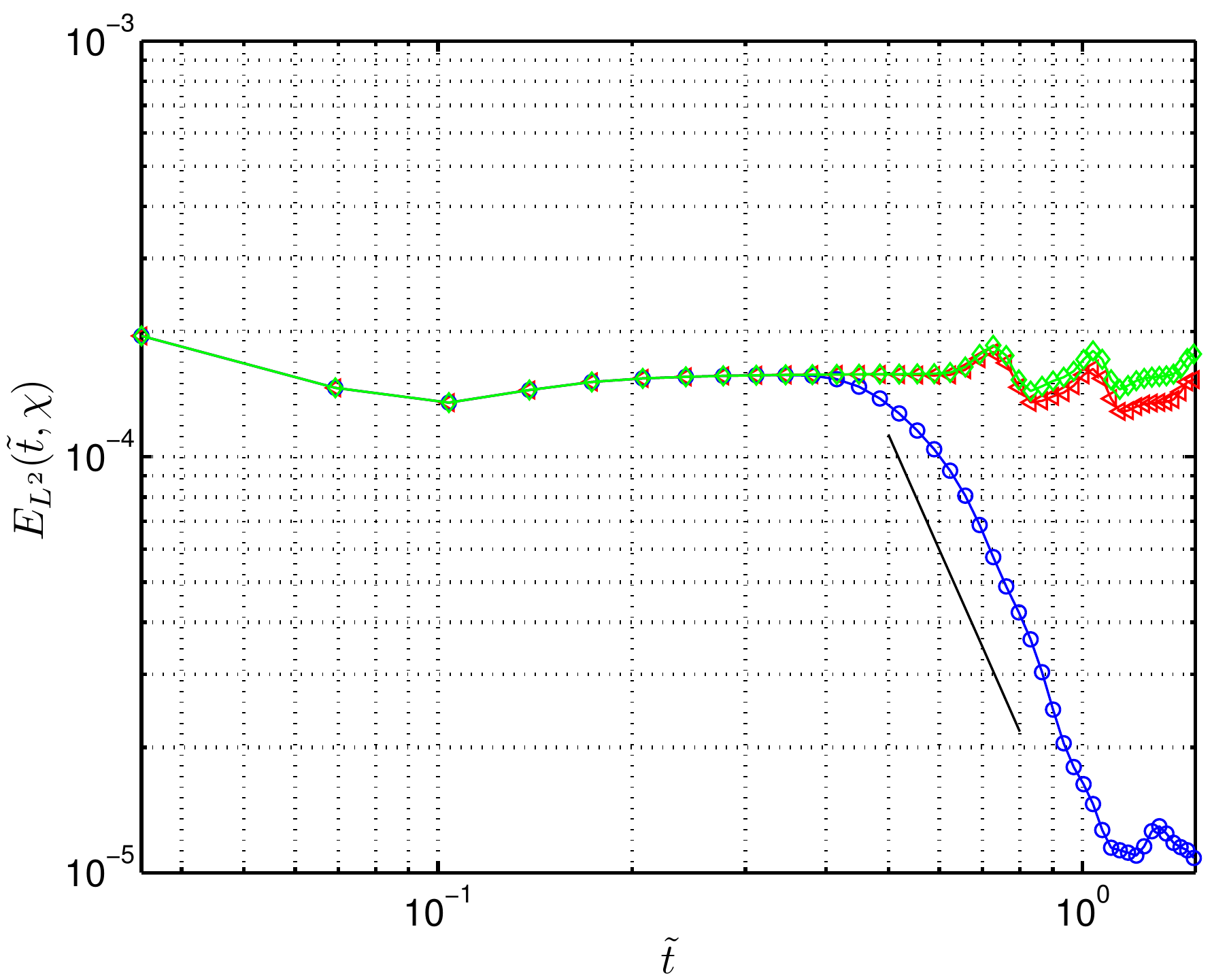}}
\caption{The $L^2$ relative error $E_{L^2}(\tilde{t},\chi)$: $-{\diamond}-$: LBS without any absorbing strategy; $-\triangleleft-$: LBS coupled with viscosity damping strategy; $-{\rm o}-$ : LBS coupled with the optimal absorbing term (\ref{app:newbgka2}).  $\rmu^f=(U_0,0)=(0.1,0)$}\label{fig:n6}
\end{center}
\end{figure}

\begin{figure}[!hbtp]
\begin{center}
\scalebox{0.6}[0.6]{\includegraphics[angle=0]{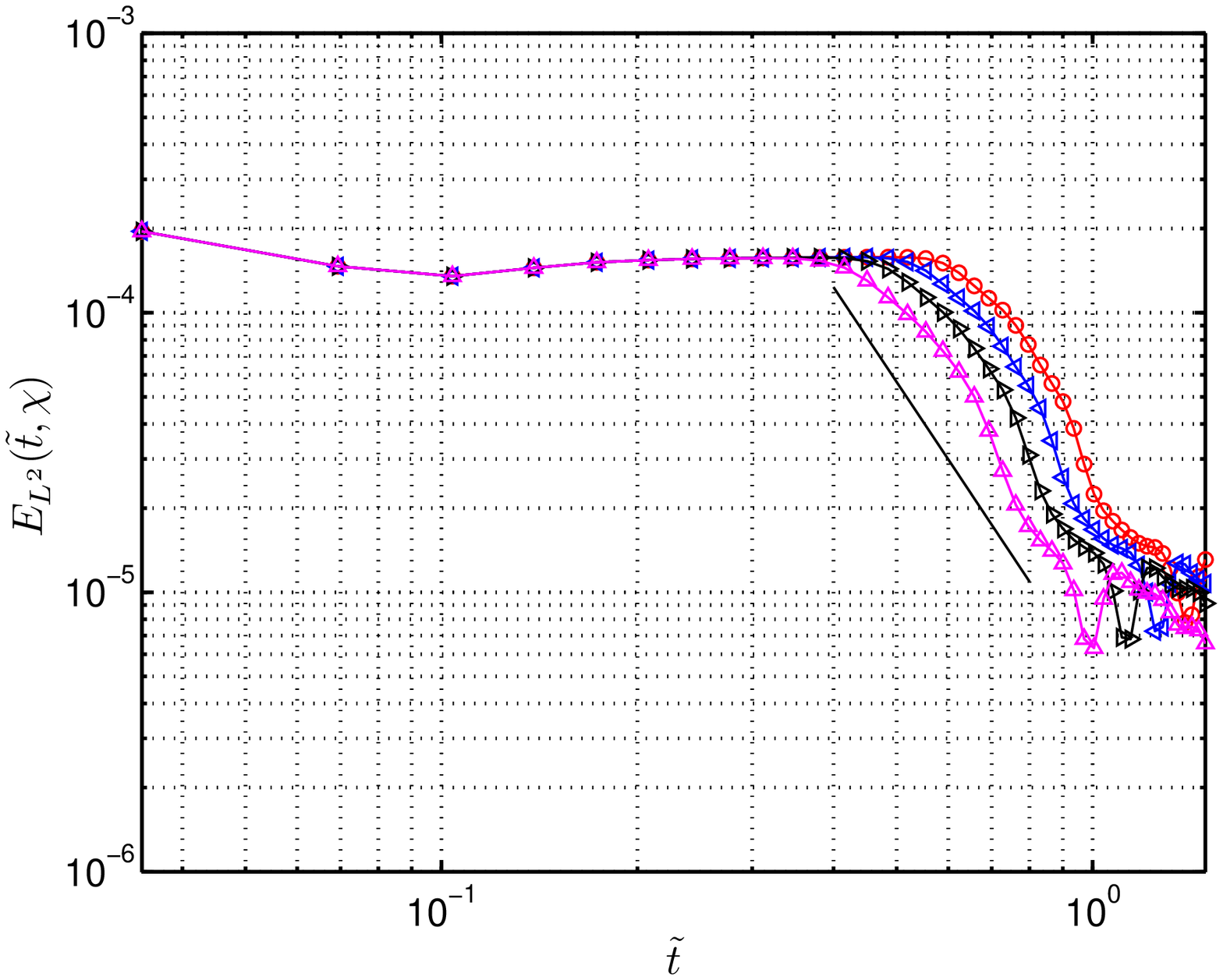}}
\caption{The $L^2$ relative error $E_{L^2}(\tilde{t},\chi)$ with respect to the thickness $m\cdot b$ of  the absorbing layers: $-{\rm o}-$: The thickness $2b$; $-\lhd-$: The thickness $2b$;  $-{\rhd}-$ :  The thickness $3b$; $-\bigtriangleup-$: The thickness $4b$. $\rmu^f=(U_0,0)=(0,0)$.}\label{fig:n7}
\end{center}
\end{figure}

In order to validate the type II absorbing term  in the presence of a uniform base flow, the computational results are given in Figs. \ref{fig:n5} and \ref{fig:n6}.  It is observed in the absorbing layer, the wave has a fast decay exponent  about -3.   In Fig.\ref{fig:n7}, the error $E_{L^2}(\tilde{t},\chi)$  is given. From this figure, it is easy to see that when the thickness of the absorbing layers increases, the error $E_{L^2}(\tilde{t},\chi)$ decreases. Meanwhile, a fast decay exponent close to -3.5 is measured.

These investigations demonstrate that the optimal absorbing term (\ref{app:newbgka2}) is very efficient in suppressing the wave reflection from the computational domain boundaries.

\subsection{Validations for time dependent acoustic line source}

\begin{figure}[!hbtp]
\begin{center}
\scalebox{0.7}[0.7]{\includegraphics[angle=0]{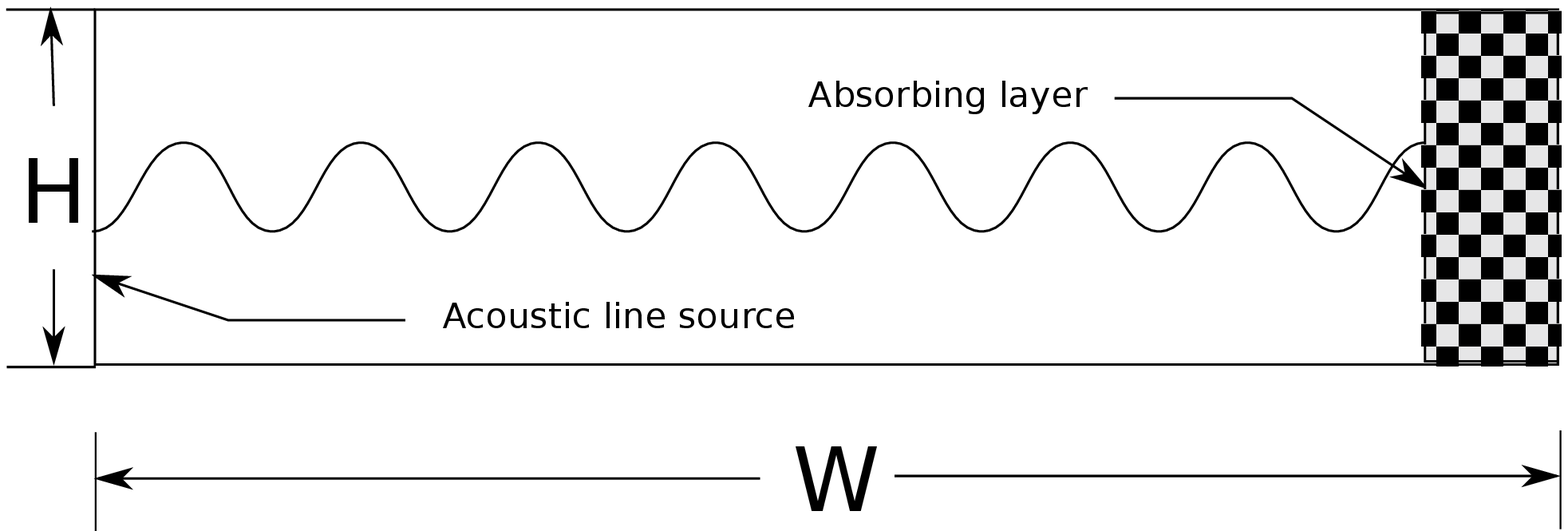}}
\caption{The schematic computational domain with the absorbing layer}\label{fig:n8}
\end{center}
\end{figure}

In order to validate the absorbing term II for a time-dependent acoustic source,  the case of an acoustic line source with and without  background flow is chosen.  The profile of the acoustic line source is described by

\begin{equation}
\slbracket{
\begin{array}{ll}
\displaystyle\rho(t) &= \rho_0 + \rho^\prime {\rm sin}(2\pi \omega t),\\[2mm]
\displaystyle\rmu(\rmx)&=(u_x,u_y),
\end{array}
}
\end{equation}

where the frequency $\omega=2$. The corresponding wave length is  $\lambda=c_s/2=0.288675$. The schematic domain with the absorbing layer is shown in Fig. \ref{fig:n8}. For present computations, the domain width is  $W=4H$ and the domain height is $H=1$, respectively, and the width of the absorbing layer is 0.8. The acoustic line source is fixed  at the left boundary $\rmx=(0,y)$. Periodic boundary conditions are enforced at the top and bottom boundaries. The outflow boundary condition is adopted for the right boundary $\rmx=(W,y)$. For the sake of convenience, we introduce two characteristic time scales for zero mean flows:

\begin{equation}
T_o=W/c_s,\ T_p=1/(\omega\cdot \delta t),
\end{equation}

where $\delta t$, $T_o$  and $T_p$ are  the time step,  the time of the wave propagation to the right boundary and the time of the wave propagation for one wave length, respectively. With the mean flow along $x$-axis, $T_o$ is defined by

\begin{equation}
T_m=W/(c_s+u_{x}^m),
\end{equation}

where $u_x^m$ is the mean flow velocity along $x$-axis. The number of grid points used for this configuration is $200^2$. The absorbing strength $\chi$ is chosen equal to $4/s-\varepsilon$, where $\varepsilon=0.001$ . $Re =1/\nu=10^6$.

\begin{figure}[!hbtp]
\begin{center}
\scalebox{0.7}[0.7]{\includegraphics[angle=0]{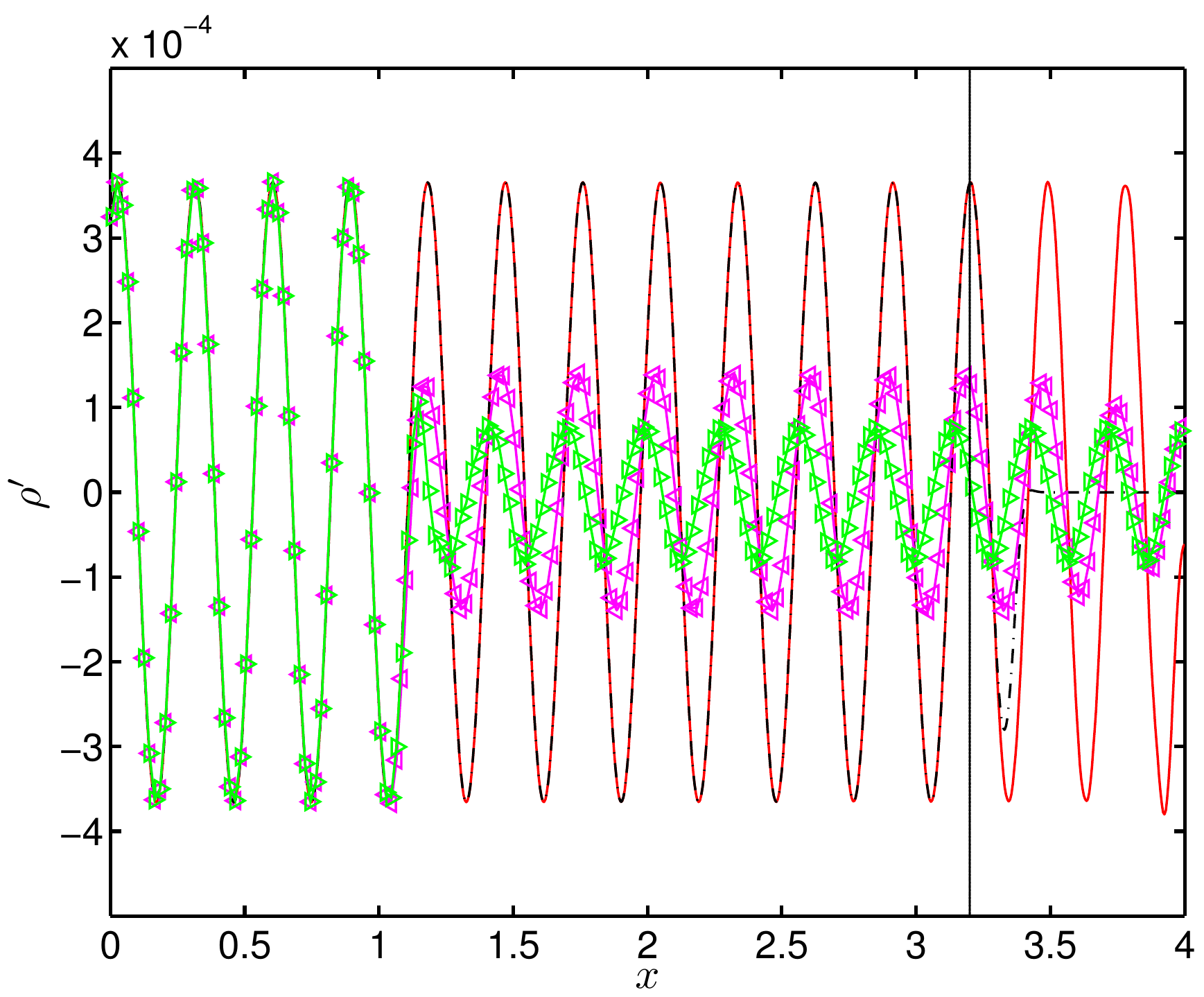}}
\caption{Comparisons among three different strategies for zero mean flow: (Red solid line) Reference $\rho^\prime$; (Dashed-dot line)  $\rho^\prime$ by LBS with the optimal absorbing strategy (\ref{app:newbgka2}); ($-\lhd-$) LBS coupled with viscosity damping strategy; ($-\rhd-$) LBS without any absorbing strategies. The sample time $t=T_o+10\cdot T_p$. The vertical line indicates the position of the absorbing layer.} \label{fig:n9}
\end{center}
\end{figure}

\begin{figure}[!hbtp]
\begin{center}
\scalebox{0.7}[0.7]{\includegraphics[angle=0]{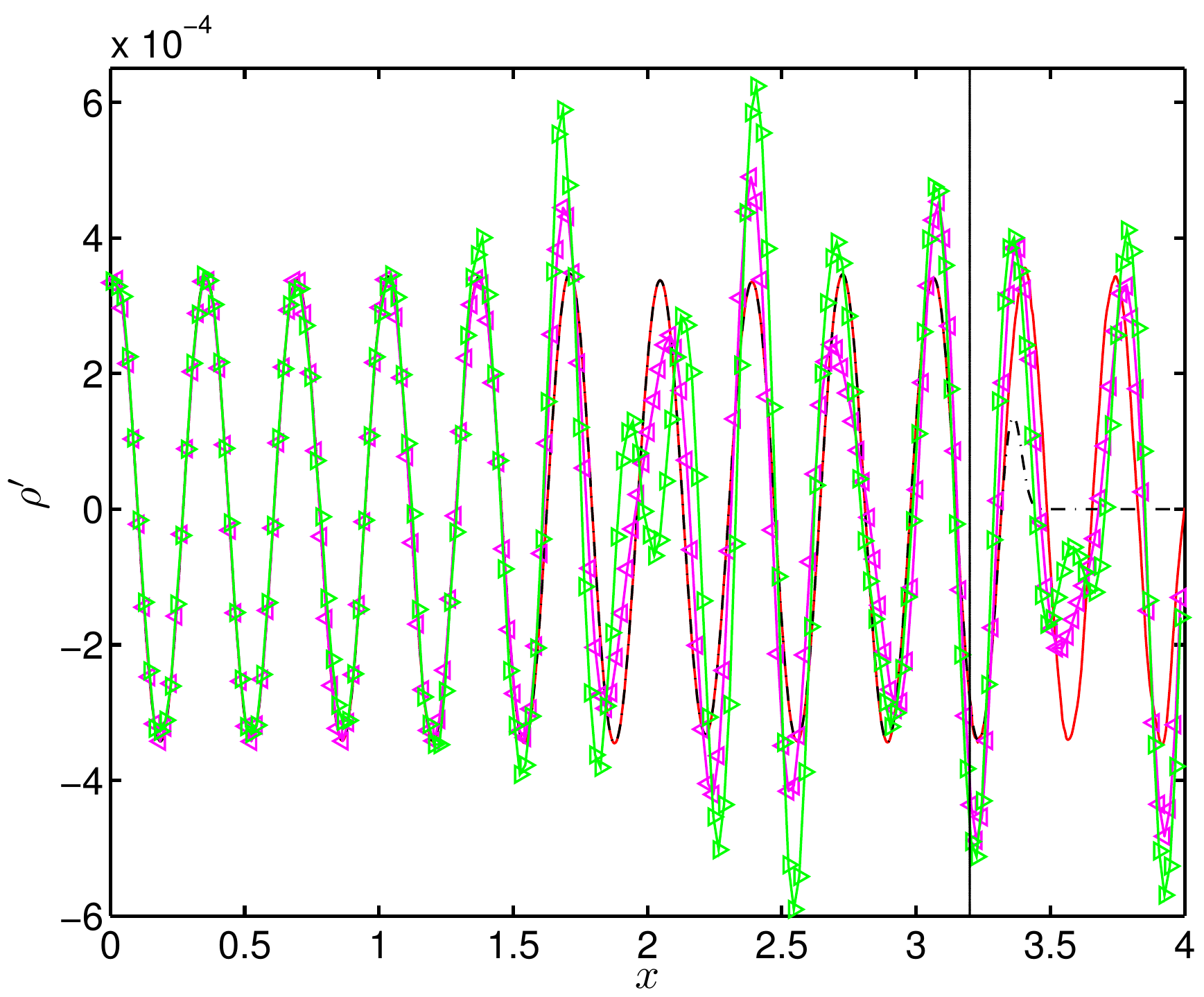}}
\caption{Comparisons among three different strategies for mean flow ($\rmu=(u_x^m,u_y^m)=(0.1,0)$): (Red solid line) Reference $\rho^\prime$; (Dashed-dot line)  $\rho^\prime$ by LBS with the optimal absorbing strategy (\ref{app:newbgka2}); ($-\lhd-$) LBS coupled with viscosity damping strategy; ($-\rhd-$) LBS without any absorbing strategies. The sample time $t=T_m+10\cdot T_p$. The vertical line indicates the position of the absorbing layer.}\label{fig:n10}
\end{center}
\end{figure}

In Figs.\ref{fig:n9} and \ref{fig:n10}, the computational results for both  zero and non-zero background flow  are shown. It is  observed that  using the optimal absorbing strategy (\ref{app:newbgka2}), the wave in the absorbing layer are almost completely damped after a short travel distance. 
Although the given thickness of the absorbing layer is equal to 0.8,   the required minimal thickness is smaller than 0.3, i.e. about one wavelength  $\lambda=c_s/2=0.288675$. Therefore, for this kind of problems, the thickness of the absorbing layer can be set to equal to the characteristic wave length or a little larger than the characteristic wave length of the acoustic wave. 


\subsection{Validations of a dipole  advected by the mean flow }

The validation of the optimal absorbing strategy (\ref{app:newbgka2})  for the vortical flows is very important for the turbulent aeroacoustic problems \cite{sagautbook}, because of the significance of the practical applications. Here, we consider the case of a dipole vortex advected by a uniform flow. The  dipole is given by 

\begin{equation}
\slbracket{
\begin{array}{ll}
\displaystyle u_x &=-\frac{1}{2}|\omega_{\rm e}|(y-y_1){\rm exp}\sbracket{\displaystyle -\sbracket{\frac{r_1}{r_0}}^2}+\frac{1}{2}|\omega_{\rm e}|(y-y_2){\rm exp}\sbracket{\displaystyle -\sbracket{\frac{r_2}{r_0}}^2}\\[2mm]
\displaystyle u_y &=-\frac{1}{2}|\omega_{\rm e}|(x-x_1){\rm exp}\sbracket{\displaystyle -\sbracket{\frac{r_1}{r_0}}^2}+\frac{1}{2}|\omega_{\rm e}|(x-x_2){\rm exp}\sbracket{\displaystyle -\sbracket{\frac{r_2}{r_0}}^2}.
\end{array}
}
\end{equation}

The vorticity distribution of the isolated monopoles has the following Gaussian form 

\begin{equation}
\displaystyle\omega_0=\omega_{\rm e}\sbracket{1-\sbracket{\frac{r}{r_0}}^2}{\rm exp}\sbracket{\displaystyle -\sbracket{\frac{r}{r_0}}^2}.
\end{equation}
If we set $\omega_{\rm e}=299.5285375226$, the initial total kinetic energy of the dipolar flow field 
$$
\displaystyle E(0)=\frac{1}{2}\int_{-1}^{1}\int_{-1}^{1}|{\rm u}(\rmx,0)|^2\rmd\rmx=2.
$$

Considering the mean flow, for the LBS, the initial field is set as

\begin{equation}\label{dipole}
\slbracket{
\begin{array}{llrl}
\displaystyle u_x &=&u_{\rm mean}&-0.1\cdot u_{\rm mean}\cdot\frac{1}{2}|\omega_{\rm e}|(y-y_1){\rm exp}\sbracket{\displaystyle -\sbracket{\frac{r_1}{r_0}}^2}+\frac{1}{2}|\omega_{\rm e}|(y-y_2){\rm exp}\sbracket{\displaystyle -\sbracket{\frac{r_2}{r_0}}^2}\\[2mm]
\displaystyle u_y &=& &-0.1\cdot u_{\rm mean}\cdot\frac{1}{2}|\omega_{\rm e}|(x-x_1){\rm exp}\sbracket{\displaystyle -\sbracket{\frac{r_1}{r_0}}^2}+\frac{1}{2}|\omega_{\rm e}|(x-x_2){\rm exp}\sbracket{\displaystyle -\sbracket{\frac{r_2}{r_0}}^2},
\end{array}
}
\end{equation}

where $u_{\rm mean}=0.1$. The computational domain is defined as $\Omega=[-1,1]\times[-1,1]$.    The parameters in Eq.(\ref{dipole}) :are taken equal to $r_0 = 0.1$, $x_1 = 0$, $x_2 = 0$, $y_1 = 0.1$ and $y_2 =-0.1$. The Reynolds number is set equal to  $Re=10^4$ and the computational grid  is equal to $400^2$. 
The computational domain with the absorbing layer is similar with the computational domain indicated by Fig.\ref{fig:abc}. The thickness of the absorbing layer is taken equal to $4r_0$. The initial pressure is obtained by solving the pressure Poisson equation and the initial distribution is initialized by the method given in  \cite{xu2}. The far field boundary condition is used. The following normalized averaged enstrophy will be investigated

\begin{equation}
\mathcal{E}(t)=\frac{1}{\delta t^2|\Omega|}\int_{\Omega}|\omega(\rmx,t)|^2\rmd\rmx.
\end{equation} 
By monitoring $\mathcal{E}(t)$, the absorbing behavior of the proposed optimal absorbing strategy will be validated for vortex flows. In all figures, if considering the physical reference velocity is equal to 1,  the following time scale is used for the LBS computations
\begin{equation}
\tilde{t}=t\cdot u_{\rm mean}.
\end{equation}

\begin{figure}[!hbtp]
\begin{center}
\scalebox{0.35}[0.35]{\includegraphics[angle=0]{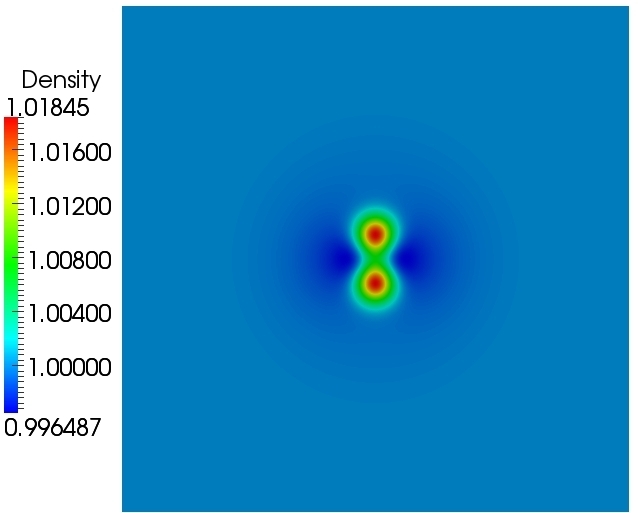}}
\scalebox{0.35}[0.35]{\includegraphics[angle=0]{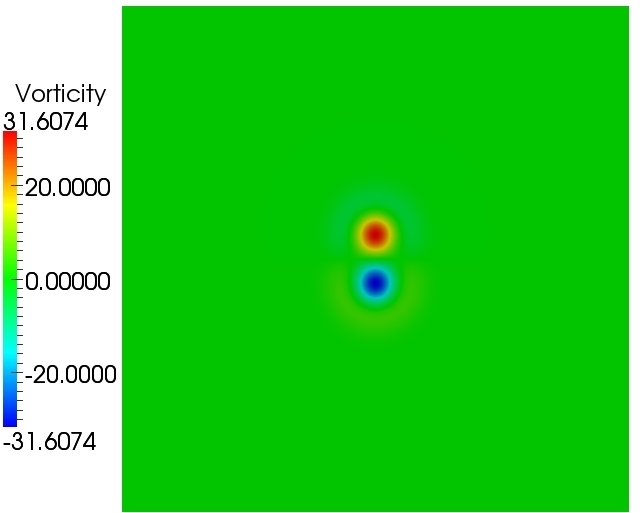}}\\
{\centering \hspace{1cm}(a)Initial density fields \hspace{5cm} (b) Initial vorticity field}
\caption{Initial density field and vorticity field (vorticity is normalized by the time step $\delta t$).}\label{fig:n11}
\end{center}
\end{figure}

\begin{figure}[!hbtp]
\begin{center}
\scalebox{0.35}[0.35]{\includegraphics[angle=0]{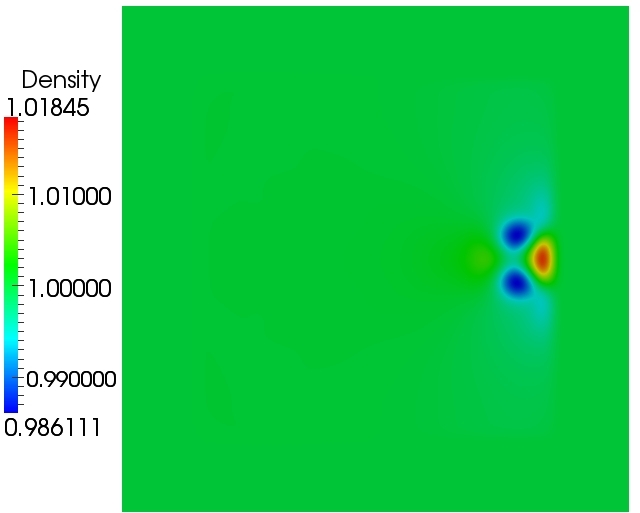}}
\scalebox{0.35}[0.35]{\includegraphics[angle=0]{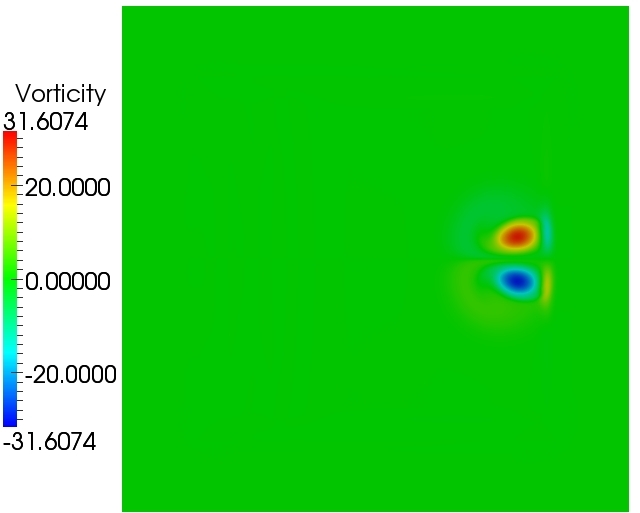}}\\
{\centering \hspace{2cm}(a-1) Density fields at $\tilde{t}=0.5$ \hspace{4cm} (b-1) Vorticity field  at $\tilde{t}=0.5$}\\
\scalebox{0.35}[0.35]{\includegraphics[angle=0]{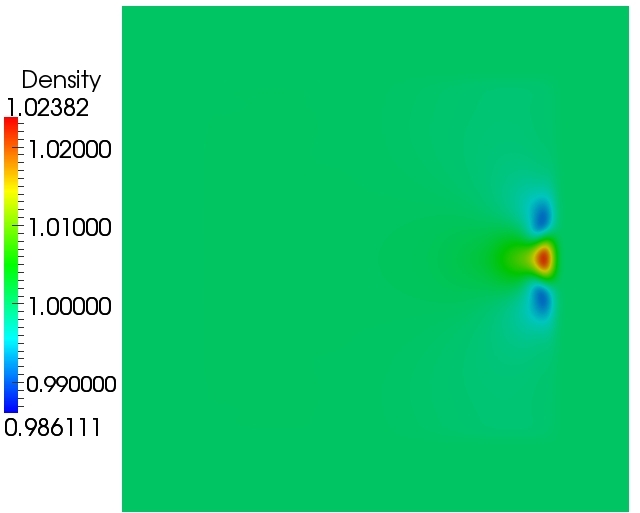}}
\scalebox{0.35}[0.35]{\includegraphics[angle=0]{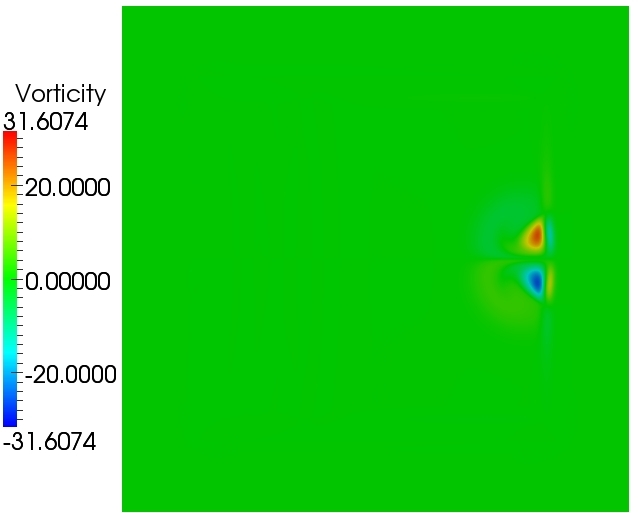}}\\
{\centering \hspace{2cm}(a-2) Density fields at $\tilde{t}=0.6$ \hspace{4cm} (b-2) Vorticity field  at $\tilde{t}=0.6$}
\scalebox{0.35}[0.35]{\includegraphics[angle=0]{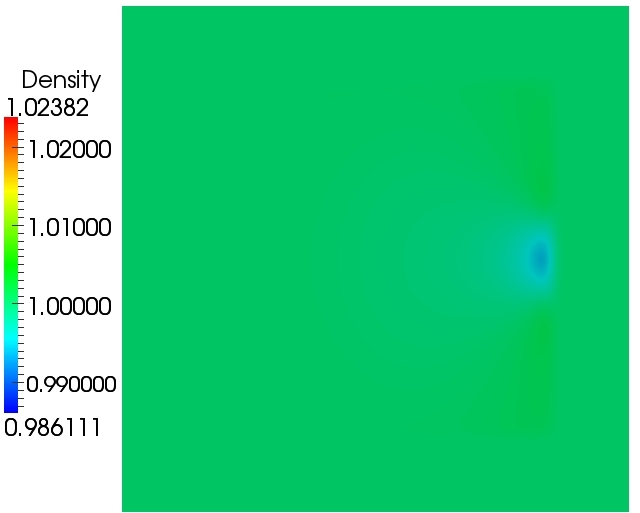}}
\scalebox{0.35}[0.35]{\includegraphics[angle=0]{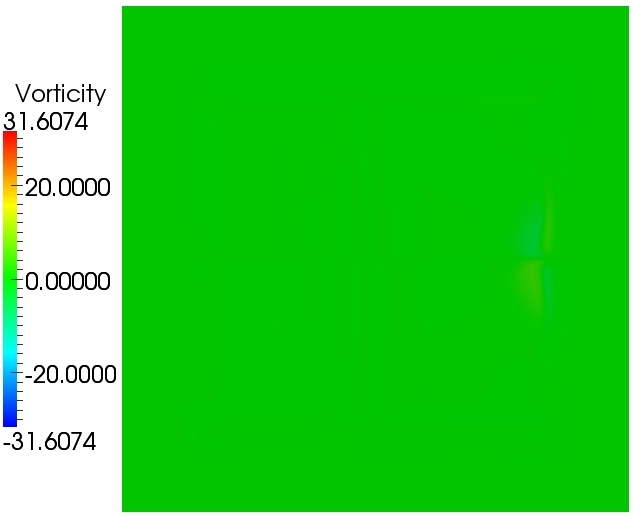}}\\
{\centering \hspace{2cm}(a-3) Density fields at $\tilde{t}=0.8$ \hspace{4cm} (b-3) Vorticity field  at $\tilde{t}=0.8$}
\caption{Evolution of density field and vorticity field (vorticity is normalized by the time step $\delta t$).}\label{fig:n12}
\end{center}
\end{figure}

\begin{figure}[!hbtp]
\begin{center}
\scalebox{0.5}[0.5]{\includegraphics[angle=0]{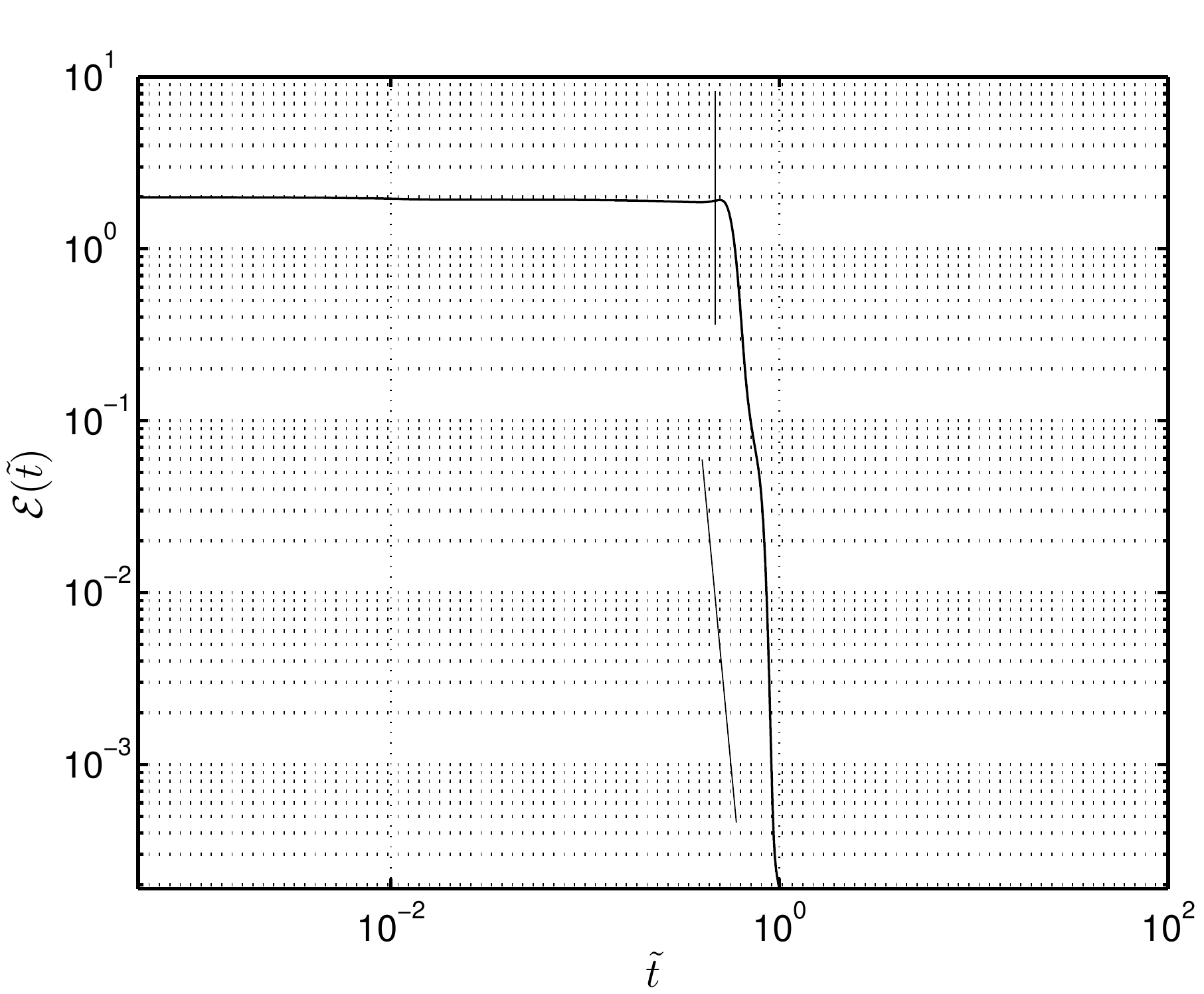}}
\caption{Time dependent normalized enstrophy $\mathcal{E}(t)$. The vertical lines indicate the time at which the dipole moves across into the absorbing layer. The slope of the reference oblique line is equal to -12. }\label{fig:n13}
\end{center}
\end{figure}

 The initial density field and vorticity field are given in Fig.\ref{fig:n11}. In Fig.\ref{fig:n12}, the evolution of the density and vorticity fields is shown with at three different rescaled times. 
 At the first time shown, the dipole is reaching the absorbing layer. At the second time, it is crossing the absorbing layer, while the third snapshot corresponds to a time at which the dipole should have left the full computational domain.
 It is seen that the outlet boundary condition does not radiate any spurious wave.  The efficiency of type II absorbing layer is measured monitoring time evolution of the enstrophy  (see Fig.\ref{fig:n13}). A very fast  decay is observed, with a decay exponent close to -12.

\subsection{Flow past two cylinders}

\begin{figure}[!hbtp]
\begin{center}
\scalebox{0.6}[0.6]{\includegraphics[angle=0]{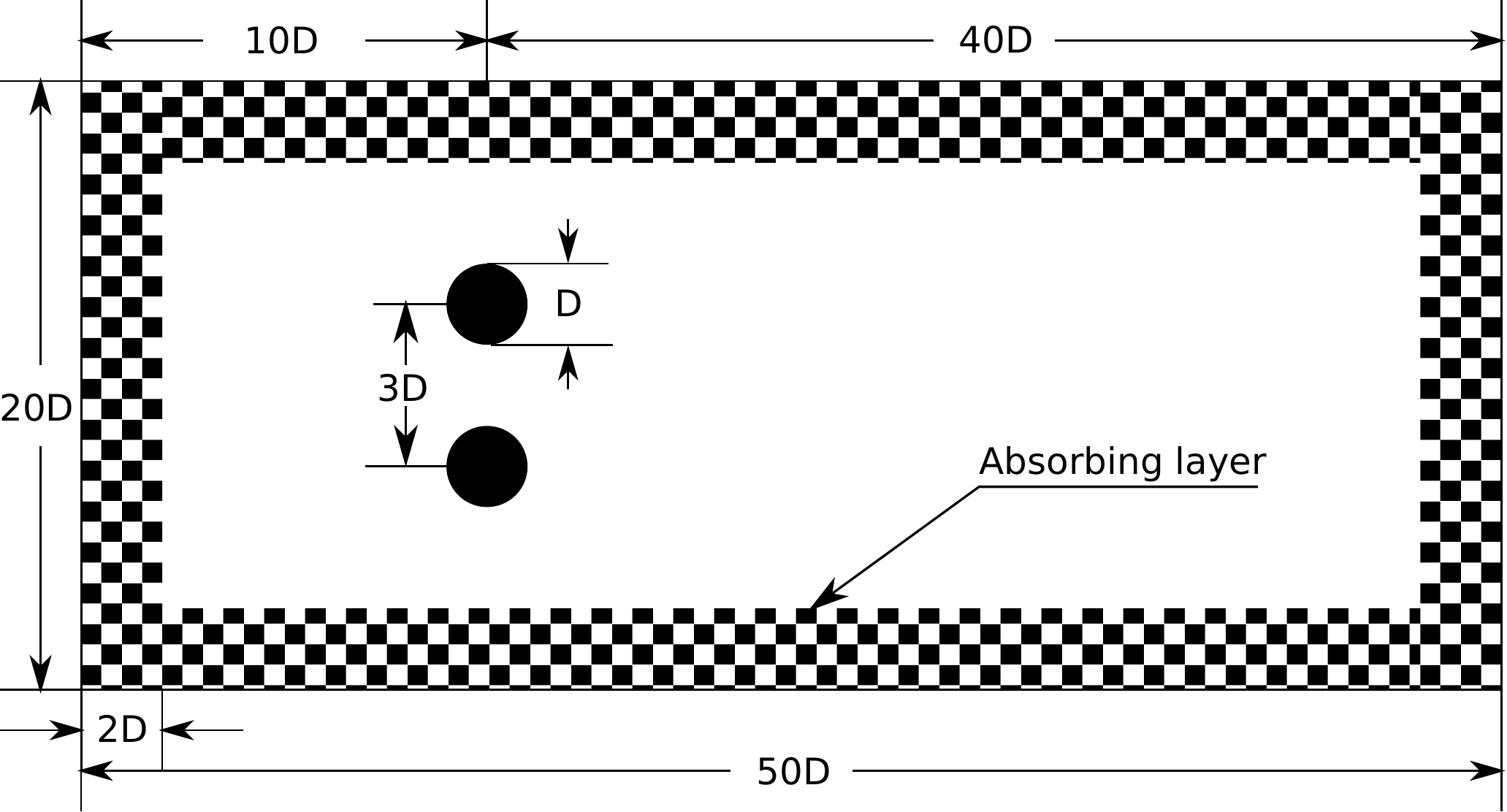}}
\caption{ The schematic computational domain with absorbing layer for the flow past two cylinders.}\label{fig:n14}
\end{center}
\end{figure}

In order to further illustrate the capability of the optimal absorbing layer, the  flow past two  cylinders with equal diameters is considered. The computational domain is drawn schematically in Fig. \ref{fig:n14}.  The centers  of the two cylinders are located at  $(10D,10D-1.5D)$ and $(10D,10D+1.5D)$ where $D=1$ is the diameter of the cylinders. The thickness of the absorbing layer is taken  equal to $2D$.  The Reynolds number is $Re=U^fD/\nu=5000$ ($U^f$ denotes the field field mean velocity magnitude). The Mach number  is $Ma=0.0714286$. The mesh resolution is equal to $D/100$. For convenience, the following dimensionless time  scale is introduced

\begin{equation}
\tilde{t}=t\cdot U^f.
\end{equation}

\begin{figure}[!hbtp]
\begin{center}
\scalebox{0.32}[0.32]{\includegraphics[angle=0]{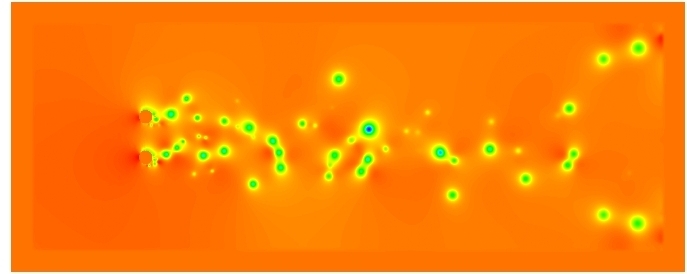}}
\scalebox{0.32}[0.32]{\includegraphics[angle=0]{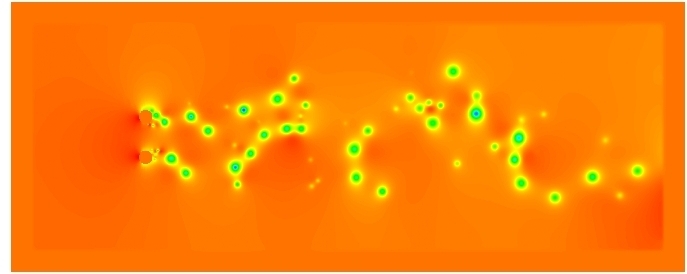}}\\
(a) $\tilde{t}=49.5$ \hspace{5cm}(b) $\tilde{t}=57.784$\\
\scalebox{0.32}[0.32]{\includegraphics[angle=0]{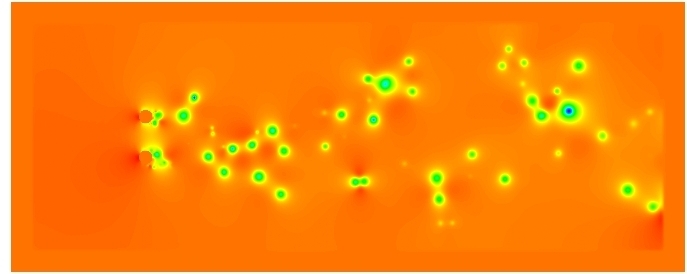}}\\
(c) $\tilde{t}=66.067$\\
\caption{ The snapshots of the density fields.}\label{fig:n15}
\end{center}
\end{figure}

\begin{figure}[!hbtp]
\begin{center}
\scalebox{0.32}[0.32]{\includegraphics[angle=0]{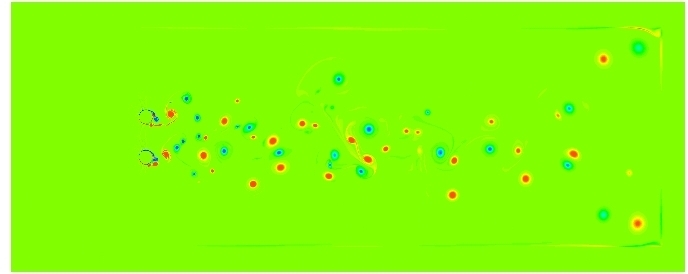}}
\scalebox{0.32}[0.32]{\includegraphics[angle=0]{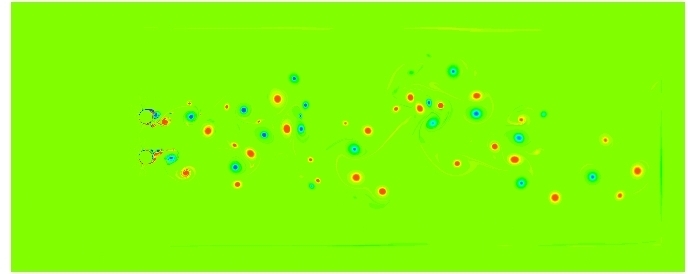}} \\(a) $\tilde{t}=49.5$ \hspace{5cm}(b) $\tilde{t}=57.784$\\
\scalebox{0.32}[0.32]{\includegraphics[angle=0]{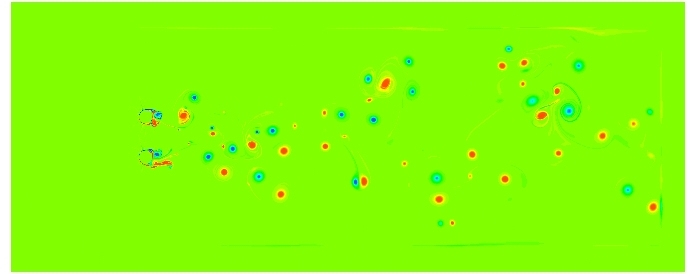}}\\
(c) $\tilde{t}=66.067$\\
\caption{ The snapshots of the dimensionless vorticity fields}\label{fig:n17}
\end{center}
\end{figure}

 The reference field velocity ${\rm u}^f=(U^f,0)$ and the reference  density $\rho^f=1$. The three snapshots of the flow fields (density, dimensionless velocity magnitude and dimensionless vorticity) are given in Figs.\ref{fig:n15}$\sim$\ref{fig:n17} for $\tilde{t}=49.5$, $57.784$ and $66.067$, respectively. The very satisfactory quality of the results show that the proposed type II absorbing layer can be used to handl complex vortical flows.

\begin{figure}[!hbtp]
\begin{center}
\scalebox{0.32}[0.32]{\includegraphics[angle=0]{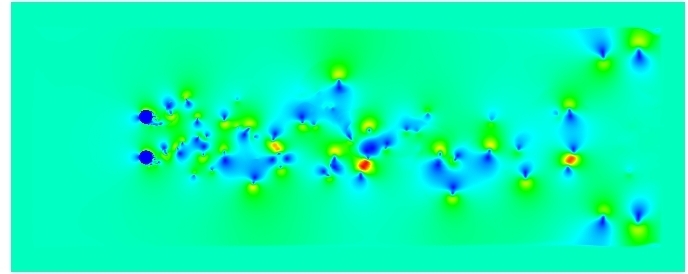}}
\scalebox{0.32}[0.32]{\includegraphics[angle=0]{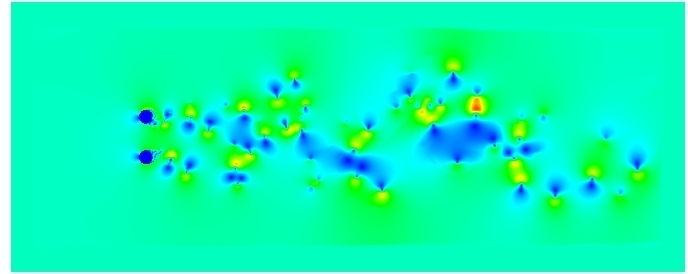}}\\
(a) $\tilde{t}=49.5$ \hspace{5cm}(b) $\tilde{t}=57.784$\\
\scalebox{0.32}[0.32]{\includegraphics[angle=0]{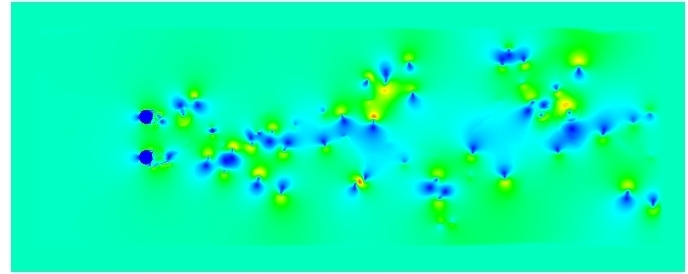}}\\
(c) $\tilde{t}=66.067$\\
\caption{ The snapshots of the dimensionless velocity magnitude fields}\label{fig:n16}
\end{center}
\end{figure}

\section{Conclusion}

A new absorbing layer technique for the definition of non-reflecting boundary conditions for Lattice-Boltzmann methods has been proposed. 
Thank to an original theoretical analysis of the associated linearized problem, dispersive and dissipative features of the method have been analyzed.
An interesting result is that the associated sponge layer equation for the associated macroscopic problem is more complicated that sponge layer equations directly defined on macroscopic quantities. By recovering the corresponding macroscopic equations, the macroscopic effect of the absorbing terms is identified and the critical absorbing strength value is  formulated. Performing  mathematical analysis in the spectral space, the influence of critical parameters are investigated. From this analysis, an optimal absorbing strategy is obtained. By simulating two basic acoustic problems, the optimal absorbing strategy is validated and the corresponding decay exponent is obtained. Numerical results show that the optimal absorbing strategy is also powerful for damping vortices. A very high damping exponent  for the enstrophy is observed. The results demonstrate that the optimal absorbing strategy Type II  is very effective for coping with vortical flows.

\section*{Acknowledgement}
This work was supported by the FUI project LaBS (Lattice Boltzmann Solver, \url{http://www.labs-project.org}).  


\appendix

\section{Transformation matrices}\label{app:t}
\subsection{For two dimensional MRT-LBM with 9 discrete velocities}\label{App:d2q9}
The transformation matrice for MRT-LBM with 9 discrete velocities is described by
\begin{equation}
M=\left[ \begin {array}{rrrrrrrrr} 1&1&1&1&1&1&1&1&1\\\noalign{\medskip}0&1&0&-1&0&1&-1&-1&1\\\noalign{\medskip}0&0&1&0&-1&1&1&-1&-1\\\noalign{\medskip}-4&-1&-1&-1&-1&2&2&2&2\\\noalign{\medskip}4&-2&-2&-2&-2&1&1&1&1\\\noalign{\medskip}0&-2&0&2&0&1&-1&-1&1\\\noalign{\medskip}0&0&-2&0&2&1&1&-1&-1\\\noalign{\medskip}0&1&-1&1&-1&0&0&0&0\\\noalign{\medskip}0&0&0&0&0&1&-1&1&-1\end {array} \right] .
\end{equation}

\subsection{For three dimensional MRT-LBM with 15 discrete velocities}\label{App:d3q15}
The transformation matrice for MRT-LBM with 15 discrete velocities is described by
\begin{equation}
M= \left[ \begin {array}{rrrrrrrrrrrrrrr} 1&1&1&1&1&1&1&1&1&1&1&1&1&1&1\\\noalign{\medskip}0&1&-1&0&0&0&0&1&-1&1&-1&1&-1&1&-1\\\noalign{\medskip}0&0&0&1&-1&0&0&1&1&-1&-1&1&1&-1&-1\\\noalign{\medskip}0&0&0&0&0&1&-1&1&1&1&1&-1&-1&-1&-1\\\noalign{\medskip}-2&-1&-1&-1&-1&-1&-1&1&1&1&1&1&1&1&1\\\noalign{\medskip}16&-4&-4&-4&-4&-4&-4&1&1&1&1&1&1&1&1\\\noalign{\medskip}0&-4&4&0&0&0&0&1&-1&1&-1&1&-1&1&-1\\\noalign{\medskip}0&0&0&-4&4&0&0&1&1&-1&-1&1&1&-1&-1\\\noalign{\medskip}0&0&0&0&0&-4&4&1&1&1&1&-1&-1&-1&-1\\\noalign{\medskip}0&2&2&-1&-1&-1&-1&0&0&0&0&0&0&0&0\\\noalign{\medskip}0&0&0&1&1&-1&-1&0&0&0&0&0&0&0&0\\\noalign{\medskip}0&0&0&0&0&0&0&1&-1&-1&1&1&-1&-1&1\\\noalign{\medskip}0&0&0&0&0&0&0&1&1&-1&-1&-1&-1&1&1\\\noalign{\medskip}0&0&0&0&0&0&0&1&-1&1&-1&-1&1&-1&1\\\noalign{\medskip}0&0&0&0&0&0&0&1&-1&-1&1&-1&1&1&-1\end {array} \right].
\end{equation}

\section{The macroscopic equations with the absorbing layer for the BGK model}\label{app:1}
\subsection{The classical absorbing layer}\label{app:1.1}
\noindent The BGK model with the absorbing layer is given by ($0\leq i,j\leq N$)
\begin{equation}\label{app:bgk}
f_i(\rmx+v_i\delta t,t+\delta
t)=f_i(\rmx,t)+s\left(f_j^{\rm(eq)}(\rho^*(\rmx,t),{\rmu^*(\rmx,t)},t)-f_j(\rmx,t)\right)+\delta t\chi\left(f_j^{\rm(eq)}(\rho^{f}(\rmx,t),{\rmu^f(\rmx,t)},t)-f_j(\rmx,t)\right),
\end{equation}
where $\rho^*(\rmx,t)$ and ${\rmu^*(\rmx,t)}$ are the equilibrium density and velocity, respectively.
Eq. (\ref{app:bgk}) can be rewritten as follows
\begin{equation}\label{app:newbgk}
f_i(\rmx+v_i\delta t,t+\delta
t)=f_i(\rmx,t)+s^\prime\left(f_i^{\rm(eq)}(\rho^*(\rmx,t),{\rmu^*(\rmx,t)},t)-f_i(\rmx,t)\right)+\delta t F_i(\rho^{f}(\rmx,t),{\rmu^f(\rmx,t)},\rho^*,{\rmu^*(\rmx,t)},t),
\end{equation}
where $s^\prime=s+\delta t\chi$ and $F_i({\rmu_f(\rmx,t)},{\rmu^*(\rmx,t)},t)$ is defined by
\begin{equation}
F_i(\rho^{f}(\rmx,t),{\rmu^f(\rmx,t)},\rho^*(\rmx,t),{\rmu^*(\rmx,t)},t)=\chi\left(f_i^{\rm(eq)}(\rho^{f}(\rmx,t),{\rmu^f(\rmx,t)},t)-f_i^{\rm(eq)}(\rho^*(\rmx,t),{\rmu^*(\rmx,t)},t)\right).
\end{equation}
Eq. (\ref{app:newbgk}) can be regarded as the new LBS with a special external force term. In order to obtain the macroscopic equation corresponding to Eq. (\ref{app:newbgk}),  the classical Chapman-Enskog derivation \cite{Guo,Buick2000} is used. The following expansions are introduced \cite{Buick2000} 
\begin{equation}
f_i=f_i^{(0)}+\epsilon f_i^{(1)}+\epsilon^2 f_i^{(0)}+\ldots,
\end{equation}
\begin{equation}
\partial_t=\epsilon\partial_{t_1}+\epsilon^2\partial_{t_2},\quad \partial_{\alpha}=\epsilon\partial_{0\alpha},\quad F_i= \epsilon F^0_i.
\end{equation}
By the multi-scale expansion, we have
\begin{eqnarray}
&&\epsilon^0 : f_i^{(0)} = f_i^{\rm(eq)}(\rho^*,{\rm u^*(x,t)},t)\label{app:eq1}\\[2mm]
&&\epsilon^1 : D_{1i}f_i^{(0)}-F_i^0 = -s^\prime/\delta t f_i^{(1)}\label{app:eq2}\\[2mm]
&&\epsilon^2 : \partial_{t_2}f_i^{(0)}+D_{1i}f_i^{(1)}+\delta t/2 D_{1i}^2f_i^{(0)}=-s^\prime/\delta t f_i^{(2)},\label{app:eq3}
\end{eqnarray}
where $D_{0i}=\partial_{t_1}+c_{i\alpha}\partial_{\alpha}$. Eq. (\ref{app:eq3}) can be expressed as follows
\begin{equation}
 \partial_{t_2}f_i^{(0)}+D_{1i}\left(f_i^{(1)}-\frac{s^\prime}{2}f_i^{(1)}+\frac{\delta t}{2}F_i^0\right)=-s^\prime/\delta t f_i^{(2)}
\end{equation}
 Here, we define the macroscopic quantities ($\rho^*$, $\rho u^*_{\alpha}$) as follows
\begin{equation}
\rho^*=\sum_{i}f_i+n\delta t\sum_{i}F_i,\quad \rho^*u^*_\alpha=\sum_{i}c_{i\alpha}f_i+m\delta t\sum_{i}c_{i\alpha}F_i.
\end{equation}
It is easy to get the following relation
\begin{equation}
\sum_{i}f_i^{(1)}=-n\delta t\sum_iF_i^0, \quad  \sum_{i}c_{i\alpha}f_i^{(1)}=-m\delta t\sum_ic_{i\alpha}F_i^0.
\end{equation}
So, $f_i^{(k)}$ has the following properties
\begin{equation}
\sum_{i}f_i^{(k)}=0,\ k>1,
\end{equation}
and
\begin{equation}
\sum_{i}c_{i\alpha}f_i^{(1)}=0,\ k>1.
\end{equation}
At the scale $t_1$, we have 
\begin{equation}\label{app:eqf1}
\partial_{t_1}\rho^*+\partial_{0\alpha}(\rho^* u^*_{\alpha}) =\sum_{i}F_i^0+n s^\prime\sum_i F_i^{0}, 
\end{equation}
\begin{equation}\label{app:eqf2}
\partial_{t_1}(\rho^* u^*_{\alpha})+\partial_{0\beta}\left(\pi_{\alpha\beta}^{(1)}\right) =\sum_{i}c_{i\alpha}F_i^0+m s^\prime \sum_{i}c_{i\alpha}F_i^0.
\end{equation}
where the second-order tensor $\pi_{\alpha\beta}^{(1)}$ is defined by
\begin{equation}
\pi_{\alpha\beta}^{(1)}=\rho^* u^*_{\alpha}u^*-_{\beta}+p^*\delta_{\alpha\beta}.
\end{equation}
At the scale $t_2$, we have
\begin{equation}\label{app:eqs1}
\partial_{t_2}\rho^*+\delta t\left(1-\frac{s^\prime}{2}\right)\left(-n\partial_{t_1}\sum_iF_i^0-m\partial_{0\alpha}\sum_{i}c_{i\alpha}F_i^0\right)+\frac{\delta t}{2}\left(\partial_{t_1}\sum_iF_i^0+\partial_{0\alpha}\sum_ic_{i\alpha}F_i^0\right)=0
\end{equation}
\begin{equation}\label{app:eqs2}
\partial_{t_2}(\rho^* u_\alpha^*)+\left(1-\frac{s^\prime}{2}\right)\left(-m\delta t\partial_{t_)}\sum_{i}c_{i\alpha}F_i^0+\partial_{0\beta}\sum_{i,\beta}c_{i\alpha}c_{i\beta}f_i^{(1)}\right)+\frac{\delta t}{2}\left(\partial_{t_0}\sum_{i}c_{i\alpha}F_i^0+\partial_{0\beta}\sum_{i,\beta}c_{i\alpha}c_{i\beta}F_i^0\right)=0
\end{equation}
Now, let us calculate $\sum_{i,\beta}c_{i\alpha}c_{i\beta}f_i^{(1)}$. From Eq. (\ref{app:eq2}), we have
\begin{equation}
-\frac{s^\prime}{\delta t}\sum_ic_{i\alpha}c_{i\beta}f_i^{(1)}=\partial_{t_0}\sum_i c_{i\alpha}c_{i\beta}f_i^{(0)}+\partial_{0\gamma}\sum_{i}c_{i\alpha}c_{i\beta}c_{i\gamma}f_i^{(0)}-\sum_i c_{i\alpha}c_{i\beta}F_i^0.
\end{equation}
Then, we have
\begin{equation}\label{app:eq3}
-\left(1-\frac{s^\prime}{2}\right)\sum_ic_{i\alpha}c_{i\beta}f_i^{(1)}=\nu\rho(\partial_{0\alpha}u^*_{\beta}+\partial_{0\beta}u^*_{\alpha})-\sigma\partial_{0\gamma}(\rho u^*_{\alpha}u^*_{\beta}u^*_{\gamma})-\sigma\sum_{i}c_{i\alpha}c_{i\beta}F_i^0.
\end{equation}
where $\nu^\prime=c_s^2\left(1/s^\prime-1/2\right)\delta t$ and $\sigma=\left({1}/{s^\prime}-{1}/{2}\right)\delta t$.\\
Eq. (\ref{app:eq3}) can be rewritten as follows
\begin{eqnarray}\label{app:eqs4}
\partial_{t_2}(\rho^* u_\alpha^*)-\partial_{0\beta}(\nu^\prime\rho(\partial_{0\alpha}u^*_{\beta}+\partial_{0\beta}u^*_{\alpha}))+\sigma\partial_{0\beta}\partial_{0\gamma}(\rho^* u^*_{\alpha}u^*_{\beta}u^*_{\gamma})+\sigma\partial_{0\beta}\sum_{i}c_{i\alpha}c_{i\beta}F_i^0+ \nonumber\\[2mm] \left(1-\frac{s^\prime}{2}\right)\left(-m\delta t\partial_{t_0}\sum_{i}c_{i\alpha}F_i^0\right)+\frac{\delta t}{2}\left(\partial_{t_0}\sum_{i}c_{i\alpha}F_i^0+\partial_{0\beta}\sum_{i,\beta}c_{i\alpha}c_{i\beta}F_i^0\right)=0.
\end{eqnarray}
Combining Eqs. (\ref{app:eqf1}), (\ref{app:eqf2}), (\ref{app:eqs1}) and (\ref{app:eqs4}), ignoring the term of $O(Ma^3), $we obtain
\begin{equation}\label{app:eqs5}
\partial_{t}\rho^*+\partial_{\alpha}(\rho^* u^*_{\alpha}) =\sum_{i}F_i+n s^\prime\sum_i F_i+\delta t\left(1-\frac{s^\prime}{2}\right)\left(n\partial_{t}\sum_iF_i+m\partial_{\alpha}\sum_{i}c_{i\alpha}F_i\right)-\frac{\delta t}{2}\left(\partial_{t}\sum_iF_i+\partial_{\alpha}\sum_ic_{i\alpha}F_i\right),
\end{equation}
\begin{eqnarray}\label{app:eqs6}
\partial_{t}(\rho^* u^*_{\alpha})-\partial_{\beta}(\nu^\prime\rho(\partial_{\alpha}u^*_{\beta}+\partial_{\beta}u^*_{\alpha}))+\partial_{\beta}\left(\rho^* u^*_\alpha u^*_\beta+p^*\delta_{\alpha\beta}\right) =\sum_{i}c_{i\alpha}F_i+m s^\prime \sum_{i}c_{i\alpha}F_i-\sigma\partial_{\beta}\sum_{i}c_{i\alpha}c_{i\beta}F_i+\nonumber\\[2mm]
 \delta t\left(1-\frac{s^\prime}{2}\right)\left(m\partial_{t}\sum_{i}c_{i\alpha}F_i\right)-\frac{\delta t}{2}\left(\partial_{t}\sum_{i}c_{i\alpha}F_i+\partial_{\beta}\sum_{i,\beta}c_{i\alpha}c_{i\beta}F_i\right).
\end{eqnarray}
From the defination of $F_i$, we have
\begin{equation}
\sum_iF_i=\chi(\rho^{f}-\rho^*),\ \sum_{i}c_{i\alpha}F_i=\chi\left(\rho^{f}u^{f}_{\alpha}-\rho^{*}u^{*}_{\alpha}\right),\ \sum_{i}c_{i\alpha}c_{i\beta}F_i=\chi\left(\rho^{f}u^{f}_{\alpha}u^{f}_{\beta}+p^{f}\delta_{\alpha\beta}-\rho^{*}u^{*}_{\alpha}u^{*}_{\beta}-p^{*}\delta_{\alpha\beta}\right).
\end{equation}
So, by Eq. (\ref{app:eqs6}), we have
\begin{equation}\label{app:eqs7}
\partial_{t}\rho^*+\partial_{\alpha}(\rho^* u^*_{\alpha}) =(1+n s^\prime)\chi(\rho^{f}-\rho^*)+\left(n\delta t\left(1-\frac{s^\prime}{2}\right)-\frac{\delta t}{2}\right)\chi\partial_{t}(\rho^{f}-\rho^*)+\chi\left(m\delta t\left(1-\frac{s^\prime}{2}\right)-\frac{\delta t}{2}\right)\partial_{\alpha}\left(\rho^{f}u^{f}_{\alpha}-\rho^{*}u^{*}_{\alpha}\right),
\end{equation}
\begin{eqnarray}\label{app:eqs8}
\partial_{t}(\rho^* u^*_{\alpha})-\partial_{\beta}(\nu^{\prime}\rho^*(\partial^*_{\alpha}u^*_{\beta}+\partial_{\beta}u^*_{\alpha}))+\partial_{\beta}\left(\rho^* u^*_\alpha u^*_\beta+p^*\delta_{\alpha\beta}\right) =(1+m s^\prime )\chi\left(\rho^{f}u^{f}_{\alpha}-\rho^{*}u^{*}_{\alpha}\right)+\nonumber\\[2mm]
\left(m\delta t\left(1-\frac{s^\prime}{2}\right)-\frac{\delta t}{2}\right)\chi\partial_{t}\left(\rho^{f}u^{f}_{\alpha}-\rho^{*}u^{*}_{\alpha}\right)-\left(\sigma+\frac{\delta t}{2}\right)\chi\partial_{\beta}\left(\rho^{ f}u^{ f}_{\alpha}u^{f}_{\beta}+p^{f}\delta_{\alpha\beta}-\rho^{*}u^{*}_{\alpha}u^{*}_{\beta}-p^{*}\delta_{\alpha\beta}\right).
\end{eqnarray}
Let $\zeta_n$ and $\zeta_m$ defined by
\begin{equation}
\zeta_n = n\delta t\left(1-\frac{s^\prime}{2}\right)-\frac{\delta t}{2},\ \zeta_m = m\delta t\left(1-\frac{s^\prime}{2}\right)-\frac{\delta t}{2}.
\end{equation}
Eqs. (\ref{app:eqs7}) and (\ref{app:eqs8}) are rewritten as

\begin{eqnarray}
\label{app:eqs9}
\partial_{t}\rho^*+\partial_{\alpha}(\rho^* u^*_{\alpha}) =(1+n s^\prime)\chi(\rho^{f}-\rho^*)+\zeta_n\chi\partial_{t}(\rho^{f}-\rho^*)+\zeta_m\chi\partial_{\alpha}\left(\rho^{f}u^{f}_{\alpha}-\rho^{*}u^{*}_{\alpha}\right),\\[2mm] \label{app:eqs10}
\partial_{t}(\rho^* u^*_{\alpha})-\partial_{\beta}(\nu^\prime\rho^*(\partial_{\alpha}u^*_{\beta}+\partial_{\beta}u^*_{\alpha}))+\partial_{\beta}\left(\rho^* u^*_\alpha u^*_\beta+p^*\delta_{\alpha\beta}\right) =(1+m s^\prime )\chi\left(\rho^{f}u^{f}_{\alpha}-\rho^{*}u^{*}_{\alpha}\right)+\nonumber\\[2mm]
\zeta_m\chi\partial_{t}\left(\rho^{f}u^{f}_{\alpha}-\rho^{*}u^{*}_{\alpha}\right)-\left(\sigma+\frac{\delta t}{2}\right)\chi\partial_{\beta}\left(\rho^{f}u^{f}_{\alpha}u^{f}_{\beta}+p^{f}\delta_{\alpha\beta}-\rho^{*}u^{*}_{\alpha}u^{*}_{\beta}-p^{*}\delta_{\alpha\beta}\right).
\end{eqnarray}

\subsection{The absorbing layer based on the equilibrium distribution functions for the BGK model}\label{app:1.2}
\noindent The BGK model with new damping terms is given by ($0\leq i,j\leq N$)
\begin{equation}\label{app:bgk_new}
f_i(\rmx+v_i\delta t,t+\delta
t)=f_i(\rmx,t)+s\left(f_j^{\rm(eq)}(\rho^*,{\rmu^*(\rmx,t)},t)-f_j(\rmx,t)\right)+\delta t F_i\left({\rmu_f(\rmx,t)},{\rmu^*(\rmx,t)},t)\right),
\end{equation}
where ${\rm u^*(x,t)}$ is the equilibrium velocity.
 $F^{\rm(eq)}_i({\rmu_f(\rmx,t)},{\rmu^*(\rmx,t)},t)$ is defined by
\begin{equation}
F^{\rm(eq)}_i(\rho^{f},{\rm u^f(x,t)},\rho^*,{\rm u^*(x,t)},t)=\chi\left(f_i^{\rm(eq)}(\rho^{f},{\rm u^f(x,t)},t)-f_i^{\rm(eq)}(\rho^*,{\rm u^*(x,t)},t)\right).
\end{equation}
From Eq. (\ref{app:bgk_new}), the obtained macro equations are similary to Eqs. (\ref{app:eqs9})-(\ref{app:eqs10}). The parameters $\nu$, $\zeta_n$ and $\zeta_m$ are defined as follows
\begin{equation}
s^\prime=s,\ \nu = c_s^2\left(\frac{1}{s}-\frac{1}{2}\right)\delta t, \ \zeta_n = n\delta t\left(1-\frac{s}{2}\right)-\frac{\delta t}{2},\ \zeta_m = m\delta t\left(1-\frac{s}{2}\right)-\frac{\delta t}{2}.
\end{equation}
\subsection{The absorbing layer based on linear damping terms}\label{app:1.3}
In \ref{app:1.1}, the absorbing terms are specified based on the equilibrium distribution functions. From Eq. (\ref{app:eqs10}), it is known the obtained damping terms in the recovered Navier-Stokes equations involves a nonlinear damping term (the second order terms of the velocity ${\rm u}^{f}$ and ${\rm u}^{*}$)  in momentum equations. The following model is propsed for  elinimating this term
\begin{equation}\label{app:bgk_no2ordor}
f_i(\rmx+v_i\delta t,t+\delta
t)=f_i(\rmx,t)+s\left(f_j^{\rm(eq)}(\rho^*,{\rmu^*(\rmx,t)},t)-f_j(\rmx,t)\right)+\delta t F^{\rm (L,\ eq)}_i\left({\rmu^f(\rmx,t)},{\rmu^*(\rmx,t)},t)\right),
\end{equation}
where ${\rm u^*(x,t)}$ is the equilibrium velocity.
 $F^{\rm(L,\ eq)}_i({\rmu_f(\rmx,t)},{\rmu^*(\rmx,t)},t)$ is defined by
\begin{equation}
F^{\rm(L,\ eq)}_i(\rho^{f},{\rm u^f(x,t)},\rho^*,{\rm u^*(x,t)},t)=\chi\left(f_i^{\rm(L,\ eq)}(\rho^{f},{\rm u^f(x,t)},t)-f_i^{\rm(L,\ eq)}(\rho^*,{\rm u^*(x,t)},t)\right).
\end{equation}
Then, along the routine in  \ref{app:1.1}, the following macroscopic equations are obtained
\begin{equation}\label{app:eqs11}
\begin{array}{c}
\partial_{t}\rho^*+\partial_{\alpha}(\rho^* u^*_{\alpha}) =(1+n s^\prime)\chi(\rho^{f}-\rho^*)+\zeta_n\chi\partial_{t}(\rho^{f}-\rho^*)+\zeta_m\chi\partial_{\alpha}\left(\rho^{f}u^{f}_{\alpha}-\rho^{*}u^{*}_{\alpha}\right),\\[2mm]
\partial_{t}(\rho^* u^*_{\alpha})-\partial_{\beta}(\nu\rho^*(\partial_{\alpha}u^*_{\beta}+\partial_{\beta}u^*_{\alpha}))+\partial_{\beta}\left(\rho^* u^*_\alpha u^*_\beta+p^*\delta_{\alpha\beta}\right) =(1+m s^\prime )\chi\left(\rho^{f}u^{f}_{\alpha}-\rho^{*}u^{*}_{\alpha}\right)+
\zeta_m\chi\partial_{t}\left(\rho^{f}u^{f}_{\alpha}-\rho^{*}u^{*}_{\alpha}\right),
\end{array}
\end{equation}
where the parameters $\nu$, $\zeta_n$ and $\zeta_m$ are defined as follows
\begin{equation}
s^\prime=s,\ \nu = c_s^2\left(\frac{1}{s}-\frac{1}{2}\right)\delta t, \ \zeta_n = n\delta t\left(1-\frac{s}{2}\right)-\frac{\delta t}{2},\ \zeta_m = m\delta t\left(1-\frac{s}{2}\right)-\frac{\delta t}{2}.
\end{equation}




\bibliographystyle{elsarticle-num}
\bibliography{<your-bib-database>}



\end{document}